\documentclass[final,authoryear,5p,times,twocolumn]{elsarticle}

\usepackage{graphicx}
\usepackage{amssymb,amsmath}
\usepackage{multirow}
\usepackage{pdflscape}
\usepackage{subcaption}
\usepackage{longtable}
\usepackage{fancyhdr}
\usepackage{array}
\usepackage[page]{appendix}

\makeatletter
\newcommand{\thickhline}{%
    \noalign {\ifnum 0=`}\fi \hrule height 1pt
    \futurelet \reserved@a \@xhline
}
\newcolumntype{"}{@{\hskip\tabcolsep\vrule width 1pt\hskip\tabcolsep}}
\makeatother

\newcommand{\beginsupplement}{%
        \setcounter{table}{0}
        \renewcommand{\thetable}{S\arabic{table}}%
        \setcounter{figure}{0}
        \renewcommand{\thefigure}{S\arabic{figure}}%
     }

\begin{document}

\begin{frontmatter}



\title{Dynamical Evolution of the Earth-Moon Progenitors - Whence Theia?}


\author[quarles]{Billy L. Quarles}
\author[quarles]{Jack J. Lissauer}

\address[quarles]{Space Science and Astrobiology Division MS 245-3, NASA Ames Research Center, 
				Moffett Field, CA 94035 (U.S.A.)}

%% This copyright statement isn't required at any stage by the Icarus
%% Editorial Office or Elsevier.  However, until you sign over the
%% copyright to Elsevier prior to publication (or negotiate with them
%% about copyright), you own the copyright to anything you create.
%% Just to keep things unambiguous, always include a copyright statement
%% or explicitly dedicate your work to the public domain.

%% ----- ELSEVIER STUFF -----
%% The commands below up to the \end{frontmatter} are commented out
%% so that we can do some Icarus-required formatting on the second and
%% third pages that is not required later on by Elsevier.  So when
%% your paper gets accepted, and you are ready to start dealing with
%% Elsevier, copy your abstract and keywords up here, uncomment these
%% lines, and comment out the ICARUS STUFF below.
%% 
%% Alternately, you might just want to move these abstract, keyword,
%% and end frontmatter commands down, and comment out the ICARUS STUFF
%% commands.  It doesn't matter.

\begin{abstract}
We present integrations of a model Solar System with five terrestrial planets {(beginning $\sim$$30-50$ Myr after the formation of primitive Solar System bodies)} in order to determine the preferred regions of parameter space leading to a giant impact that resulted in the formation of the Moon.  {Our results indicate which choices of semimajor axes and eccentricities for Theia (the proto-Moon) at this epoch can produce a late Giant Impact, assuming that Mercury, Venus, and Mars are near the current orbits.}  We find that the likely semimajor axis of Theia, at the epoch when our simulations begin, depends on the assumed mass ratio of Earth-Moon progenitors (8/1, 4/1, or 1/1).  The low eccentricities of the terrestrial planets are most commonly produced when the progenitors have similar semimajor axes at the epoch when our integrations commence.  Additionally, we show that mean motion resonances among the terrestrial planets and perturbations from the giant planets can affect the dynamical evolution of the system leading to a late Giant Impact. 
\end{abstract}

% %% Keywords should appear after the abstract. 
\begin{keyword}
Moon; Planetary dynamics; Origin, Solar System; resonances, orbital
\end{keyword}

%% ----- END ELSEVIER STUFF -----

\end{frontmatter}

%% ----- ICARUS STUFF -----
%% Some formatting on the first, second, and third pages are required
%% by the Icarus Editorial Office that are not required by Elsevier.
%% This section contains those things.  When you are ready to transition
%% to ``Elsevier'' mode, copy your abstract and keywords up into
%% the ELSEVIER STUFF section, and then you can just delete everything
%% in this section.

%% We need to list the number of manuscript pages, figures, and tables. 
%%
%% Rather than manually count these things out, we'll use a little
%% trick here from Paul.  All you have to do is place three \label{}
%% tags on the last page, the last table, and the last figure, that
%% way these values are automatically updated (as long as you remember
%% to move the lasttable and lastfig labels when you add or remove
%% tables and figures).

\begin{flushleft}
\vspace{1cm}
Number of pages: \pageref{lastpage} \\
Number of tables: \ref{lasttable}\\
Number of figures: \ref{lastfig}\\
\end{flushleft}

%% Don't worry about finding the various last* tags and deleting them
%% when you go to ``Elsevier'' mode if you don't want to, they should be
%% silently ignored.

% %% Keywords should appear after the abstract. 
\begin{keyword}
Moon; Planetary dynamics; Origin, Solar System; resonances, orbital
\end{keyword}

\lfoot{\scriptsize
Copyright \copyright\ 2014 Billy L. Quarles and Jack J. Lissauer}

%% ----- END ICARUS STUFF -----

%main text
\section{Introduction}
Significant effort has been placed in determining the origins of the bodies within our Solar System.  One of the most perplexing areas of study is the formation of Earth's moon, and more generally, of the Earth-Moon system.  Several theories have been explored, including five scenarios that have garnered serious study by the scientific community over the past few decades.  These scenarios include a fission wherein the Moon split from a rapidly rotating Earth, co-accretion of the Earth and Moon as a binary pair, capture of the Moon as a renegade planet, precipitation of the Moon from the Earth caused by a intense bombardment of small planetesimals, and a Giant Impact resulting from a collision of a Mars-sized or larger object with the Earth.

The reigning explanation is that the Moon comes from a Giant Impact on the Earth from a Mars-sized \citep{Hartmann1975,Cameron1976} or larger object \citep{Cameron1997,Cameron2000,Canup2012}, although a smaller impactor may also be possible \citep{Cuk2012}.  This theory rises to the top as it provides a sufficient explanation to many characteristics of the Earth-Moon system, most notably the amount of angular momentum residing in their mutual orbit and in the Earth's rotation, differences in mean densities of the two bodies together with compositional similarities between the Moon and the Earth's mantle \citep[cf.][]{Herwartz2014}, and variations in comparative radioisotopic ratios that all suggest a formation during the late stage of planetary accretion.  During this late stage, it is very likely that the terrestrial region was fairly clear of large objects based upon numerical models of the duration of terrestrial planet growth \citep{Chambers2013}.  \cite{Chambers2013} demonstrated that $3-5$ terrestrial planets could have formed in the Solar System based on a new framework considering the effects of fragmentation and hit-and-run collisions.  Specifically, \citeauthor{Chambers2013} shows that a 5 terrestrial planet system can persist through a full planetary growth simulation \cite[Figure 3 of][]{Chambers2013}.  Other works \citep{Jacobson2014a,Izidoro2013,Walsh2011,Brasser2011,Chambers2007,OBrien2006} have also shown that the number of terrestrial planets possible is consistent with the $3-5$ estimate. In the case of a 5 planet model, the extra planet could have formed between the orbits of present day Venus and Mars.

Theories on the details of the Giant Impact hypothesis continue to be innovated and investigated further.  Recent scenarios include: a hit-and-run scenario wherein a $30^\circ - 40^\circ$ collision angle is preferred \citep{Reufer2012}, variations on the angular momentum of the Earth-Moon system following the impact \citep{Cuk2012}, and variations upon the scaled impact parameter \citep{Canup2012}.  The newest scenarios \citep{Cuk2012,Canup2012} invoke special conditions that allow for a Moon-forming impact, but the conditions to arrive at these scenarios may prove constraining.  \cite{Cuk2012} requires that the proto-Earth be nearly formed ($\sim$0.99 M$_\oplus$) and spinning at a rate near the breakup threshold to allow a smaller projectile to produce the protolunar disk.  The alternate scenario proposed by \cite{Canup2012} invokes a collision between similar-sized progenitors and requires that the impact angle to be less oblique than previously indicated.  

Other previous inquiries \citep{Wetherill1986,Chambers1998,Chambers2001} suggest that planetary accretion is largely completed in a few tens of millions of years, with the early heavy bombardment lasting about 100 Myr.  The effects of giant impacts are largely stochastic and typically produce a large rotational angular momentum \citep[and references therein]{Safronov1966,Lissauer1991,Lissauer2000}.  Terrestrial planet formation and the consequences of large impacts have been active areas of inquiry that have produced interesting and ingenious solutions to specific problems \citep{Agnor1999,Kokubo2006,Kokubo2007,Kokubo2010,Raymond2006,Raymond2009,Morishima2008,Morishima2010,Hansen2009,Elser2011,Walsh2011}.  Early simulations with a SPH (smooth particle hydrodynamics) code to characterize the impact suggested a mass ratio of the colliding bodies of 7:3 \citep{Cameron1997,Cameron2000}.  More recent studies using SPH simulations indicate a wider range of impacts could lead to successful Moon forming events \citep{Canup2001,Canup2004,Canup2013}.

Several studies based upon radiogenic dating \citep{Brandon2007,Halliday2008,Borg2011,Bottke2014,Jacobson2014a} suggest that the Moon-forming impact was late in the accretionary sequence, implying that at least five terrestrial planets persisted for tens of millions of years prior to a collision reducing the number.  The best known observable to constrain the possible solutions is the dating of lunar samples.  We place special emphasis on this constraint as the early estimates of this indicate the age of the lunar melt at $60-120$ Myr \citep{Taylor1975} after the the formation of Calcium Aluminum Inclusions (CAIs) in the Solar System asteroids and updated measurements that obtain an age of $70-110$ Myr \citep{Touboul2007,Brandon2007,Halliday2008,Borg2011}.  {However, other works \citep{Yin2002,Jacobsen2005,Yu2011} argue for a Moon-forming event earlier than 40 Myr.  On the other hand, recent works \citep{Jacobson2014a,Bottke2014}, which coupled dynamical simulations with geochemical constraints and impact age distributions on meteorites, concluded that the Moon formed $70-130$ Myr after the CAIs.}  

Terrestrial planet formation simulations through the growth of planetesimals \citep{Chambers2001,Chambers2013} show that most planetary embryos are cleared in $30-50$ Myr after the CAIs, typically leaving of $3-5$ terrestrial bodies surviving.  Radiometric dating of the Earth using $^{182}$Hf-$^{182}$W suggests the bulk Earth to have formed $\sim$$30-50$ Myr after the CAIs \citep{Kleine2009}.  {While it is possible that more than five terrestrial planetary embryos were present during this time, dynamical formation simulations show this to be unlikely \citep{Chambers2001,Raymond2006,Chambers2013}.  Simulations also show that a total mass of 0.02 -- 0.2 M$_\oplus$ in (small) planetesimals could be expected $\sim$$30-50$ Myr after CAIs \citep{Jacobson2014b}.  Thus, from all these considerations it is likely that there was a significantly long timespan before the Moon-forming event, during which the inner Solar System contained five planetary bodies and a planetesimal population with a small total mass.}

\begin{figure}[!ht]
\centering
\includegraphics[width = \linewidth]{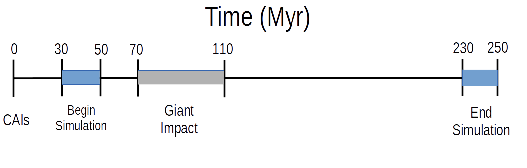}
\caption{Timeline illustrating our windows of interest with respect to the beginning of the Solar System.  Our simulations begin subsequent to the bulk formation of the terrestrial planets, which is indicated to be at $30-50$ Myr.  The time range from $70-110$ Myr represents the timing of the Giant Impact (from other studies), and the window of $20-80$ Myr from the beginning of our simulations corresponds to the full range of allowed times of the bulk formation of proto-Earth and the Giant Impact.}
\label{fig:time}
\end{figure}

Following \cite{Rivera2001,Rivera2002}, we model the late stage formation of the Solar System with five inner terrestrial planets and four outer giant planets whose dynamical evolution leads to a Giant Impact.  Based on the dating of early Solar System events discussed above, we favor simulations that lead to a Giant Impact after $20-80$ Myr have elapsed. This relative time window of $20-80$ Myr considers the maximum range that is consistent with both the estimate of $30-50$ Myr for our starting epoch {(after the inner Solar System is reduced to five planetary bodies and a population of left-over planetesimals of negligible mass)} and the $70-110$ Myr range as the expected timing of the Giant Impact (Figure \ref{fig:time}).

The Solar System epoch that we are considering is subsequent to the dissipation of the gaseous component of the protoplanetary disk (which is estimated to have occurred a few million years after the beginning of planet formation), so we consider neither gas drag nor planetary migration in our simulations.  However, in the Nice model, the Giant Impact occurs prior to the rearrangement of the giant planets induced by interactions with the disk of planetesimals in the Kuiper belt. Therefore, we perform some of our integrations using a configuration of the giant planets commensurate with the Nice model.  Through these considerations, we seek to determine likely masses and orbital properties of the Earth-Moon progenitors at the epoch when our simulations begin.  We outline our methodology in Section 2, present  and interpret our results in Section 3, and provide our conclusions in Section 4.

\section{Methodology}
\subsection{Starting parameters}
\label{sec:ip}
In most of our integrations, the major planets (excluding the Earth-Moon system) begin with orbital elements from a well-defined recent epoch.  Following \cite{Rivera2002}, we use the orbital elements given in Table \ref{tab:orbelem}.  Our Nice model simulations use different parameters for the giant planets.  We make the following assumptions about certain properties of the proto-Earth and proto-Moon:

\begin{enumerate}
	\item The sum of the masses of the proto-Earth and proto-Moon is equal to the present Earth-Moon system.  The sum of angular momenta of the proto-Earth and proto-Moon (primarily in their motions about the Sun) is equal that of the current Earth-Moon system. 
	\item {A relationship of equipartition of orbital excitation energy} exists to describe the eccentricities of the Earth-Moon progenitors.
	\item The proto-Moon originated from the {general} neighborhood of the proto-Earth.  Specifically, in most of our simulations we place the starting orbit of the proto-Moon between the orbit of Venus and slightly exterior to the orbit of Mars.  However, we also present some simulations in which the proto-Moon begins as close to the Sun as 0.44 AU and as distant as 2.18 AU.
\end{enumerate}

%\begin{landscape}
\begin{table*}[!ht]
\footnotesize
\begin{center}
\begin{tabular}{lccccccc}
\hline 
\hline
Planet  & ${\rm JD_{peri}}$	& $\omega$ ($^\circ$)	& $\Omega$ ($^\circ$) & $i$ ($^\circ$)	&    $e$ 			 & $q$ (AU) \\
\hline
Mercury & 2449127.113714  	& 	29.1042601 				& 	48.3388908 			 &  7.0054188	 		& 0.2056278714 &  0.3074995003 \\
Venus		& 2449041.819991  	& 	54.8629740 				& 	76.6956157 			 &  3.3946460	 		& 0.0067943065 &  0.7184249114 \\
Earth		& 2448990.652003		&	 102.4671799				&		 0.4100818			 &  0.0007676			& 0.0166964471 &  0.9833131797 \\
Mars		& 2448759.979478		&  286.4396588				&		49.5749701			 &	1.8503321			& 0.0934143237 &  1.3813660973 \\
Jupiter & 2446987.110797		&  275.2009899				&	 100.4690203			 &  1.3046385			& 0.0482824683 &  4.9518484496 \\
Saturn	& 2452837.155266		&  339.7172216				&  113.6715191			 &	2.4866595			& 0.0535612527 &  9.0171884206 \\
Uranus	& 2439607.450219		&	  99.3953701				&	  74.0292053			 &  0.7723997			& 0.0475786139 & 18.3118586561 \\
Neptune & 2467331.295541		&  266.9801732				&  131.7531606			 &  1.7720163			& 0.0063426402 & 29.9323591806 \\

\hline
\end{tabular}
\caption[Orbital elements for the eight planets in the Solar System at epoch JD 2449101.0]	% The class file doesn't do 
										% anything with the square 
										% bracketed short caption, 
										% but I always put one in.
	{
	Orbital elements in the Solar System at epoch JD 2449101.0.	The parameters presented for the Earth are those of the Earth-Moon barycenter.
	%\label{lasttable}		% notice the second label for counting
	}
	\label{tab:orbelem}
\end{center}
\end{table*}

%\end{landscape}

These assumptions are driven by observational evidence (e.g., dating of Apollo lunar samples and isotopic ratios) and current theories pertaining to the formation of the Solar System.  The most general set of possible parameters is large, and we investigate only a small fraction in order to determine the general trends and processes present.  We use the work of \cite{Rivera2002} to begin our investigation, and we expand his results by considering much larger regions of parameter space for the semimajor axis and eccentricity of the proto-Moon, incorporating updated constraints and techniques that are now available.  

We begin our simulations in the era of late stage formation when the vast majority of {planetesimals} are expected to have already accreted onto the surviving embryos (see Fig. \ref{fig:time}) {and only five planetary bodies are still present in the inner Solar System.  We do not include planetesimals in our simulations, so our results represent starting conditions consistent with the expected lower limit ($\sim$0.02 M$_\oplus$) of summed planetesimal mass indicated by \cite{Jacobson2014b}.}  \emph{We present our collision time relative to the beginning of our integrations, and any comparison of our results with the dating of the Moon should adjust for this difference.}    

Our starting conditions, given in Table \ref{tab:orbelem}, come from \cite{Rivera2002}, where \citeauthor{Rivera2002} obtained the values through private communication from E. Bowell.  We use these values so that we can make a qualitative comparison between our results and those of \cite{Rivera2002}.  Since we use a different numerical integration package, substantially newer software \& hardware (e.g., 64-bit words vs. 32-bit words), and a newer compiler with different optimization options, we don't expect to be able to reproduce exactly the results as given in \cite{Rivera2002}.  However, we should obtain statistically similar results.  We assume that the sum of the initial angular momenta of the Earth and Moon progenitors is equal to that of the Earth-Moon system at the present epoch (in both cases, most of the angular momentum resides in the heliocentric orbit), and we assume an {equipartition of excitation energy relative to circular orbits} for the starting eccentricities, $m_S e_{S}^2 = m_L e_{L}^2$, where the subscript $L$ and $S$ refer to the proto-Earth (Large) and proto-Moon (Small), respectively.  These constraints can uniquely define an starting semimajor axis, $a_L$, given the chosen values of $m_S$, $a_S$, and $e_S$.  Table \ref{tab:IC1} represents an example of the proto-Earth's starting conditions for a specific mass ratio given our assumptions, and Table \ref{tab:IC2} shows the general ranges for which we have evaluated $a_L$.  

\begin{table}[!ht]
\small
\begin{center}
\begin{tabular}{cc}
\hline 
\hline
$a_S$ (AU) & $a_L$ (AU)	\\
\hline
0.76 & 1.2722531834616 \\
0.78 & 1.2466743326366 \\
0.80 & 1.2216761752564 \\
0.82 & 1.1972370663153 \\
0.84 & 1.1733366729743 \\
0.86 & 1.1499558658252 \\
0.88 & 1.1270766214724 \\
0.90 & 1.1046819350257 \\
0.92 & 1.0827557412981 \\
0.94 & 1.0612828436737 \\
0.96 & 1.0402488497524 \\
0.98 & 1.0196401129983 \\
1.00 & 0.9994436797214 \\
1.02 & 0.9796472408081 \\

\hline
\end{tabular}
\caption[Starting conditions for S and L given $M_L/M_S$ = 1/1, $e_L = e_S = 0.00$, and $i_L = i_S = 0^{\circ}$.]	
										% The class file doesn't do 
										% anything with the square 
										% bracketed short caption, 
										% but I always put one in.
	{
	Starting conditions for S and L given $m_L/m_S = 1/1$, $e_L = e_S = 0.00$, and $i_L = i_S = 0^{\circ}$.	
	%\label{lasttable}		% notice the second label for counting
	}
	\label{tab:IC1}
\end{center}
\end{table}

Following \cite{Rivera2002}, we explore three different mass ratios, $m_L/m_S$ = 8/1, 4/1, 1/1, to evaluate whether a mass dependence exists on the orbital elements of our ``success'' and ``pseudo-success'' cases.  These mass ratios are especially pertinent for comparison to the recent results of \cite{Cuk2012} and with those of \cite{Canup2012}, as they assume different masses of the impactor.  We use the same value for the argument of perihelion and ascending node as given by \cite{Rivera2002} for consistency, and we compute the mean anomaly from the time of periastron passage.  Thus we have a unique state vector $\left\{a,e,i,\omega,\Omega,M\right\}$ for each planet within our simulation.  From one run to the next, we vary only the initial state vector for the proto-Moon and update the corresponding values for the proto-Earth due to our angular momentum constraint.

\begin{table}[t]
\begin{center}

\begin{tabular}{ccccc}
\hline 
\hline
Case & Mass ratio	& $a_S$ range  & $a_L$ range  & Number \\
Name     &  (L/S)     & (AU)      & (AU)      & of runs\\
\hline
SS8M & 8/1 & 0.760-1.550  & 0.940-1.032 & 880 \\
SS1 & 1/1 & 0.760-1.550  & 0.555-1.261 & 880 \\
SS4 & 4/1 & 0.760-1.550  & 0.870-1.054 & 8109 \\
Nice4 & 4/1 & 0.760-1.550  & 0.870-1.054 & 8109 \\
\hline
SS8I & 8/1 & 0.440-0.750 & 1.033-1.086 & 352 \\
SS8E & 8/1 & 1.550-2.180 & 0.884-0.939 & 704 \\
\thickhline
QL8 & 8/1 & 0.76-1.54 & 0.94-1.03 & 40 \\
QL4 & 4/1 & 0.76-1.54 & 0.91-1.05 & 40 \\
QL1 & 1/1 & 0.76-1.08 & 0.91-1.22 & 34 \\
\hline
Riv8 & 8/1 & 0.76-1.40 & 0.95-1.03 & 33 \\
Riv4 & 4/1 & 0.76-1.40 & 0.91-1.05 & 33 \\
Riv1 & 1/1 & 0.76-1.00 & 1.00-1.22 & 13 \\

\hline
\end{tabular}
\caption[Initial conditions for S and L given $M_L/M_S$ = 1/1, $e_L = e_S = 0.00$, and $i_L = i_S = 0^{\circ}$.]	
										% The class file doesn't do 
										% anything with the square 
										% bracketed short caption, 
										% but I always put one in.
	{
	Summary of starting parameters used in our simulations.  Four groupings are shown: our primary study whose results are shown in Table \ref{tab:gmcomp} and Figure \ref{fig:gm}; extended regions of SS8 used in Table \ref{tab:gm8comp}; our simulation similar to the cases studied by \cite{Rivera2002}; and Rivera's simulations.  The parameters associated with Fig. \ref{fig:gm} assume the proto-Earth and proto-Moon to be initially coplanar and vary in eccentricity as described in $\S3$.  The Rivera results and our reproduction (QL) evaluate two values of eccentricity and a small (${2/3^\circ}$) inclination for the proto-Moon.  Further details of the Rivera and QL results can be found in the Supplementary Tables \ref{tab:Col11All0} - \ref{tab:Col81Ecc05Inc23}.	
	%\label{lasttable}		% notice the second label for counting
	}
	\label{tab:IC2}
\end{center}
\end{table}

We have also produced a set of simulations that use a Nice model configuration of the giant planets \citep{Gomes2005,Tsiganis2005}.  {The Nice model assumes that migration in a disk of gas brings the giant planets into mean motion resonances early, and planetesimal-induced migration leads to the giant planets scattering each other from these resonances into the present day configuration at a much later epoch.}  Our simulations do not incorporate this much later ($\sim$650 Myr after CAIs) event as the Giant Impact is constrained to occur $70-110$ Myr after the start of planetary formation.  Many multiresonant configurations \citep{Morbidelli2007,Batygin2010} have been investigated and shown to produce a variety of results, likely due the chaos in the Solar System.  Following the suggestion of A. Morbidelli (private communication 2013), we have chosen a configuration where the period ratios of Jupiter:Saturn, Saturn:Uranus, and Uranus:Neptune are near 3:2, 3:2, and 4:3 resonances, respectively.  This configuration places the semimajor axes of the giant planets at near 5.4, 7.2, 9.6, and 11.6 AU during the epoch that we study \citep[Fig. 6 of][]{Morbidelli2007}.

\begin{table}[ht]
\footnotesize
\begin{center}
\begin{tabular}{lcccc}
\hline 
\hline
Planet  & $a$ (AU)	& $e$	&  $\omega$ ($^\circ$)		 & M ($^\circ$) \\
\hline
Jupiter &  5.43012707 & 0.00497661 & 115.56319685 &   6.80435139 \\
Saturn	&  7.29928758 & 0.00987140 & 291.25675049 & 191.70506305 \\
Uranus	&  9.64081698 & 0.04981294 & 268.25089842 & 204.85820795 \\
Neptune & 11.61323534 & 0.01061949 &  47.45382836 & 277.32751572 \\

\hline
\end{tabular}
\caption[Orbital elements for the eight planets in the Solar System at epoch JD 2449101.0]	% The class file doesn't do 
										% anything with the square 
										% bracketed short caption, 
										% but I always put one in.
	{
	Orbital elements used for our Nice model simulations.	These orbital parameters use a multiresonant configuration given by \cite{Morbidelli2007} where the period ratios of Jupiter:Saturn, Saturn:Uranus, and Uranus:Neptune are near 3:2, 3:2, and 4:3 resonances, respectively.  Our Nice4 simulations begin with the giants planets orbiting in the ecliptic plane, $i=0^\circ$.
	%\label{lasttable}		% notice the second label for counting
	}
	\label{tab:niceelem}
\end{center}
\end{table}

\subsection{Integrations and Collision Tracking}

The evolution of the early Solar System bodies is calculated using a modified version of the hybrid symplectic integrator in the \texttt{mercury} package developed by \cite{Chambers1999}.  The Sun and planets are treated as spherical, rigid bodies, and the orbital evolution is calculated subject to Newtonian gravitational interactions.  Collisions between planets are treated as completely inelastic.  We make similar assumptions as \cite{Rivera2002} regarding the density and radius of the Earth-Moon progenitors (see Table \ref{planetprop}) that determine the collisional radius, i.e., the distance between the centers of the two bodies at the time of impact, $r_{col}$.  Our module simulates the system using an initial timestep $\epsilon = 0.015$ yr = 5.48 days to determine if and when a collision occurs within 200 Myr from when our simulations begin.  This choice of timestep has been shown to be appropriate because the fractional errors in energy ($\sim10^{-9}$) and angular momentum ($\sim10^{-12}$) remain small prior to the collision.  

\begin{table}[!ht]
\begin{center}

\begin{tabular}{ccccc}
\hline 
\hline

Mass  			& Planet	& 	Mass 					& 		Density 			& 	Radius	\\
ratio				&  				&	(M$_\oplus$) 	& 	(g cm$^{-3}$)		& 	(km)	\\
\hline
		8/1			& 	L			& 	 	0.90012			& 			5.47				&		6167.7 \\			
		8/1			& 	S			& 	 	0.11252			& 			4.05				&		3408.8 \\	
		4/1			& 	L			& 	 	0.81012			& 			5.47				&		5954.8 \\	
		4/1			& 	S			& 	 	0.20252			& 			4.05				&		4146.6 \\	
		1/1			& 	L			& 	 	0.50631			& 			4.76				&		5332.8 \\	
		1/1			& 	S			& 	 	0.50631			& 			4.76				&		5332.8 \\	

\hline
\end{tabular}
\caption[Planetary properties of L and S]	% The class file doesn't do 
										% anything with the square 
										% bracketed short caption, 
										% but I always put one in.
	{
	Planetary properties of L and S.
	
	}
	\label{planetprop}
	
\end{center}
\end{table}

{For cases where we wish to analyze the collision parameters, we have used a modified version of the \textit{close6} program that accompanies the \texttt{mercury} package to determine the state vectors, $\textbf{x}_i = \left\{a,e,i,\omega,\Omega,M\right\}_i$, of each mass one timestep prior to collision.  Since we want to know the state of the system on the order of seconds prior to collision, we have implemented a python script to continue the integration of the colliding bodies under the 2-body approximation up to the collision time.  This dual integration approach enables us to use the well-tested \texttt{mercury} package without unnecessarily frequent output of data, and our benchmarks indicate that it is substantially faster in terms of wall clock time than is a single integration with sufficiently frequent outputs to allow us to adequately determine the collision parameters that we are seeking. The final step in our algorithm uses the state vectors just prior to the collision to determine the energy and orbital angular momentum of the colliding masses relative to the respective center of mass.}  Then conservation of energy and momentum is used to calculate the final collision parameters.  In addition to the collision parameters, we determine the rotational period and obliquity of the merged mass assuming a perfectly inelastic collision.

\subsection{System Characterization}
\label{sec:Sys_char}
Following \cite{Rivera2001,Rivera2002}, we define a set of terms to characterize the final state of each simulation.  The goal of this work is to explore further the plausibility of an additional terrestrial planet existing in the Solar System for 8 -- 200 Myr after most of the small {planetesimals} in the terrestrial planet region have been accreted, and this planet impacts the proto-Earth producing a distribution of terrestrial planet mass and eccentricity consistent with the reality of today.  Our time window of collision is broader (than 20 -- 80 Myr) to increase our collisional statistics and to assess more widely the question of large, late impacts.  In order to accomplish this goal, we have evaluated three different assumptions on the mass ratio between the progenitors (1/1, 4/1, 8/1) as well as a representative case considering a different giant planet arrangement (4/1 Nice model).  We denote these cases as SS1, SS4, SS8, and Nice4 to differentiate easily between results, and then distinguish between different categorical outcomes within each case.  The SS8 case is divided into three ranges for $a_S$ (interior, middle, and exterior) that we distinguish by the labels: SS8I, SS8M, and SS8E, respectively.

We define the terms ``\textit{success}'', ``\textit{pseudo-success}'', ``\textit{non-SS mass}'', and ``\textit{early}'' to characterize the outcome of the simulation based upon the timing of a collision relative to when we begin ($\sim$$30-50$ Myr after CAIs) and the mass of the resultant body.  For a simulation to be deemed a ``success'', we require that a collision occur between L and S after at least 8 Myr of simulation time has elapsed.  If a collision occurs prior to 8 Myr between an Earth-sized body (Venus or proto-Earth) and a smaller body (Mercury, proto-Moon, or Mars) other than the L-S combination (5 possible pairings), the simulation is regarded as a ``pseudo-success''.  For the SS1 runs, we restrict the definition of a ``pseudo-success'' to encompass only a collision of Venus and Mars or Venus and Mercury (2 possible pairings).  When the mass of the merged body is significantly different from an Earth mass and the system has evolved for at least 8 Myr, we designate the result as ``non-SS mass''.  This category could result from a collision between the Earth-sized bodies (Venus and proto-Earth) or two of the smaller bodies (proto-Moon, Mars, or Mercury) for the SS4, Nice4, and SS8 mass ratios (4 possible pairings).  For the SS1 mass ratio, this category is expanded to include collisions of the proto-Moon or the proto-Earth with any of the other planets (bringing the total to 7 distinct pairings).  An ``early'' category is placed upon systems when two terrestrial planets collide before the threshold of 8 Myr without regard to the resultant mass distribution.  An additional category, ``ejection'', describes the outcomes where a terrestrial body collides with the Sun or reaches a distance greater than 100 AU where it is assumed to be ejected.  The final category of no collisions ``NC'' represents those simulations that are stable for 200 Myr, i.e., without collision or ejection of any terrestrial body.

We use the angular momentum deficit (AMD) of the terrestrial planets to further characterize the ``success'' cases for all the simulations performed and ``pseudo-success'' cases for the SS8 runs.  Through the AMD, we identify which post-collision systems are dynamically cold as is the current state of the Solar System.  \cite{Laskar1997} performed a long-term (25 Gyr) evaluation of the variations of the AMD for the Solar System and found that the maximum variation of the terrestrial planet AMD did not exceed twice the mean value.  We calculated the sum of the terrestrial planet AMD for the epoch in Table \ref{tab:orbelem} and use this to scale our results, defining 
\begin{equation}
\label{eqn:AMD}
{\rm AMD}_{\rm tp} \equiv \sum\limits_{i=1}^N {{\rm AMD}_i \over 5.673969\times 10^{-8}~{\rm M}_\odot {\rm AU}^2{\rm yr}^{-1}},
\end{equation}
where $i=1 \ldots N$ represents the heliocentric ordering of the terrestrial planets.  For these runs we calculate the instantaneous AMD$_{\rm tp}$ at one year after the collision and its mean value over an additional 10 Myr of evolution, $\left\langle{\rm AMD_{\rm tp}}\right\rangle$, to assess the dynamical {excitation} of the system.  Simulations that approximate reality in terms of AMD $\left(\left\langle{\rm AMD_{\rm tp}}\right\rangle <1.5 \right)$, relative timing of the collision (20 -- 80 Myr after the simulations begin), heliocentric ordering (collision produces the third planet from Sun), and the mass distribution of the terrestrial planets are considered as ``\textit{Solar System-like}'', which is a subcategory of either ``success'' or ``pseudo-success''.

We examine the collision characterization for the late impacts ($8-200$ Myr after our simulations begin) using SS4 runs.  The two key factors in determining the outcome of a collision are the ratio of the impact speed, $v_{col}$, to the mutual escape speed of the two bodies, $v_{esc}$ and the scaled parameter, $b_{col}/r_{col}$, where $b_{col}/r_{col} = 0$ or 1 refers to a head-on or grazing collision, respectively.  These parameters are compared to successful initial conditions in detailed models of various type of impacts that may produce the Earth-Moon system, namely ``canonical'', ``large impactor'', ``small impactor'', or ``hit-and-run''.  The ``canonical'' impact scenario refers to the more grazing ($b_{col}/r_{col} \approx 0.8$) impact \citep{Canup2001,Canup2004} with the collision velocity, $v_{col}$, restricted to a value less than 1.1$v_{esc}$.  In contrast, the ``hit-and-run'' scenario requires the collision velocity ratio to be slightly larger ($v_{col}/v_{esc}$ = $1.2 - 1.3$) and a smaller impact parameter ($b_{col}/r_{col}$ = $0.5 - 0.64$) for the smooth particle hydrodynamic (SPH) models to produce a body with a composition similar to the Moon.  However, some large impactor masses (near 1/1) that overlap with the ``hit-and-run'' in terms of the scaled impact parameter and encompass a broader range in the collision velocity ratio, $v_{col}/v_{esc}$ = 1.0 -- 1.6, have been recently considered \citep{Canup2012}. The ``small impactor'' scenario \citep{Cuk2012} considers a different region of the parameter space where the impact parameter ($b_{col}/r_{col}$ = $0.0 - 0.15$) is close to head-on and the collision velocity is substantially higher ($v_{col}/v_{esc}$ = $1.35 - 1.80$). By performing this additional comparison, we provide a qualitative likelihood between the different models. 

\section{Results}
\label{sec:stat_riv}

We began our study by running six small sets of simulations that are analogous 
to those included in the study of \cite{Rivera2002}.  The parameter ranges 
studied and summaries of the results are presented in Table \ref{tab:comprivera}, and 
lists of individual collisional outcomes are given in the Supplementary Tables \ref{tab:Col11All0} -- \ref{tab:Col81Inc23}.  Table \ref{tab:Col81Ecc05Inc23} is included to show the results of runs that were prescribed in \cite{Rivera2002}, even though he did not present the individual outcomes corresponding to this set of runs.  In Table \ref{tab:comprivera} we find good agreement when comparing our reproduction with Rivera's results, within the statistical uncertainties of the small numbers of simulations run.

\begin{table}[!ht]
\centering

\begin{tabular}{|r|l|cc|cc|}
\hline 
\hline
&Category  & {$\rm QL_R$} & \citeauthor{Rivera2002} &  {$\rm QL_R$} & \citeauthor{Rivera2002}  \\
\hline
&					 & \multicolumn{2}{c|}{\textit{Circular}}	&  \multicolumn{2}{c|}{\textit{Eccentric}} \\
 \parbox[t]{2mm}{\multirow{5}{*}{\rotatebox[origin=c]{90}{\textbf{1/1}}}}&
Survived 200 Myr  				  &  7   	& 	7  		& 	4   &  4	 \\
&Ejection 									&  0	  & 	0  		& 	1	  &  0	 \\
&Any collision						  &  6  	& 	6  		&   8	  &  9	 \\
&``Success'' 							  &  0 		& 	1	 		& 	1	  &  1	 \\
&``Pseudo-success'' 				&  0 		& 	0  		& 	0	  &  0	 \\
\hline
\parbox[t]{2mm}{\multirow{5}{*}{\rotatebox[origin=c]{90}{\textbf{4/1}}}}&
Survived 200 Myr  				  & 19   	  & 	19 		& 	3   &  0	 \\
&Ejection 									&  1	  	& 	 0 		& 	0	  &  0	 \\
&Any collision							& 13  		& 	13 		&  30	  &  33	 \\
&``Success'' 							  &  2 			& 	 2		& 	4	  &  4	 \\
&``Pseudo-success'' 				&  1 			& 	 7		& 	6	  &  3	 \\
\hline
&     & \multicolumn{2}{c|}{\textit{Circular}}	&  \multicolumn{2}{c|}{\textit{Inclined}} \\
\parbox[t]{2mm}{\multirow{5}{*}{\rotatebox[origin=c]{90}{\textbf{8/1}}}}&
Survived 200 Myr  				  & 18   & 	18 & 	21    &  15	 \\
&Ejection 								 	&  0	 & 	 0 & 	 0	  &   1	 \\
&Any collision 						  & 15   & 	15 & 	12	  &  17  \\
&``Success'' 							  &  2 	 & 	 2 & 	 4	  &   2  \\
&``Pseudo-success'' 				&  2 	 &   3 & 	 1	  &   5	 \\
\hline

\hline
\end{tabular}
\caption[Comparison of the results of this work to \cite{Rivera2002}]	% The class file doesn't do 
										% anything with the square 
										% bracketed short caption, 
										% but I always put one in.
	{
	Counting statistics derived from the Supplementary Tables \ref{tab:Col11All0} -- \ref{tab:Col81Inc23} for the 1/1, 4/1, and 8/1 mass ratios and equivalent statistics found in \cite{Rivera2002}.  The first column ({$\rm QL_R$}) of each subset shows our results using the same starting conditions as \citeauthor{Rivera2002}.  The counts of the specific categories that lead to an Earth-like mass, ``success'' and ``pseudo-success'', are also given.  The headings ``Circular'' and ``Eccentric'' denote coplanar starting conditions for the progenitors.  The 4/1 and 1/1 simulations in the ``Eccentric'' column begin with an eccentricity of 0.05 for the proto-Moon.  The 8/1 simulations with the ``Inclined'' heading begin with an inclination of $2/3^\circ$ for the proto-Moon and $e_S=0$.
	%\label{lasttable}		% notice the second label for counting
	}
	\label{tab:comprivera}

\end{table}

%\section{Results and Discussion}
\label{sec:RD}

Following the motivation from the study by \cite{Rivera2002}, we investigate the general parameter space at much higher resolution, considering the starting (initial) conditions in a similar manner but limiting the initial inclination of the Earth-Moon progenitors to zero to reduce the dimensionality of the phase space of possible initial values.  We simulated the SS1 and SS8M cases at a ``low'' resolution of (11$\times$80) in initial eccentricity-semimajor axis phase space, whereas the SS4 and Nice4 cases were studied at ``high'' resolution (51$\times$159).  Each case considers a range of semimajor axis ($a_S=0.76-1.55$ AU) and eccentricity ($e_S=0.0-0.1$) for the proto-Moon.  

The inner limit on $a_S$ was chosen to include all orbits sufficiently exterior to Venus that they might be able to avoid close encounters with Venus at very early times.  The exterior limit on $a_S$ allows close encounters of the proto-Moon with Mars.  We included this outer region because Mars is smaller than Venus in both mass and size, and thus it is not highly unlikely that the system survives for many millions of years after the initial close encounters.  

{When an encounter involving the proto-Moon with either Venus or Mars occurs, both orbits are displaced due to an exchange of angular momentum.  Consequently, the final orbit of Venus or Mars will be different from the orbit of corresponding real planet.  However, the final orbits of these planets are not considered as a criterion for ``success'' and enter the criteria of ``Solar System-like'' only through the planet ordering and their contribution to $\left\langle{\rm AMD_{tp}}\right\rangle$.}  We note that choosing values of $a_S$ beyond the orbit of Mars together with our angular momentum condition leads to the proto-Earth initially orbiting interior to Venus in some of the SS1 simulations.  Simulations were also performed for extended ranges in the SS8 case (see Table \ref{tab:IC2}), as our angular momentum condition places the proto-Earth near 1 AU for a greater range of $a_S$.  Moreover, the possibility of swapping orbits between the proto-Moon and Mars was of interest because of the similarity in mass for the SS8 runs.

{\renewcommand\arraystretch{0.15}
\begin{figure*}[!ht]
\centering
\begin{tabular}{c}
	\includegraphics[width = 0.95\linewidth]{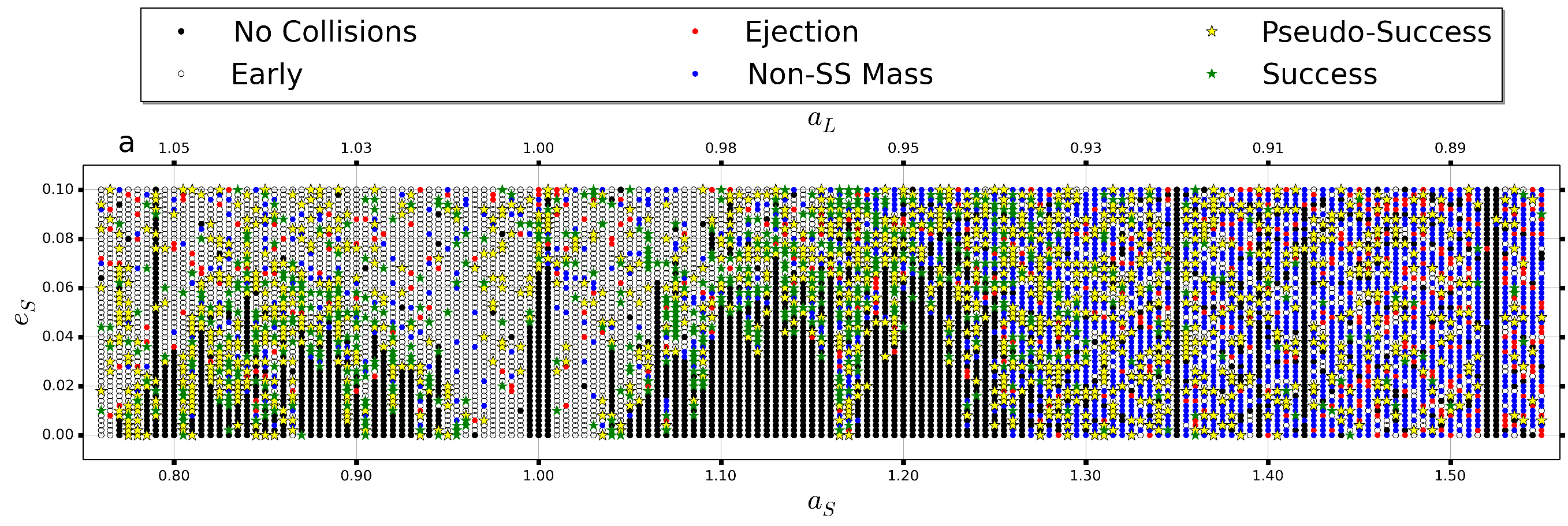}\\
	\includegraphics[width = 0.95\linewidth]{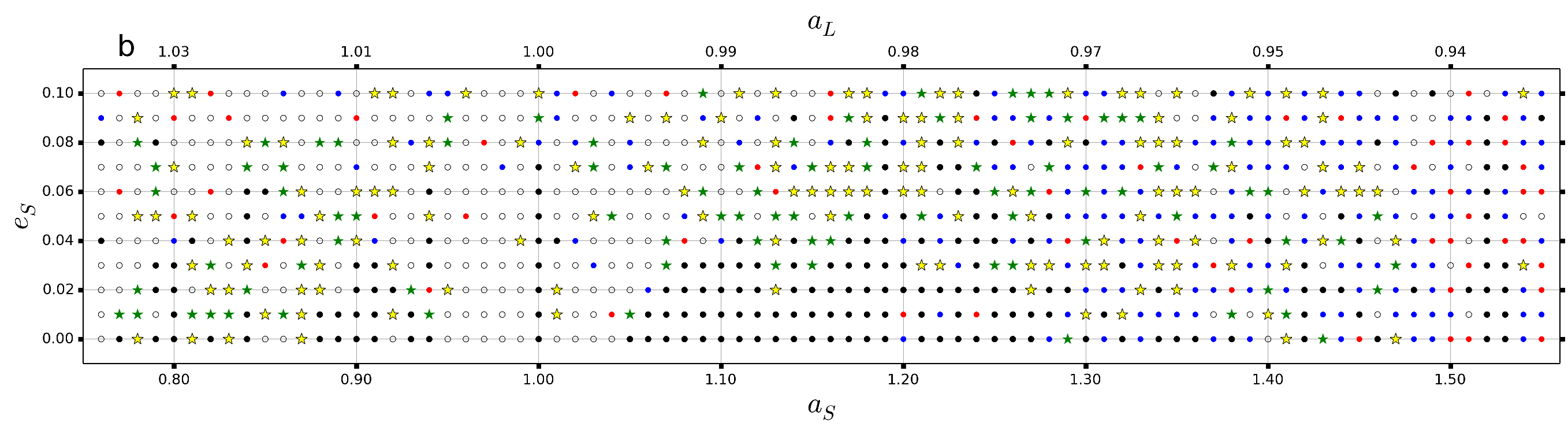}\\
	\includegraphics[width = 0.95\linewidth]{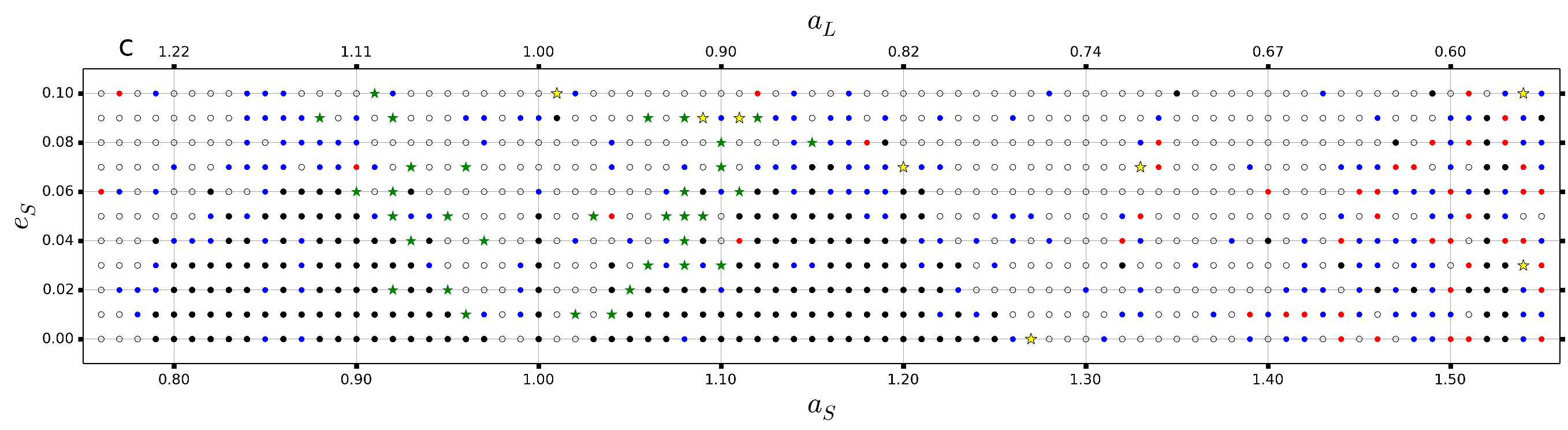}\\
	\includegraphics[width = 0.95\linewidth]{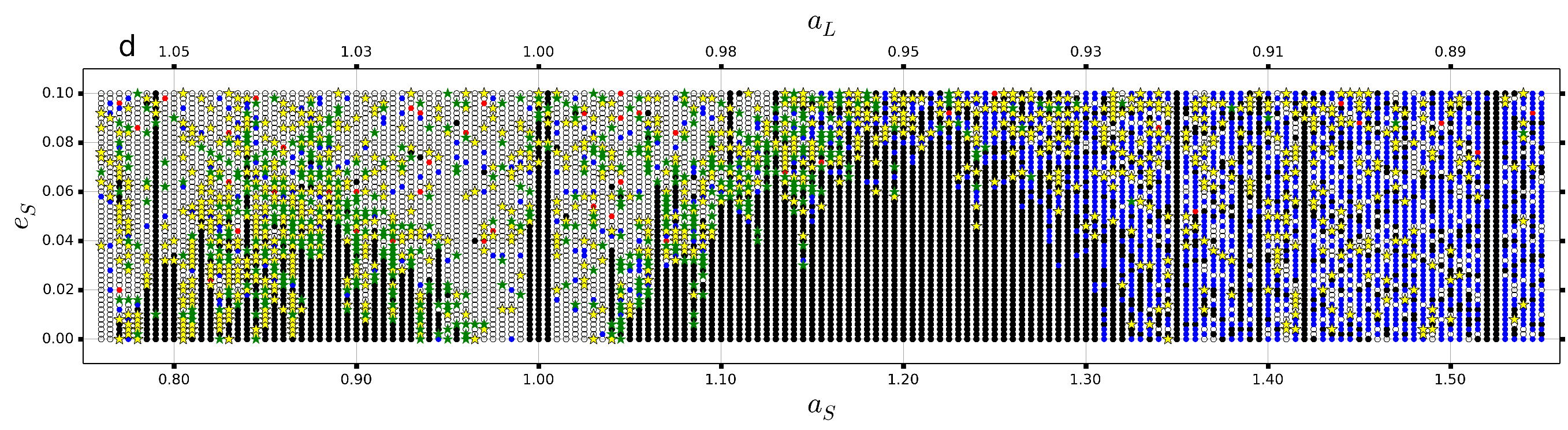}
\end{tabular}
\caption{Classification of results for our primary simulation sets: a) SS4, b) SS8M, c) SS1, and d) Nice4.  We denote the starting semimajor axis of the proto-Earth (L), $a_L$, along the top axis and the starting semimajor axis of the proto-Moon (S), $a_S$,  on the bottom axis. Two low resolution (b,c) and two high resolution (a,d) maps show the global results of those simulations.  The black dots indicate where the system survived the entire 200 Myr integration without a collision/ejection.  The open circles show runs that suffered a collision within the first 8 Myr of the integration.  Runs represented by a red dot suffered an ejection, where a planetary body exceeded 100 AU from the Sun.  The collisions occurring between 8 and 200 Myr that produce an Earth-like resultant mass are denoted by stars, where green indicates a success (collision between L and S) and yellow represents a pseudo-success impact.  Blue dots depict collisions occurring between 8 and 200 Myr that produce a resultant mass much smaller or larger than the Earth.}
\label{fig:gm}
\end{figure*}
}

Figure \ref{fig:gm} presents the results of all four sets of runs with $a_S=0.76-1.55$ AU.  Figure \ref{fig:cm} has been created using the collision/ejection times as seeds for a cubic spline interpolation with respect to the time of collision/ejection (color scale).  The color scale reflects the collision/ejection time, with dark blue corresponding to simulations where no collisions or ejections occur and dark red corresponding to early collisions/ejections.  The yellow and green regions that appear between the extreme regions indicate collision times of a late impact.  We have over-plotted the stars (white) for our ``success'' characterization to illustrate where the probability of this outcome is highest.  

The contour maps reveal the dynamics of each case {as well as structures} caused by possible resonances between the terrestrial planets as mean motion resonances (MMRs) and secular effects from the dynamics of their Jovian counterparts.  To this end we have labeled the locations of the nominal first-order MMRs with respect to interactions between terrestrial planets on the top axis of each plot.  Our notation denotes the ratio of periods between another terrestrial planet and S unless otherwise noted.  For example, the 4L:3 signifies that the proto-Moon (S) orbits the Sun three times for every four orbits of the proto-Earth (L).  The 4L:3, 5L:4, and 6L:5 mean motion resonances (MMRs) are clearly correlated with instability strips (Figure \ref{fig:cm}), whereas other MMRs (5M:6) indicate strips of increased stability.  The co-rotational resonances are clearly manifest between the proto-Moon with either the proto-Earth (1L:1) or Mars (1M:1) at $a_S=1.0$ and $a_S=1.523$, respectively.  

{\renewcommand\arraystretch{0.15}
\begin{figure*}[!ht]
\centering
\begin{tabular}{c}
	\includegraphics[width = \linewidth]{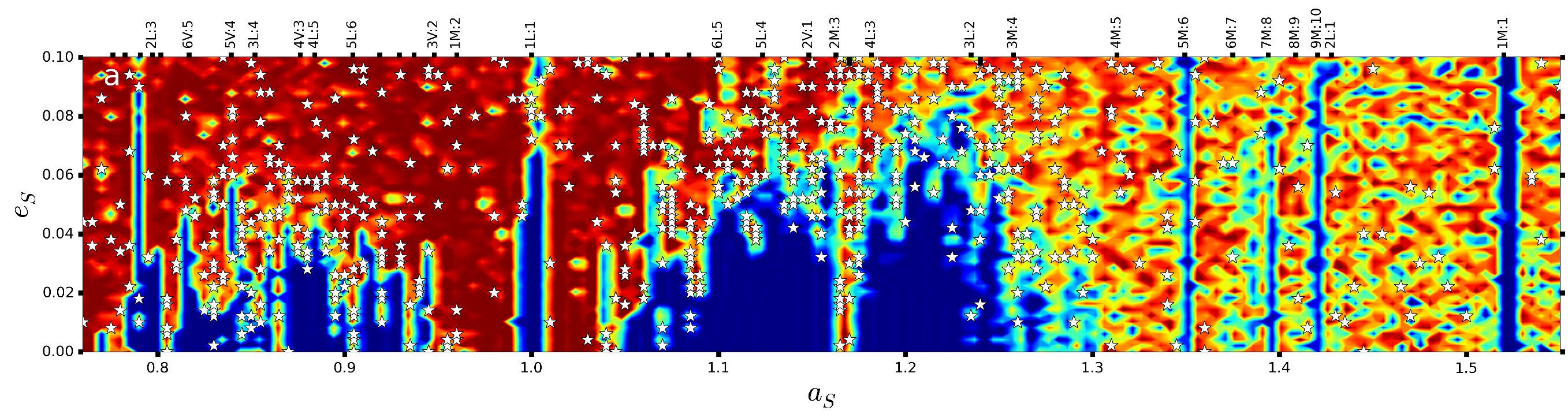}\\
	\includegraphics[width = \linewidth]{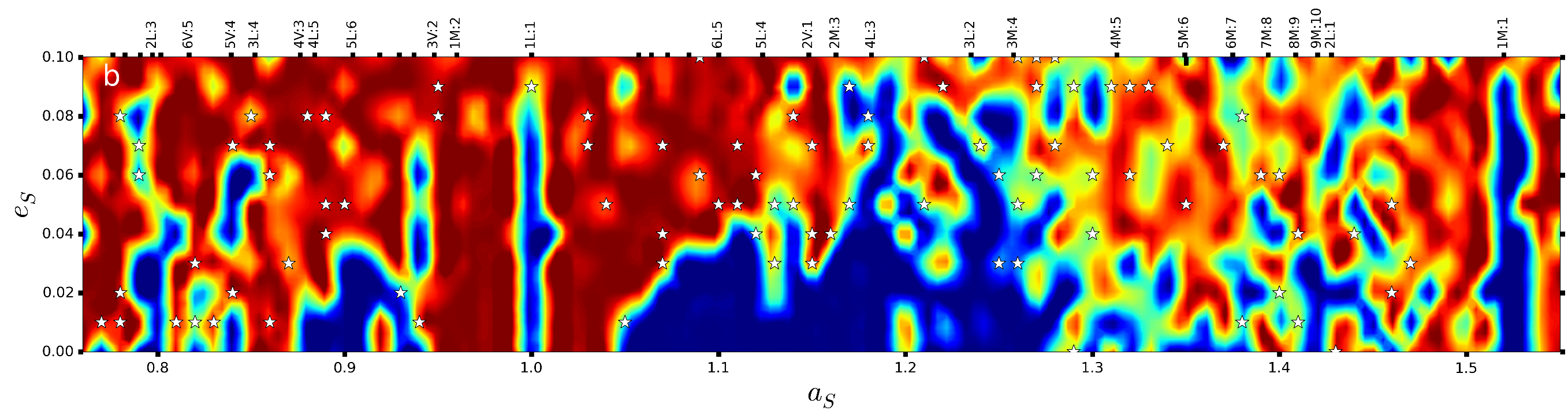}\\
	\includegraphics[width = \linewidth]{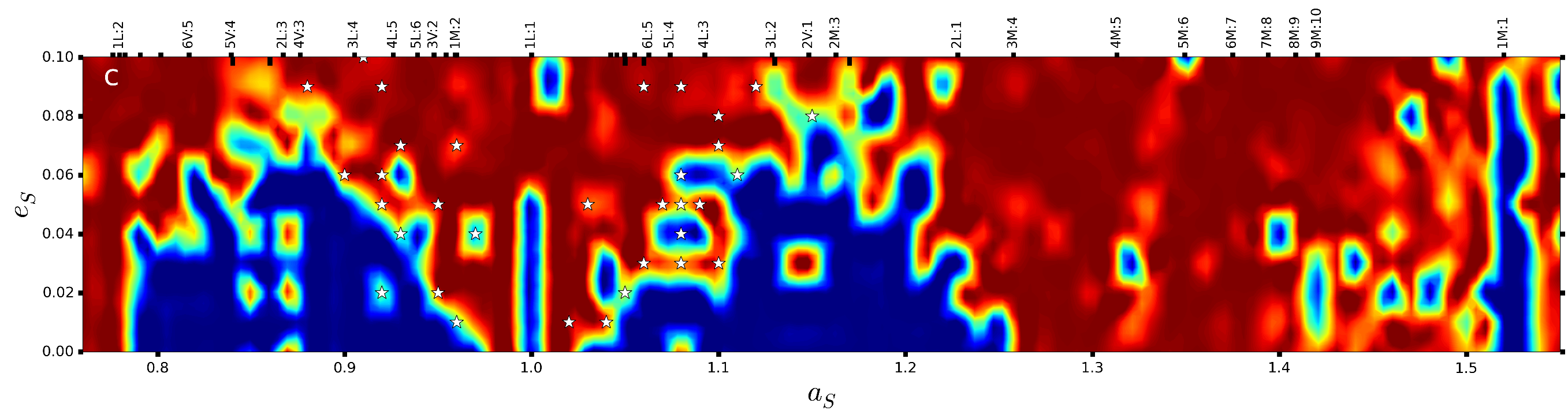}\\
	\includegraphics[width = \linewidth]{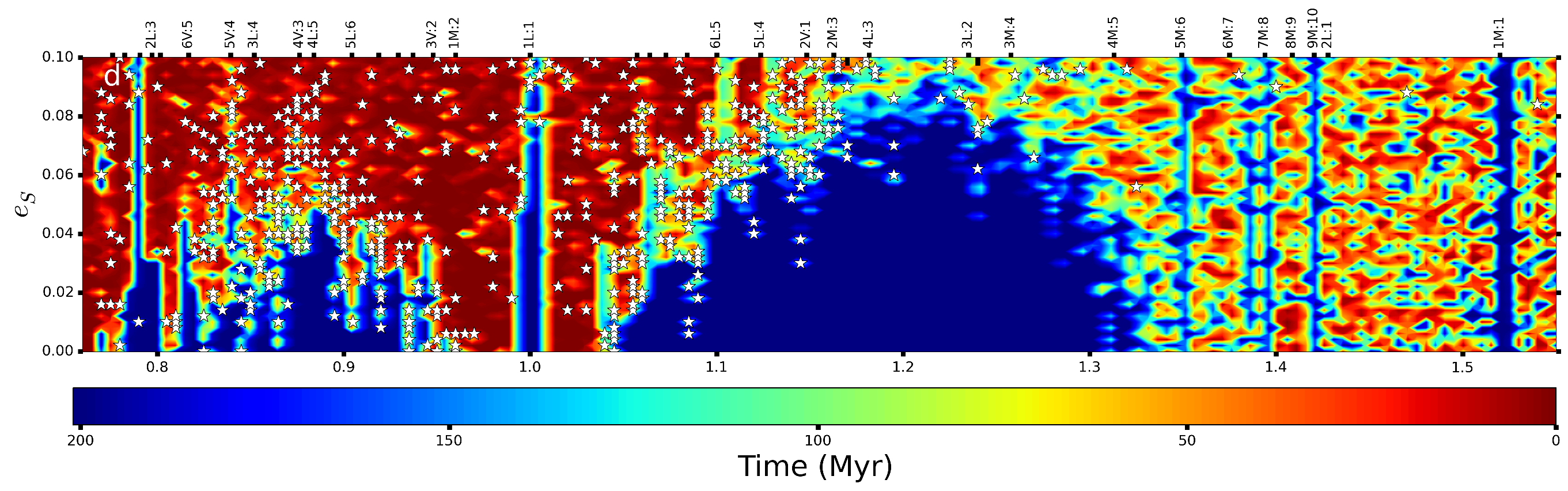}
\end{tabular}
\caption{Contour maps of results for our primary simulation sets: a) SS4, b) SS8M, c) SS1, and d) Nice4 in the initial semimajor axis ($a_S$) and eccentricity ($e_S$) parameter space of the proto-Moon.  The color scale is determined by the collision/ejection epoch, with blue representing simulations with no collisions (NC) and dark red representing early ($<$8 Myr) collisions/ejections.  First-order mean motion resonances (MMRs, top axis) of the proto-Moon with Venus (V) or the proto-Earth (L) are labeled up to the 5th degree, with higher degree first-order MMRs denoted by only upward tick marks.  The corresponding MMRs between the proto-Moon and Mars (M) are labeled up to 10th degree.  Downward ticks on the top axis indicate the initial semimajor axis ($a_S$) of the proto-Moon when the proto-Earth has a first order MMR with Venus or Mars.  The white stars indicate ``success'' outcomes, where the proto-Earth and proto-Moon collide within the time interval of $8-200$ Myr from when our simulations begin.}
\label{fig:cm}
\end{figure*}
}

\begin{figure}[!ht]
\centering
\begin{tabular}{c}
	\includegraphics[width=\linewidth]{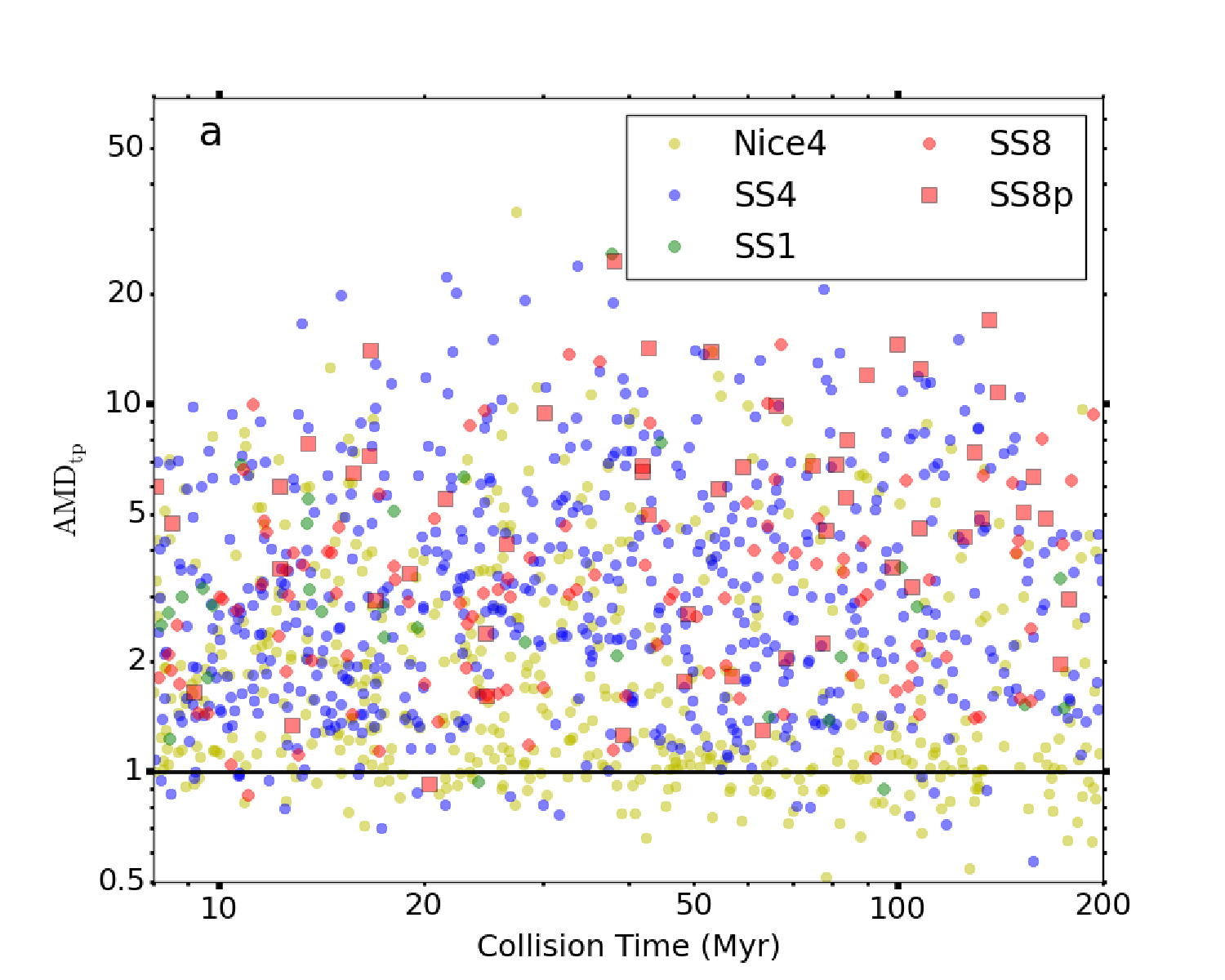}\\
	\includegraphics[width=\linewidth]{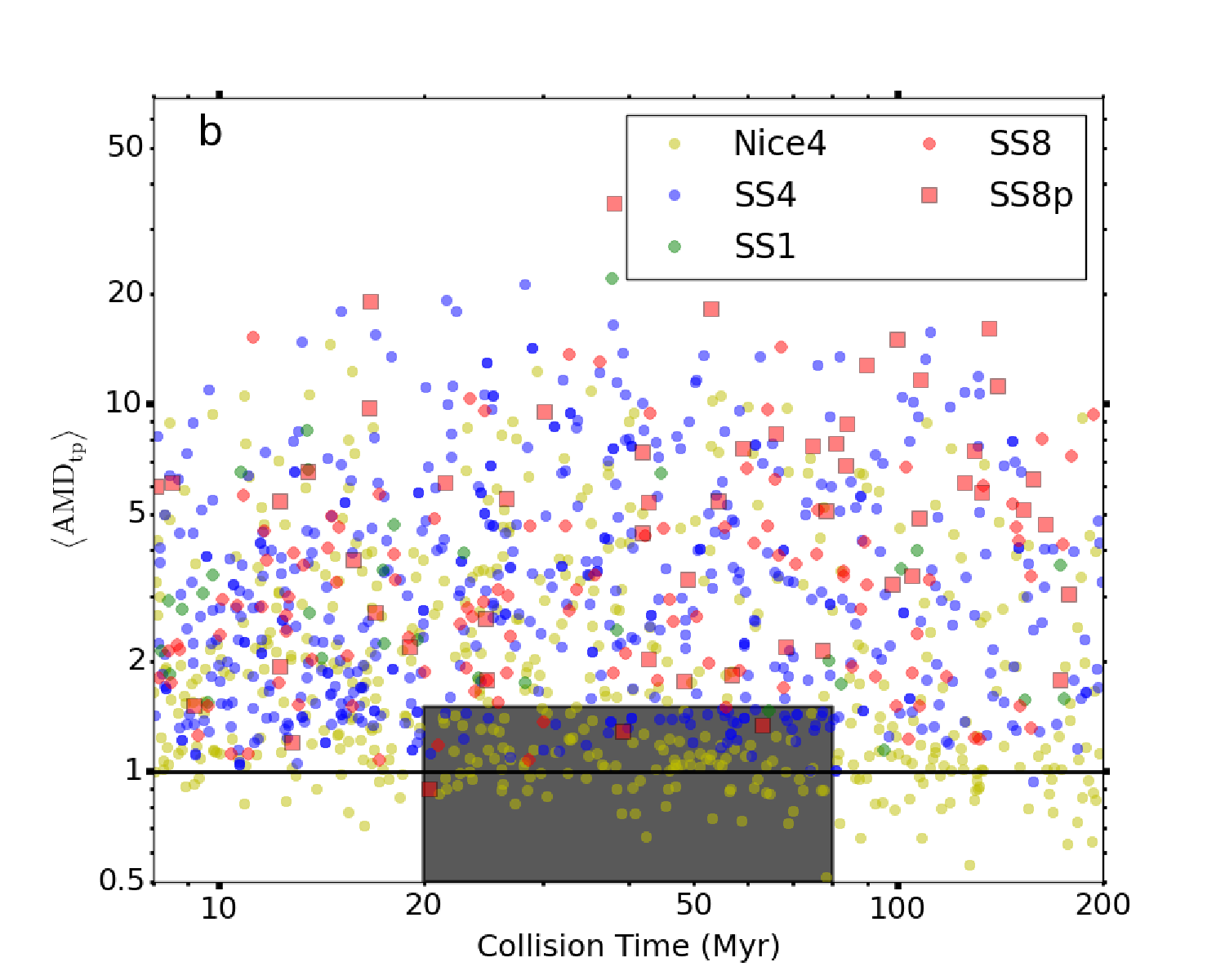}
\end{tabular}
\caption{Characterization of the ``success'' outcomes of each set of runs with respect to both (a) the 1-year ${\rm AMD_{tp}}$ and (b) the 10-Myr mean $\left\langle{\rm AMD_{tp}}\right\rangle$ after the collision; note the logarithmic scale.  The horizontal line denotes a value of 1.0 which corresponds to the ${\rm AMD_{tp}}$ of the present day Solar System.  The shaded region (b) identifies the conditions that represents a ``success'' most consistent with current estimates for the age of the Earth-Moon system. The square points, SS8p, show the ``pseudo-success'' outcomes for the 8/1 case that can mimic ``success'' outcomes due possible switching of the proto-Moon and Mars.  Tables \ref{tab:gm8comp} and \ref{tab:gmcomp} show the statistics (counts) of each dataset represented.  Note: The vertical axis is scaled by the value of AMD for the inner Solar System at the reference epoch of JD 2449101.0 (see Eqn. \ref{eqn:AMD}).  The collision produced the third planet from the Sun in the vast majority of runs denoted herein, including all of those in the shaded box.}
\label{fig:AMDgm}
\end{figure}

We have evaluated the summed angular momentum deficit of the terrestrial planets both 1 year after the collision (instantaneous), AMD$_{\rm tp}$, and averaged over 10 Myr following the collision, $\left\langle{\rm AMD_{tp}}\right\rangle$, for each of the $\sim$1270 simulations that was classified as a ``success'' and the $\sim$210  simulations deemed a ``pseudo-success'' in the full SS8 results.  Cases with high $\left\langle{\rm AMD_{tp}}\right\rangle$ are expected to be unstable within the lifetime of the Solar System.  For stable systems, higher $\left\langle{\rm AMD_{tp}}\right\rangle$ typically implies that Earth's eccentricity reaches higher values than in the actual Solar System.  Figure \ref{fig:AMDgm} shows the results of this analysis considering both the AMD$_{\rm tp}$ and $\left\langle{\rm AMD_{tp}}\right\rangle$ on a logarithmic scale with the respective simulation dataset color coded.  The most striking aspect of these plots is that the vast majority of points in all four sets yield systems that have AMD$_{\rm tp}$ larger that the actual terrestrial planets, as found by \cite{Raymond2009}.  We note that strong similarities exist between the results in Fig. \ref{fig:AMDgm} of this work and Fig. 9 of \cite{Raymond2009}.  {We also note that a population of remnant planetesimals of a few percent of an Earth mass (required to deliver the late veneer to the Earth after the Moon-forming event; \cite{Raymond2013}), that we neglect in our study, could reduce the final $\left\langle{\rm AMD_{tp}}\right\rangle$.}

In the following subsection, we discuss the SS4 case in detail.  Subsequently, the results for the SS8 (which we have performed over the extended region in semimajor axis 0.44 $\geq$ $a_S$ $\geq$ 2.18 AU), SS1, and Nice4 cases are presented and compared with the SS4 results.

\subsection{SS4 Results}
\label{sec:SS4}
Inspecting Figs. \ref{fig:gm}a and \ref{fig:cm}a, we see  a majority of the ``early'' collision category occurs when the proto-Moon is initially placed near the proto-Earth or Venus, with many values of $a_S$ spanning the region between Venus and 1.1 AU leading to early collisions for high starting eccentricity.  The early ($<8$ Myr from our starting epoch) collisions are expected for $a_S$ near the proto-Earth from Hill stability and the overlap of first-order resonances.  However, there exists a stable region when the Earth-Moon progenitors are placed at or very near the same semimajor axis but separated in longitude.  In almost symmetric locations around 1.0 AU, there are nominally stable regions where the simulations progressed to the full duration of 200 Myr.  The ``ejection'' cases appear at random outside the stable regions in Figure \ref{fig:gm}.  The ejected body is either Mars or Mercury, the least massive planets, which are the easiest to perturb and thus possibly eject from the system.  It is less clear how correlated the remaining ``success'' and ``pseudo-success'' outcomes are with respect to the parameter space, but the following statistical inquiry is performed to resolve this issue.  

\begin{figure}[!ht]
\centering
\includegraphics[width=\linewidth]{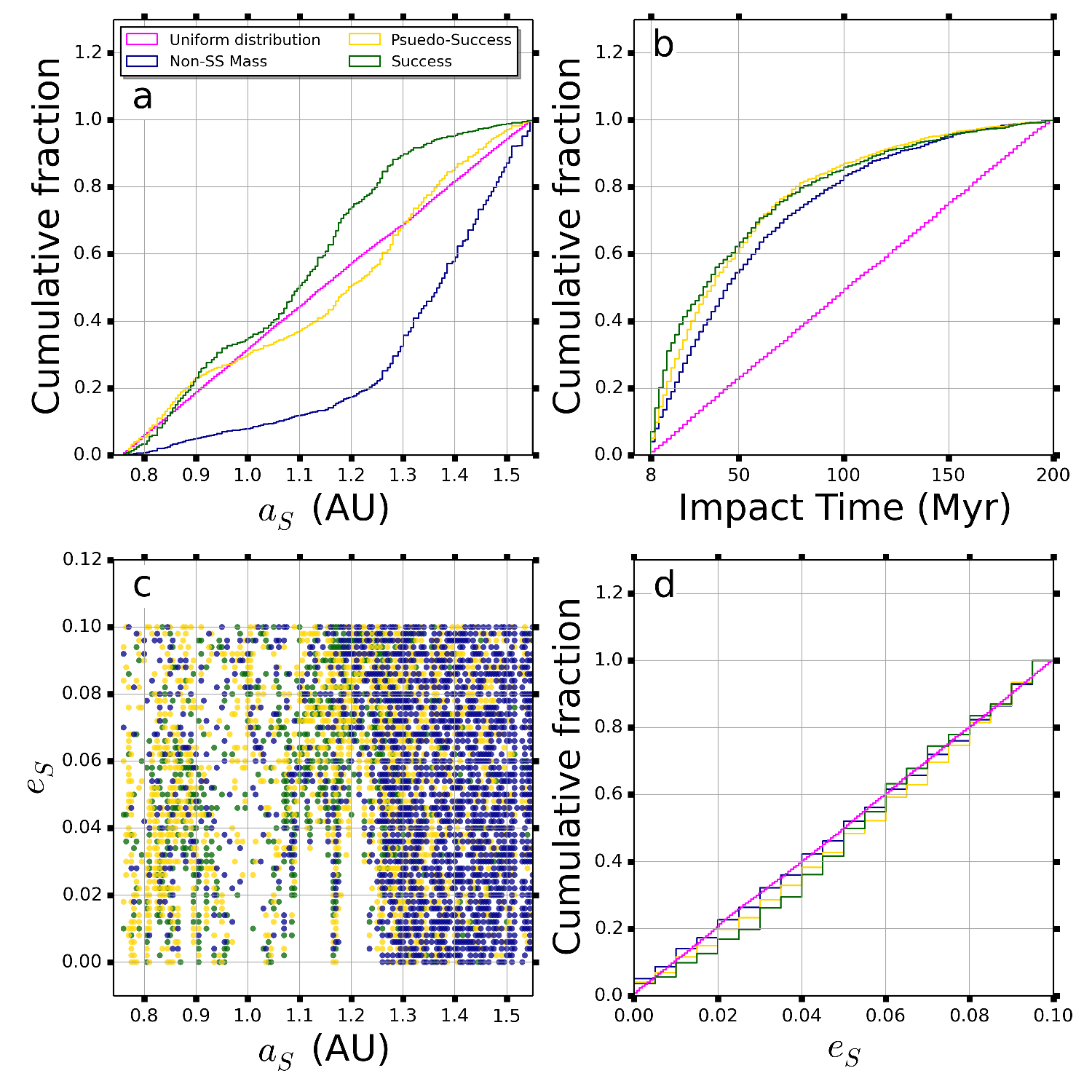}
\caption{The cumulative distributions for the late (8--200 Myr) collisions in the SS4 runs, which probe (a) the starting semimajor axis of the proto-Moon ($a_S$), (b) the time of the collision for the respective category, and (d) the eccentricity ($e_S$) of the proto-Moon at the start of the simulation.  A general map (c) is given to show the location of these collisions within the parameter space; points have been color coded by the collision characterization to match Figure \ref{fig:gm} and its associated legend.}
\label{fig:jd4}
\end{figure}

From the results shown in Figs. \ref{fig:gm}a and \ref{fig:cm}a, we can determine preferred locations within the parameter space for a late (8 Myr $< t_{col}\leq$ 200 Myr) collision to occur for the SS4 case.  Figure \ref{fig:jd4} examines the statistics of these late collision outcomes.  There is a depletion of all late collision types in the highly unstable regions near the orbits of the proto-Earth, Venus, and perhaps Mars.  This can also be seen in the cumulative distribution functions (Fig. \ref{fig:jd4}a) as the ``success'' cases flatten in slope in the region near 1.0 AU and as a scarcity of points in the general joint distributions (Fig. \ref{fig:jd4}c).  For this mass ratio (SS4), we see the distributions of ``success'' and ``pseudo-success'' have similar shapes, while the ``non-SS mass'' distribution is skewed towards higher values of $a_S$, specifically with $80\%$ of the population beyond 1.2 AU, for which the proto-Moon begins close to Mars and the proto-Earth begins closer to Venus.  Distinct excesses from uniform for the ``success'' category are seen for $a_S$ in the ranges $\sim$$0.8-0.94$ and $\sim$$1.06-1.28$.  

\begin{figure}
\centering
\includegraphics[width=\linewidth]{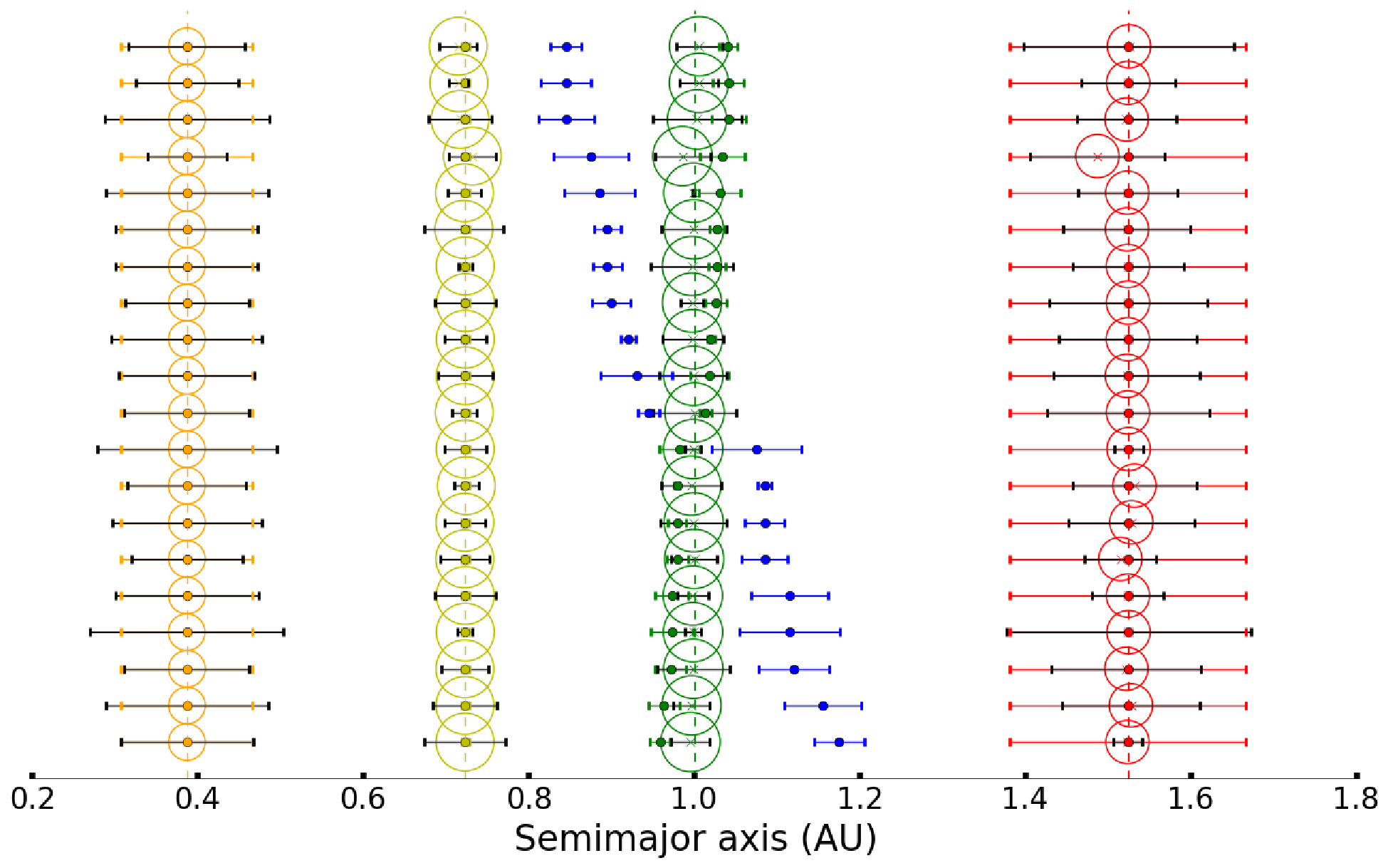}
\caption{Starting and final semimajor axes of the terrestrial planets for the  ``success'' 4/1 Solar System results listed in Table \ref{tab:ColSS_super}.  The simulations are ordered to correspond with the results in Table \ref{tab:ColSS_super} by the starting semimajor axis of the proto-Moon $a_S$.  The starting values in semimajor axis for Mercury (black), Venus (yellow), proto-Earth (green), proto-Moon (blue), and Mars (red) are indicated by points, with the ends of the bars denoting the periastron and apastron values of each respective body.  The open circles correspond to the post-impact states of the resulting planets, where the size of each circle is scaled to the relative size of the respective planet.  Vertical dashed lines are also shown to guide the eye and illustrate the deviation between the post-impact states and the orbits of the real Solar System planets.}
\label{fig:ex_sim}
\end{figure}

Figure \ref{fig:ex_sim} illustrates the starting and final states for the SS4 cases given in Table \ref{tab:ColSS_super}. Note the small deviation from the initial semimajor axes for the non-progenitor masses, which supports our use of current orbital elements of the terrestrial planets apart from the Earth-Moon progenitors as initial conditions beginning at $30-50$ Myr after CAIs to arrive at a configuration similar to that of the current Solar System.

The colliding bodies and timing of collision are sensitive to the starting eccentricity and semimajor axis of the proto-Moon and the proto-Earth.  For the SS4 case, the interactions of a 4L:3 mean motion resonance (MMR) between the Earth-Moon progenitors and a secular effect with Jupiter induce enough perturbations uniformly with eccentricity to suggest a possible mode of production for Earth-Moon type systems through secular chaos.  The significance of the secular interactions are illustrated in \ref{sec:colres} through a comparison with a Nice model configuration where only the giant planet architecture of the system has changed. 

The cumulative distribution functions with respect to the collision time for each collision category (Fig. \ref{fig:jd4}b) are all weighted towards earlier times and do not resemble a uniform distribution.  They are reasonably well fit by either a power law distribution or an exponential decay with a half-life $\sim$$30$ Myr relative to the bulk formation of the proto-Earth (the starting time of our simulations) and are weakly dependent on the collision category.  The ``success'' and ``pseudo-success'' categories are slightly more weighted toward earlier times than the ``non-SS mass'' outcome.  The distributions of eccentricity (Fig. \ref{fig:jd4}d) are similar to the uniform distribution.  This is because the tendency for low eccentricity cases to be stable is roughly balanced by the excess of early collisions for high eccentricity.

Figure \ref{fig:AMDgm} shows a wide range in AMD$_{\rm tp}$ and $\left\langle{\rm AMD_{tp}}\right\rangle$ for the SS4 runs (blue dots).  The SS4 results depicted in Figure \ref{fig:AMDgm}a show a substantial number of cases with instantaneous AMD$_{\rm tp}< 1.5$ and even cases with AMD$_{\rm tp}< 1.0$, i.e., smaller AMD$_{\rm tp}$ than the present day Solar System (see $\S$\ref{sec:Sys_char}).  However, this observation doesn't provide an accurate portrayal of the long term evolution.  Through inspection of the $\left\langle{\rm AMD_{tp}}\right\rangle$ (Fig. \ref{fig:AMDgm}b) shows only 1 of the 612 SS4 ``success'' runs having $\left\langle{\rm AMD_{tp}}\right\rangle<1.0$.  Nonetheless, a substantial number remain in the regime of $\left\langle{\rm AMD_{tp}}\right\rangle$ $<$ 1.5 and within the 20 -- 80 Myr time window.

\begin{figure}[!ht]
\centering
\includegraphics[width=\linewidth]{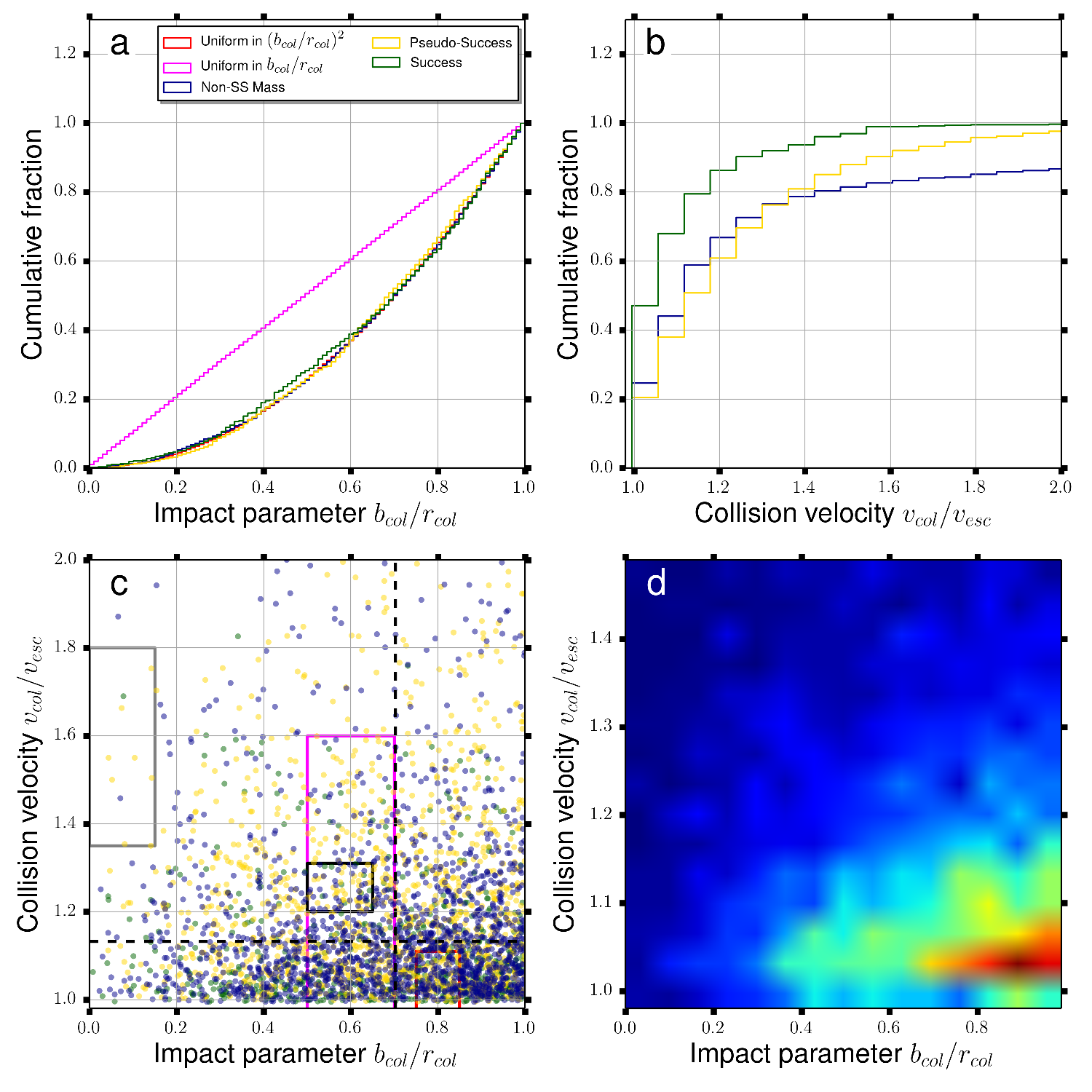}
\caption{Cumulative distributions for the collisions in the SS4 runs with respect to  (a) the collision parameter and (b) the collision velocity.  An additional curve in (a) is given comparing to a uniform distribution in the square of the collision parameter.  A general map (c) is shows the location of the collisions within the parameter space.  The  (black) box centered at $b_{col}/r_{col}\approx0.57$ indicates the region of parameter space for a hit-and-run type of collision, and the larger rectangle (magenta) in the same region denotes conditions for a large impactor.  An additional box (red) centered at $b_{col}/r_{col}\approx0.80$ indicates the canonical type of collision. A 2D histogram (d) illustrates the density of late collisions, with the color scale representing the gradient from high (red) to low (blue) density.}
\label{fig:coljd4}
\end{figure}

Finally, we characterize the collisions for the SS4 cases within the parameter space of the impact parameter ($b_{col}/r_{col}$) and the collision velocity ($v_{col}/v_{esc}$).  Figure \ref{fig:coljd4}c shows the joint distribution of the late collision outcomes for the SS4 runs while Figure \ref{fig:col} illustrates the the collision characteristics for the other mass ratios.  The median values of the impact parameter (0.70) and collision velocity (1.13) are shown as dashed lines. The highest density of points is in the lower right corner of the parameter space, i.e., slow collisions with large impact parameter, with most outcomes having a collision velocity ratio less than 1.13.  This is also evident in the respective cumulative distributions (Figs. \ref{fig:coljd4}a and \ref{fig:coljd4}b).  The collision cases considered in Fig. \ref{fig:coljd4}a appear to follow a uniform distribution in the square of the impact parameter, which we attribute to the weak gravitational focusing for most collisions. 

We have indicated the region of parameter space representative of a hit-and-run scenario (see $\S$\ref{sec:Sys_char} for details) on Figure \ref{fig:coljd4}c by a small black box centered at $b_{col}/r_{col} \approx 0.58$ with a collision velocity from 1.2 -- 1.3. { The domain contains a fair number of each outcome considered, but the ``success'' outcome that would describe the Earth-Moon system occurs least frequently.}

The successful conditions for a large (0.40 -- 0.45 M$_\oplus$) impactor akin to \cite{Canup2012}, the magenta box in Fig. \ref{fig:coljd4}c, is centered at $b_{col}/r_{col} \approx 0.60$, but with a collision velocity from 1.0 -- 1.6, where the evection resonance as used by \cite{Cuk2012} would redistribute the excess angular momentum into the heliocentric orbit of the Earth-Moon system.  The lower portion ($v_{col} \lesssim 1.1v_{esc}$) contains a higher density of points than the hit-and-run scenario.  We have denoted these regions of parameter space of the non-canonical impact scenarios to highlight the successful regions of parameter space representative of successful SPH initial conditions.

The red box (Fig. \ref{fig:coljd4}c) centered at $b_{col}/r_{col} \approx 0.8$ encloses the runs that could be described by a canonical impact \citep{Canup2004}.  The canonical hypothesis suggests a collision parameter approximately equal to 0.8, where we have considered the range $0.75 - 0.85$ and a collision velocity less than 1.10.  There is a high density of successes in the region of the $b_{col}-v_{col}$ plane consistent with the canonical impact scenario.

An alternative scenario considers a rapidly spinning Earth with a small impactor \citep{Cuk2012}; this is best approximated by our SS8 runs.  The domain of such a scenario within the collisional parameter space would have $v_{col}/v_{esc} = 1.35 - 1.80$ and $b_{col}/r_{col} = 0.00 - 0.15$, which for our case is very sparsely populated as indicated by the gray box in Fig. \ref{fig:coljd4}c.  In addition, these high velocity, small impactors would most likely have originated from beyond the orbit of Mars in order to account for the higher impact velocity and would tend to produce a system with a large $\left\langle{\rm AMD_{tp}}\right\rangle$.

\subsection{SS8 Results}
\label{sec:SS8}
We have produced similar maps for the SS8 set of runs as in the SS4 case, but at low resolution and over a substantially larger range in $a_S$.  We first describe results (SS8M) for the same range in $a_S$ as used for the other mass ratios (0.76 -- 1.55 AU) and then consider regions of initial $a_S$ interior and exterior to that range.  

\begin{figure*}[!ht]
\centering
\begin{tabular}{l}
	\includegraphics[width = 0.429\linewidth]{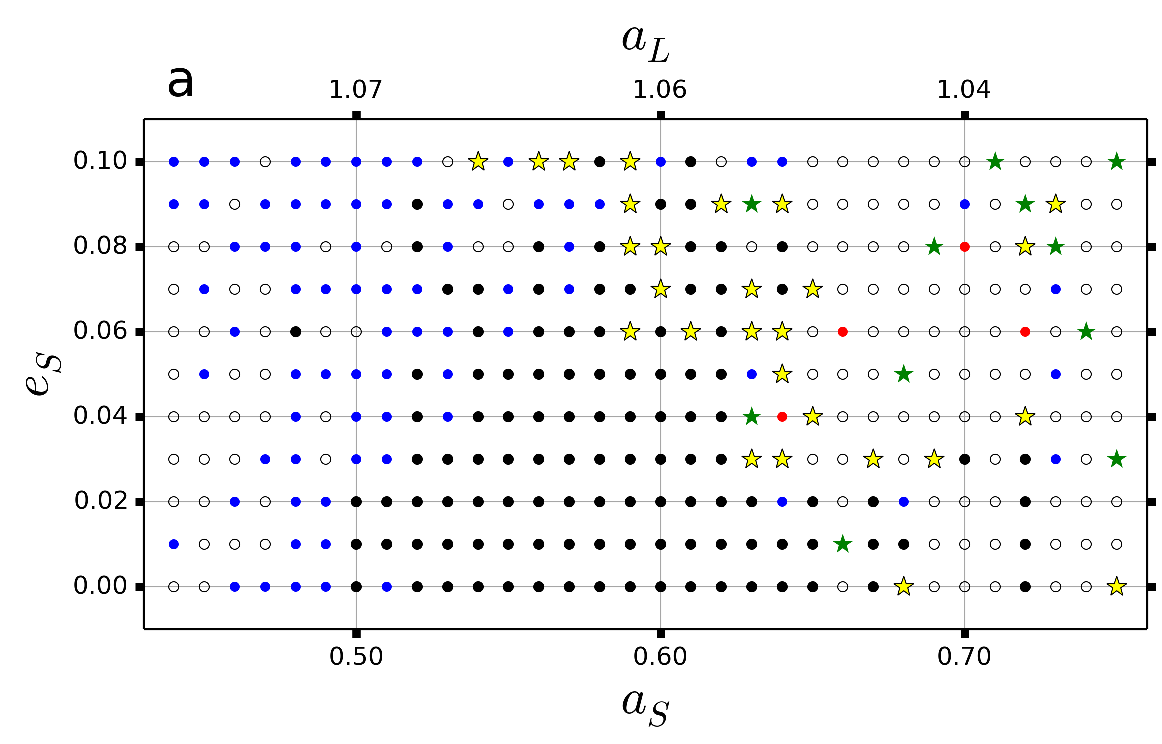}\\
	\includegraphics[width = 0.8\linewidth]{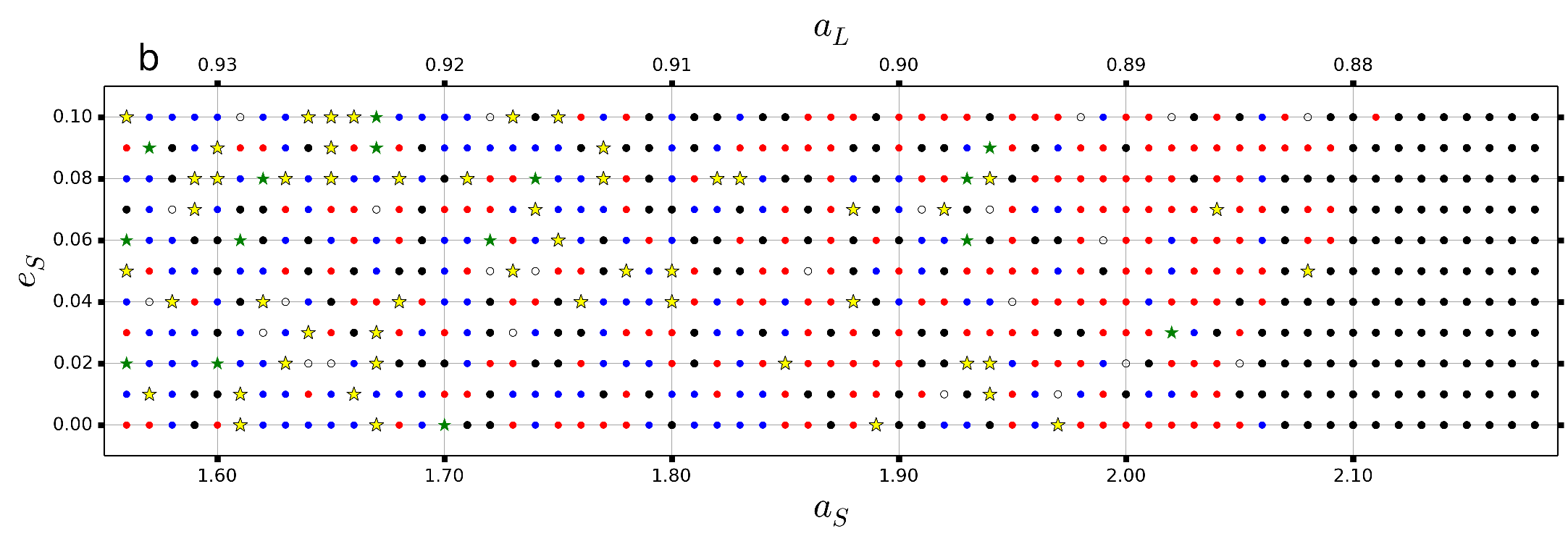}
\end{tabular}
\caption{Similar to Figure \ref{fig:gm}, but for our extended simulation sets: a) SS8I and b) SS8E.}
\label{fig:gm8IE}

\end{figure*}

In the characterization map (Fig. \ref{fig:gm}b), the distributions of ``early'' and ``non-SS mass'' in initial semimajor axis-eccentricity phase space is very similar to the SS4 results.  Our angular momentum constraint causes the ranges of initial $a_L$ and $e_L$ to differ for each mass ratio (see Table \ref{tab:IC2}).  The interpolated contour map (Fig. \ref{fig:cm}b) has lower resolution, and therefore it doesn't resolve the possible resonant structures as well.  Despite this limitation, Fig. \ref{fig:gm}b shows a wider instability region for the proto-Moon semimajor axis at $a_S\approx 0.86$ than is present in the SS4 runs (Fig. \ref{fig:gm}a).  There are regions that seem to be correlated with interplanetary resonances.  The most defined resonances correspond to the co-rotational ones that protect the proto-Moon from significant perturbations.  Formation at these locations is improbable within current theories of solid planet formation, and hence we consider these to be less likely scenarios.

Figure \ref{fig:gm8IE} shows characterization maps for the interior and exterior regions of SS8.  For the SS8I runs (Fig. \ref{fig:gm8IE}a), there is a stable region from a perihelion distance of $\sim$0.5 AU to aphelion distance of $\sim$0.65 AU.  Runs with the proto-Moon initially near either Mercury or Venus typically result in ``early'' collisions, while ``non-SS mass'' cases occur between Mercury and the stable region.  Cases resulting in a ``pseudo-success'' appear between the stable region and Venus.  Some ``success'' outcomes occur, but these cases have $\left\langle{\rm AMD_{tp}}\right\rangle > 2.0$.  Figure \ref{fig:gm8IE}b shows a much different set of outcomes when considering the SS8E runs.  There is a stable region ranging from 2.08 -- 2.18 AU.  Interior to 2.08 AU, most runs result in either “ejection” or “non-SS mass” outcomes, with other characterizations appearing  less frequently and without an apparent pattern, in contrast to the SS8I and SS8M results.

The SS8 runs roughly approximate the terrestrial planet interactions with a test particle and are consistent with such investigations \citep{Evans1999,Robutel2001}.  Specifically, the broad blue regions in Fig. \ref{fig:cm}b that we find to be stable correspond to regular orbits, whereas other areas may exhibit chaos as previously indicated \cite[Figs. 2a, 2b, \& 2c in][]{Robutel2001}. These regions are present in the other Solar System mass ratios as well, but they have differing  sizes because the test particle approximation is no longer as appropriate and because the proto-Earth is more displaced as a result of our assumption of conservation of total angular momentum of the Earth plus the Moon. 

\begin{figure}[!ht]
\centering
\includegraphics[width=\linewidth]{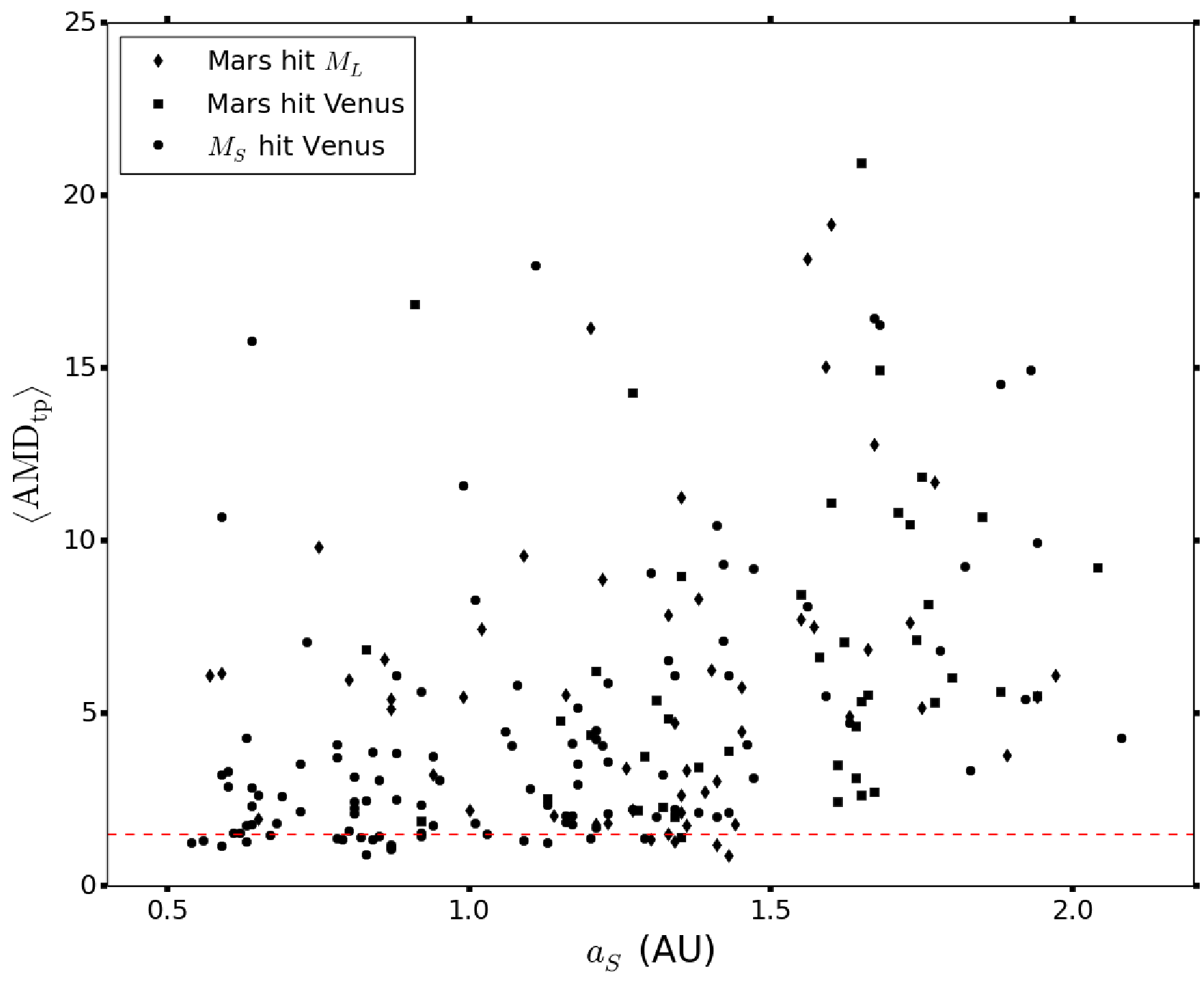}
\caption{$\left\langle{\rm AMD_{tp}}\right\rangle$ results considering the SS8 ``pseudo-success'' outcomes for the full range (0.44 AU -- 2.18 AU).  This set illustrates the wide range of $\left\langle{\rm AMD_{tp}}\right\rangle$ possible where the shape of each point represents each impact scenario as indicated.  The red dashed line indicates $\left\langle{\rm AMD_{tp}}\right\rangle=1.5$.  All ``pseudo-success'' simulations that begin with $a_S>1.5$ result in systems with $\left\langle{\rm AMD_{tp}}\right\rangle > 2$. }
\label{fig:pseudoAMD}
%\label{lastfig}
\end{figure}

\begin{figure}[!ht]
\centering
\includegraphics[width=\linewidth]{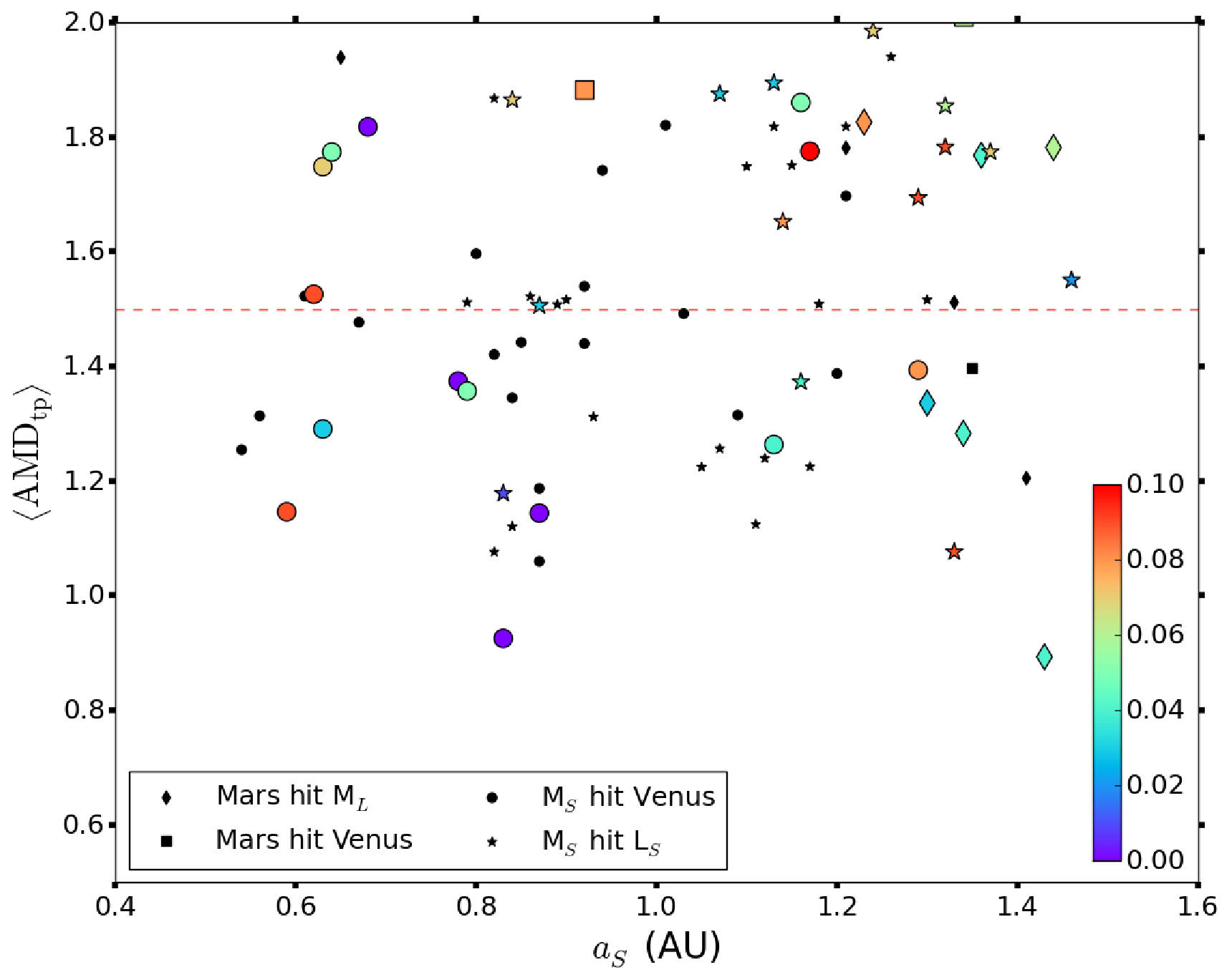}
\caption{$\left\langle{\rm AMD_{tp}}\right\rangle$ results considering the SS8 ``success'' and ``pseudo-success'' outcomes.  The colored points have collisions within the time range, $20-80$ Myr, and the color indicates the starting eccentricity, $e_S$, of the proto-Moon.  The red dashed line indicates $\left\langle{\rm AMD_{tp}}\right\rangle=1.5$ where colored cases below the line are considered to be similar to the Solar System.}
\label{fig:pseudoAMD_cut}

\end{figure}

\begin{figure}[!ht]
\centering
\includegraphics[width=\linewidth]{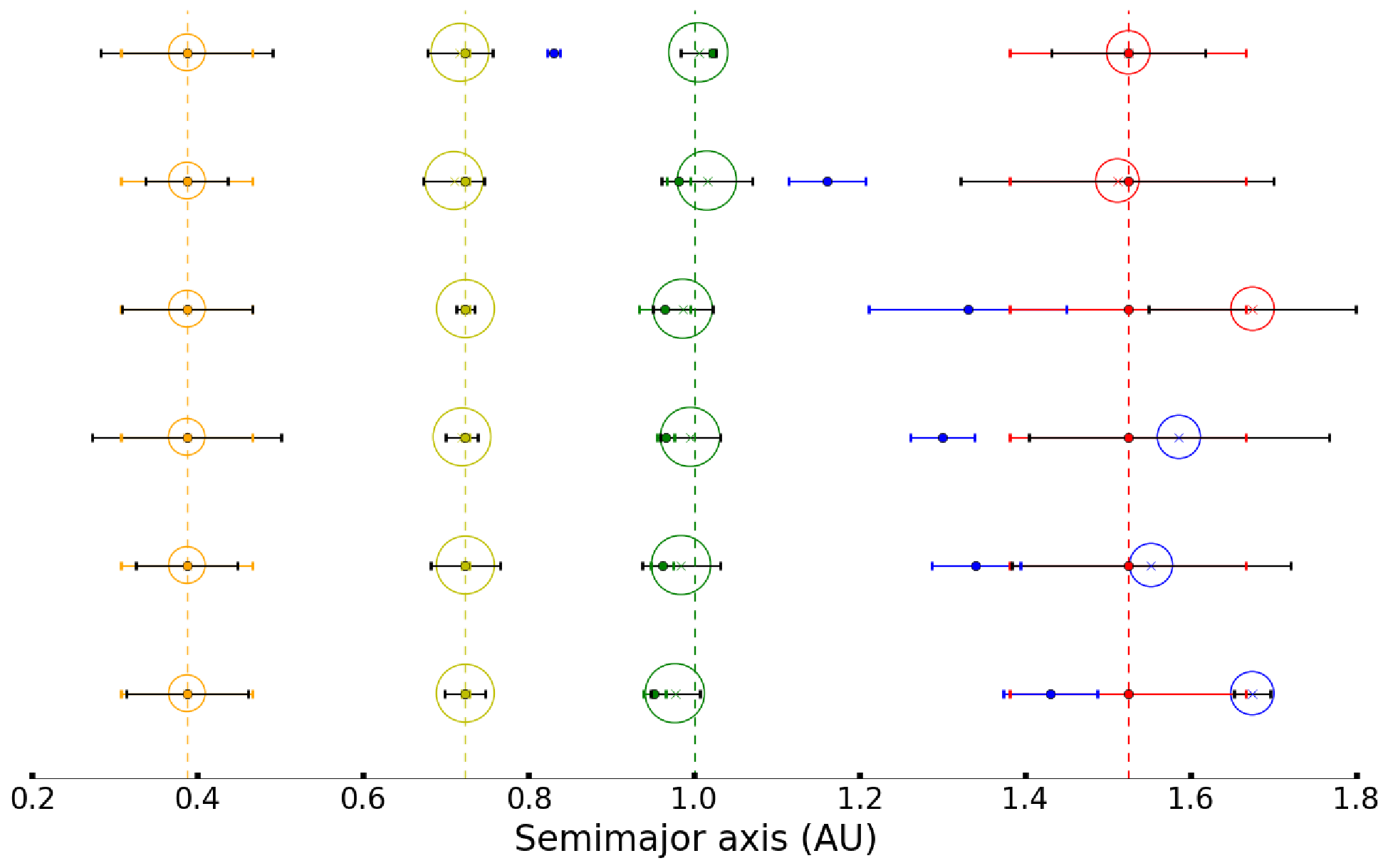}
\caption{Similar to Figure \ref{fig:ex_sim} but for the  8/1 Solar System ``success'' results listed in Table \ref{tab:ColSS_super} and the ``pseudo-success'' cases where Mars collides with the proto-Earth and the proto-Moon resides in an orbit consistent with present-day Mars.  The top 3 rows (``success'') and the bottom 3 rows (``psuedo-success'') have been selected by the collision time, $20-80$ Myr, and the $\left\langle{\rm AMD_{tp}}\right\rangle < 1.5$.}
\label{fig:8ex_sim}
\end{figure}

The SS8 case allows for an alternative scenario wherein a collision between Mars and the proto-Earth occurs leaving the proto-Moon at a semimajor axis consistent with present-day Mars.  This scenario originates from a mass degeneracy between the proto-Moon and Mars.  Figure \ref{fig:pseudoAMD} demonstrates that a wide range of $\left\langle{\rm AMD_{tp}}\right\rangle$ values are possible within three different collision scenarios of the ``psuedo-success'' category.  More importantly, Figure \ref{fig:pseudoAMD_cut} shows the ``psuedo-success'' and ``success'' runs that have an $\left\langle{\rm AMD_{tp}}\right\rangle$ value low enough to be considered similar to our Solar System.  The colored points that have a collision within a $20-80$ Myr timescale illustrate a small number of ``psuedo-success'' cases with low $\left\langle{\rm AMD_{tp}}\right\rangle$ values, roughly similar to our own, as indicated by the final states in the bottom 3 rows of Figure \ref{fig:8ex_sim}.  Figure \ref{fig:AMDgm} shows a similar trend in the SS8 results as compared to the SS4.  Figure \ref{fig:8ex_sim} demonstrates that ``success'' cases where a proto-Moon with an starting $a_S$ near Mars can result in a different final semimajor axis of Mars.

\begin{figure}[!ht]
\centering
\includegraphics[width=\linewidth]{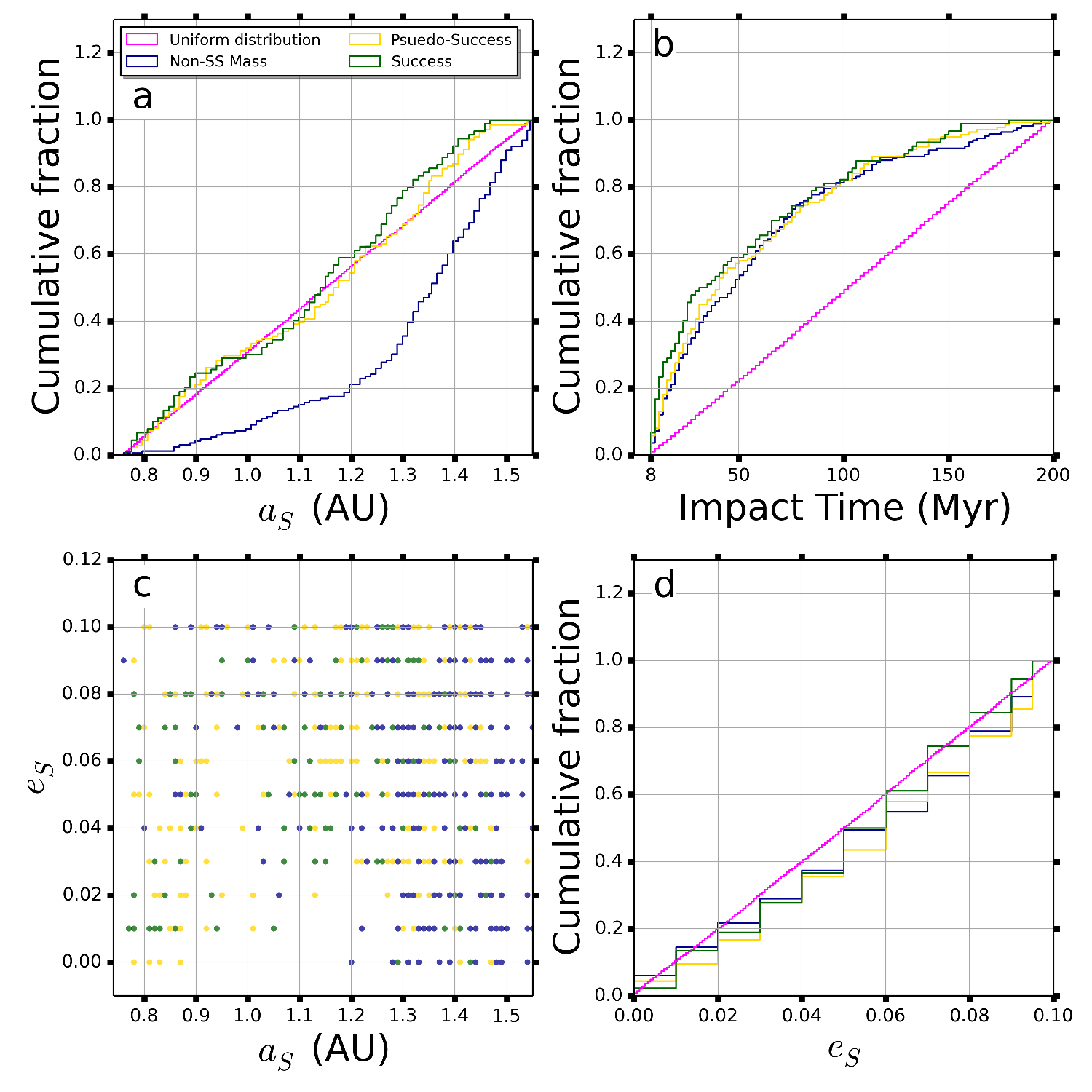}
\caption{Similar to Fig. \ref{fig:jd4}, but showing the results for the SS8 runs.  Results from the SS8I and SS8E regions have been excluded as they are dominated by outcomes (``early'' \& ``non-SS mass'') substantially different from the Solar System.}
\label{fig:jd8}

\end{figure}

The statistics of this case are similar to those in the SS4 with respect to the cumulative distributions of collision time and eccentricity (Figs. \ref{fig:jd8}b vs. \ref{fig:jd4}b and \ref{fig:jd8}d vs. \ref{fig:jd4}d).  The main difference is that the semimajor axis distribution (Fig. \ref{fig:jd8}a) for ``success'' and ``pseudo-success'' categories are more uniform, although regions with $a_S$ near 1.0 AU and 1.5 AU still have few collisions.  The lower mass (inertia) of the proto-Moon allows it to be more easily tossed around by the other planets, leading to this greater level of uniformity.

In the extended region of $a_S$ that we consider in the SS8 case (0.44 AU -- 2.18 AU), we see a more varied occurrence of outcomes, as shown in Table \ref{tab:gm8comp}.  For a proto-Moon initially close to or interior to Venus (SS8I), ``early'' collisions occur more frequently than in our standard region of interest and ejections are less common.  Most collisions in the 8 -- 200 Myr time window were ``non-SS mass''.  Neither ``success'' nor ``pseudo-success'' appear to occur very often and none of our simulations in this region could be considered as similar to the Solar System due to our constraints.  In contrast, the region beyond Mars, SS8E (1.55 AU -- 2.18 AU), shows that the primary characterizations occur in approximately equal quantities.  This region is expected to exhibit more chaos from simulations of test particles \citep{Robutel2001}.  When a collision does happen, the outcome of ``non-SS mass'' dominates, specifically with collisions between the proto-Moon and Mars.  Once more the ``success'' or ``pseudo-success'' outcomes do not reflect final systems similar to our own, as they have values of $\left\langle{\rm AMD_{tp}}\right\rangle$ $>$ 2, as shown in Figure \ref{fig:pseudoAMD}.

\begin{figure}[!ht]
\centering
\includegraphics[width=\linewidth]{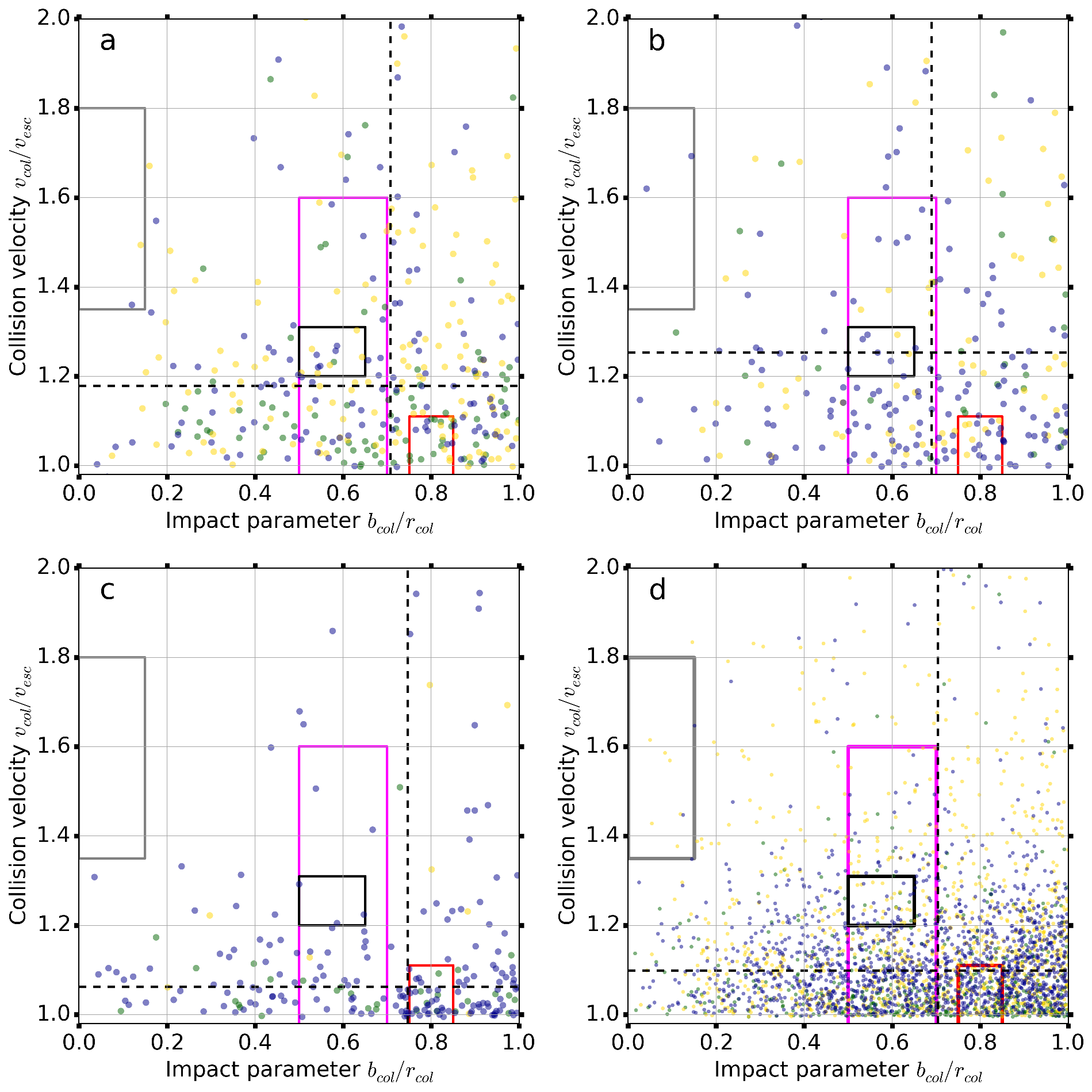} 
\caption{Collision parameter space similar to Fig. \ref{fig:coljd4}c, but for the (a) SS8M, (b) combined SS8I \& SS8E, (c) SS1, and (d) Nice4 runs.  The points are color coded by the collision outcome with ``success'' (green), ``pseudo-success'' (yellow), and ``non-SS mass'' (blue).}
\label{fig:col}
\end{figure}

{In Fig. \ref{fig:coljd4}c (SS4), most runs produced slow, grazing collisions and that trend continues within SS8 cases shown in Figs. \ref{fig:col}a and \ref{fig:col}b.  The median scaled impact parameter ($b_{col}/r_{col}$) (vertical dashed line) in Fig. \ref{fig:col}a is similar to Fig. \ref{fig:coljd4}c, but the median normalized collision velocity (horizontal dashed line) has increased to 1.18, despite the larger vlaue of $v_{esc}$.  Nonetheless, collisions corresponding to a small impactor scenario remain fairly rare in this parameter space. The hit-and-run, large impactor, and canonical regions are similarly populated compared to Fig. \ref{fig:coljd4}c.  For the combined SS8I and SS8E runs (Fig. \ref{fig:col}b), the median collision velocity has increased to 1.25 as a result of the proto-Moon starting from a more distant location relative to the proto-Earth.  The ``pseudo-success'' outcomes are of greater interest for this mass ratio (due to possibility of a Mars-Theia swap) and are absent from the domain for a small impactor.  They do appear in regions corresponding to the other impact scenarios, but the $\left\langle{\rm AMD_{tp}}\right\rangle$ is much higher from those collisions (Fig. \ref{fig:pseudoAMD}), possibly leading to a future instability.  Overall the SS8 runs tend to prefer a canonical impact scenario over the small impactor, as demonstrated by the ``success'' results in Fig. \ref{fig:col}a and the combination of ``success'' with ``psuedo-success'' outcomes in Fig. \ref{fig:col}b.}

\begin{table*}[ht]
\footnotesize
\centering
\begin{tabular}{|l|cr|cr|cr|cr|}
\hline 
\hline
Category  & \multicolumn{2}{c|}{Interior}	& \multicolumn{2}{c|}{Middle} & \multicolumn{2}{c|}{Exterior}  &  \multicolumn{2}{c|}{Total} \\
\cline{2-9}
					&  Counts & Percent &  Counts & Percent &  Counts & Percent &  Counts & Percent \\
\hline
Survived 200 Myr  				&  112 	& 31.8  &  195 	& 22.2 &	 240  & 34.1  & 547	 & 28.3	\\
Ejection 									&   4	  &  1.1  &   59	&  6.7 &	 210  & 29.8  & 273	 & 14.1  \\
Any Collision             &  236  & 67.1  &  626  & 71.1 &   254  & 36.1  & 1116 & 57.6  \\
\hline
``Early''						      &  123  & 35.0 &  232   & 26.4   &  24  &  3.4  &  379 & 19.6  \\
``Success'' 							&  11 	& 3.1  &   90 	& 10.2   &  16  &  2.3  &  117 &  6.0  \\
\hspace{0.5cm}``SS-like'' &   (0)  &  (0.0)  &  (3)   & (0.34) &  (0)  &  (0.0)   & (3)   &  (0.15) \\
``Pseudo-success'' 				&  27	  & 7.7  &  138 	& 15.7   &  53  & 7.5   &  218 & 11.3  \\
\hspace{0.5cm}``SS-like'' &   (0)  &  (0.0)  &  (3)   & (0.34) &  (0)  &  (0.0)   & (3)   &  (0.15) \\
``Non-SS mass''						&  75	  & 21.3 &  166	  & 18.9   & 161  &	22.9  &  402 & 20.7  \\
\hline
Mean survival time   & \multicolumn{2}{c|}{81.67}  & \multicolumn{2}{c|}{73.82}  & \multicolumn{2}{c|}{107.17} & \multicolumn{2}{c|}{87.92} \\
Median survival time & \multicolumn{2}{c|}{26.64}  & \multicolumn{2}{c|}{35.83}  & \multicolumn{2}{c|}{100.84} & \multicolumn{2}{c|}{56.73}\\
Standard deviation   &  \multicolumn{2}{c|}{88.02} &  \multicolumn{2}{c|}{78.59} & \multicolumn{2}{c|}{80.13}  & \multicolumn{2}{c|}{82.34} \\

\hline
\end{tabular}
\caption[Comparison of the results of this work for the 1/1 mass ratio to \cite{Rivera2002}]	% The class file doesn't do 
										% anything with the square 
										% bracketed short caption, 
										% but I always put one in.
	{Counts of the results for three different regions within the SS8 runs.  The interior (0.44 AU $\leq$ $a_S$ $\leq$ 0.75 AU), middle (0.76 $\leq$ $a_S$ $\leq$ 1.55 AU), and exterior (1.56 AU $\leq$ $a_S$ $\leq$ 2.18 AU) regions included 352, 880, and 704 runs, respectively.  The final column (total) combines all the counts of the SS8 runs from 0.44 AU -- 2.18 AU and presents the percentage for the larger parameter space.  The subcharacterization of ``SS-like'' requires that the time of collision occur between 20-80 Myr from the start of the simulation, the colliding bodies include the proto-Earth with either the proto-Moon or Mars, and the $\left\langle{\rm AMD_{tp}}\right\rangle$ be less than 1.5 times the mean value of the current Solar System terrestrial planets.  The mean and median survival times (in Myr) are given considering all outcomes from each region.}
	
	\label{tab:gm8comp}
	
\end{table*}

\subsection{SS1 Results}
\label{sec:SS1}
The SS1 set of runs (Fig. \ref{fig:gm}c) have very different distributions of mass than the SS4 and SS8 runs.  In this case, comparisons between the ``early'' and ``non-SS mass'' outcomes with previous results cannot be performed in a similar manner.  The semimajor axis of the ``proto-Earth'' (in this case equal in mass to the ``proto-Moon'')  can range in values from $0.555 - 1.261$ AU (see Table \ref{tab:IC2}) as prescribed by our angular momentum assumption and therefore includes Venus-crossing orbits.  So a large number of ``early'' outcomes are indicated for proto-Moon semimajor axes greater that 1.2 AU (1.25 AU for low eccentricity) due to the proximity of the proto-Earth to Venus.  The SS1 runs are also degenerate to the impactor (i.e., proto-Earth vs. proto-Moon) as the two bodies have the same mass in the SS1 runs.  

The resonances that involve L lie at different locations of S for different mass ratios.  These differences result from the relative values of $a_S$ and $a_L$ that scale with the mass ratio and are correlated with our constraint on angular momentum (i.e., the range of semimajor axes $a_L$ that the proto-Earth can occupy).  The SS1 runs thus show the more dramatic change in resonance locations compared to the SS4 and SS8M runs.  For example, compare the location of the 2L:3 MMR in Figs. \ref{fig:cm}a, \ref{fig:cm}b, and \ref{fig:cm}c. 

Despite these significant differences, there are similarities to the previously presented cases.  Large regions of stability exist in symmetric regions around 1.0 AU (Fig. \ref{fig:cm}c).  All the ``success'' outcomes are anchored around 1.0 AU as well.  These features speak to the plausibility of forming the Moon from similar mass impactors.  Namely, the progenitor pair are likely to have {orbited} close to one another at the epoch when our simulations begin, which may have implications concerning the expected isotopic compositions \citep{Wetherill1994,Herwartz2014}.

\begin{figure}[!ht]
\centering
\includegraphics[width=\linewidth]{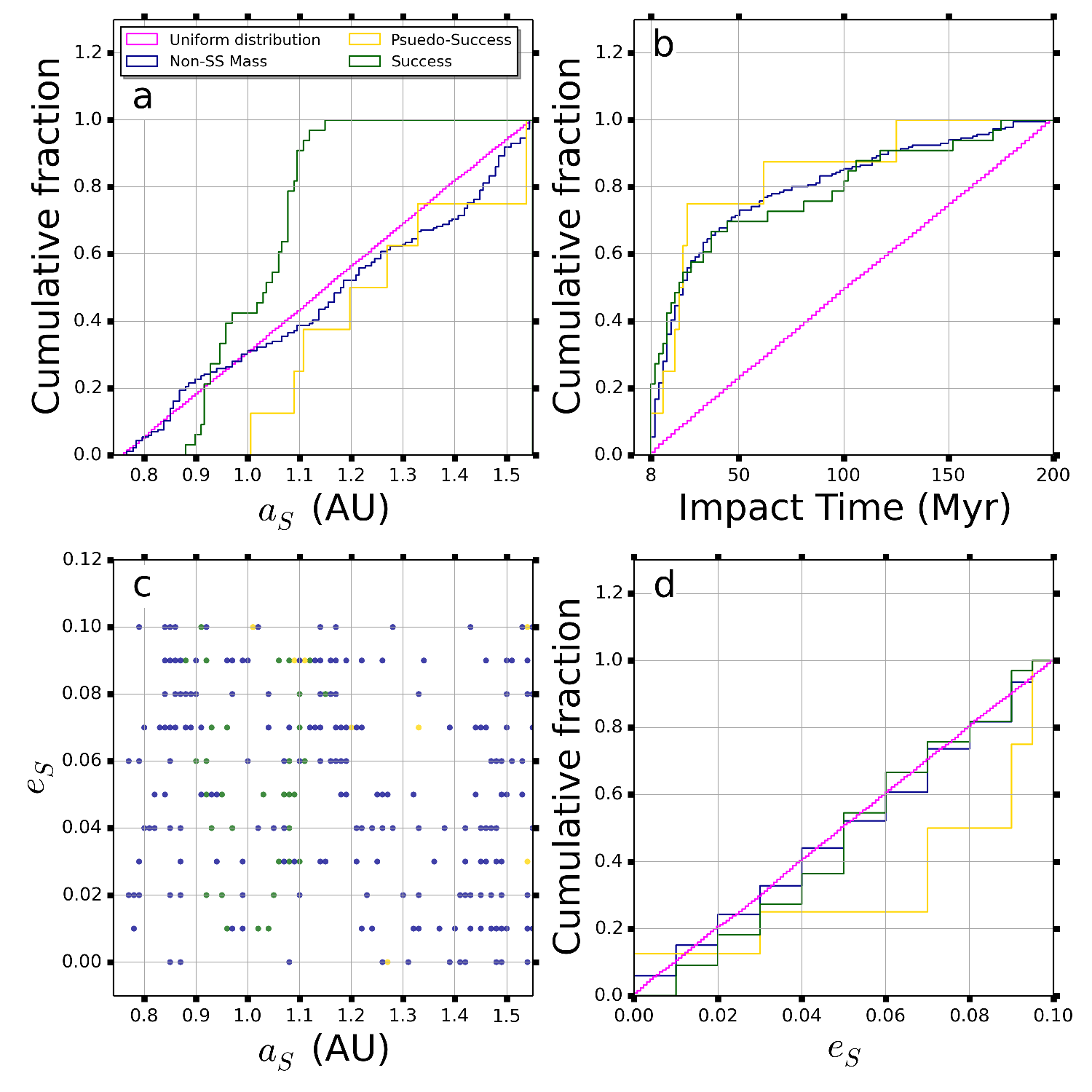}
\caption{Similar to Fig. \ref{fig:jd4}, but showing the results for the SS1 runs.}
\label{fig:jd1}
\end{figure}

The SS1 runs must be considered a little differently because the are far fewer ``pseudo-success'' outcomes as that category is significantly more difficult to produce.  Thus, we look primarily at the ``success'' outcomes as a group and the remaining late collision outcomes (``pseudo-success'' + ``non-SS mass'') as a separate group (Figure \ref{fig:jd1}).  As a result, the SS1 results have a clearer dichotomy in the resultant merged mass within the late collision outcomes as compared to the SS4 and SS8 results.  The distribution of the ``success'' outcomes has 2 populations approximately symmetric about 1.0 AU (as expected from the symmetry in masses), and the full population is reached at $a_S = 1.15$ AU as shown in Fig. \ref{fig:jd1}a.  In contrast, Fig. \ref{fig:jd1}a shows the other late collision outcomes are more uniformly distributed.  The collision time distribution (Fig. \ref{fig:jd1}b) resembles the SS4 case, but the ``success'' and ``non-SS mass'' outcomes are more tightly intertwined except for a small depletion of the ``success'' category at $50-100$ Myr.  The eccentricity distribution is approximately uniform, as in the previous cases (Fig. \ref{fig:jd1}d).

{The collisional characteristics for the SS1 runs (Fig. \ref{fig:col}c) display a more dramatic shift to a lower median normalized collision velocity (1.06) and a higher scaled impact parameter (0.75).  This is a result of the starting distance between the progenitors being smaller as compared to the SS4/SS8M runs and the difficulty in scattering more massive planets to orbits with higher eccentricities and inclinations.  Simulations that result in a ``success'' often have collision parameters similar to those called for in the large impactor and canonical impact scenarios.}

\subsection{Nice4 Results}
\label{sec:Nice4}
The terrestrial planets in our Solar System are perturbed by the giant planets, and the orbits of the giant planets may have been different during the early Solar System era that we are simulating than they are at the current epoch.  Thus we consider a Nice model giant planet arrangement (Table \ref{tab:niceelem}) for the 4/1 mass ratio, Nice4.  We have produced high resolution characterization and interpolation maps (Figs. \ref{fig:gm}d and \ref{fig:cm}d) to illustrate where a different giant planet model would become important.  Comparing Fig. \ref{fig:gm}d with the SS4 case in Fig. \ref{fig:gm}a, we see a similar (almost identical) landscape for a starting proto-Moon semimajor axis less than 1.1 AU.  In this region, the gravitational forces of Venus and/or the proto-Earth dominate over the perturbations of the much more distant giant planets.  But this is not the case for a semimajor axis greater than 1.1 AU.  A region of mostly stable systems at low eccentricity begins near 1.05 AU in both sets of simulations, but it extends to beyond 1.3 AU for Nice4 whereas it peters out near 1.25 AU for SS4.  The instability due to the 4L:3 MMR (Fig. \ref{fig:cm}a) is suppressed in the Fig. \ref{fig:cm}d; however other resonances (6L:5 and 5L:4) are present that induce similar perturbations, but to a lesser degree.  In the SS4 runs, there is a strip of ``success'' and ``pseudo-success'' outcomes near 1.165 AU that is missing from the Nice4 runs.  Also the ``success'' outcomes are rare beyond 1.3 AU for the Nice4 runs and more uniform in SS4 case.  Both cases have ``success'' peninsulas starting just exterior to 1.0 AU and extending up to 1.165 AU.  Fig. \ref{fig:cm}d illustrates the full extent of these differences in collision outcomes and the similarities with resonance structures between 1.25 and 1.4 AU. 

\begin{figure}
\centering
\includegraphics[width=\linewidth]{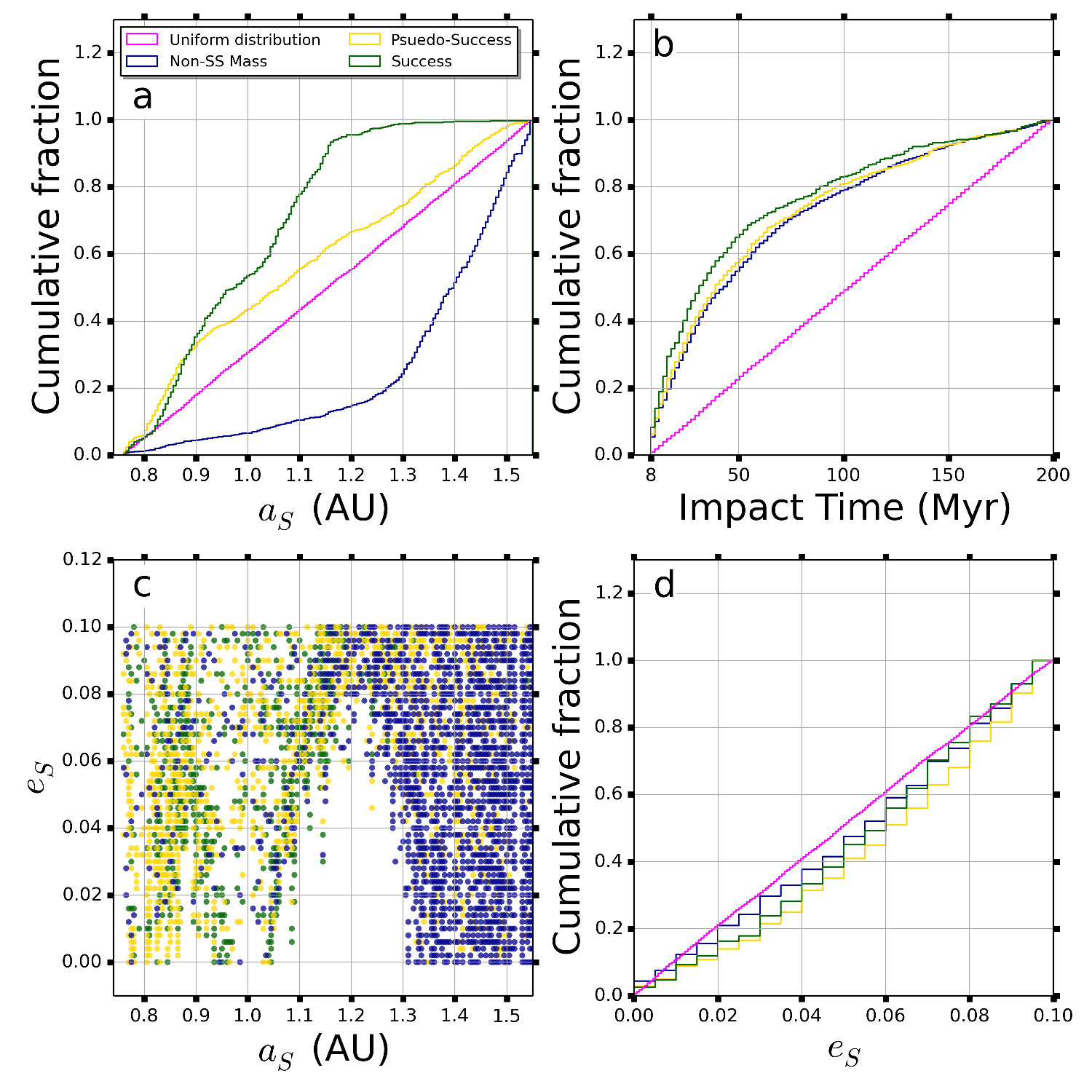}
\caption{Similar to Fig. \ref{fig:jd4}, but showing the results for the Nice4 runs.}
\label{fig:jd4N}

\end{figure}

The semimajor axis cumulative distributions (Fig. \ref{fig:jd4N}a) for the Nice4 case are similar to the SS4 set (Fig. \ref{fig:jd4}a), but the ``success'' outcomes rise more steeply due to the significantly lower occurrence for large values of $a_S$.  The ``pseudo-success'' outcomes follow a similar trend in the $0.76 - 0.9$ AU region and become more uniform beyond $0.9$ AU, which is quite a different trend than in the SS4 case.  For the cases that have collisions, the collision time and eccentricity distributions follow similar trends in the two sets of runs (Figs. \ref{fig:jd4N}b and \ref{fig:jd4N}d vs. Figs. \ref{fig:jd4}b and \ref{fig:jd4}d).

{The collision characteristics of Fig. \ref{fig:col}d vs. Fig. \ref{fig:coljd4}c depict a similar landscape, but with a lower median collision velocity (1.09).  This is likely due to a smaller interaction with the giant planets.  The hit-and-run, large impactor, and canonical impact scenarios contain substantially more dense domains that the small impactor and a stronger preference towards impact scenarios with a lower collision velocity (large impactor and canonical).}

Figure \ref{fig:AMDgm}a displays a similar distribution of the instantaneous post-impact AMD$_{\rm tp}$ for the Nice4 runs (yellow dots) as the previous runs considering the Solar System giant planets, albeit offset to slightly lower typical values.  A key difference occurs in Figure \ref{fig:AMDgm}b, where the upward shift from ${\rm AMD_{tp}}$ to the $\left\langle{\rm AMD_{tp}}\right\rangle$ that is observed in results for the Solar System runs does not occur nearly as strong for the Nice runs.  In particular, Figure \ref{fig:AMDgm}b shows $\sim$90 Nice4 simulations where $\left\langle{\rm AMD_{tp}}\right\rangle$ $<$ 1, but only one run with the Solar System giants has such low terrestrial planet AMD.  One would expect that if the interactions with the giant planets did not substantially affect the terrestrial planets, then these distributions would be similar.

There is a small but significant secular interaction between the giant planets and the inner Solar System.  From the basic setup of the Nice model, the giant planets are initialized on near circular, coplanar orbits with Jupiter's and Saturn's eccentricities dramatically reduced, $\left({e_{\rm SS} \over e_{\rm Nice}}\right)_{\rm Jup} \approx 9.61$ and $\left({e_{\rm SS} \over e_{\rm Nice}}\right)_{\rm Sat} \approx 4.21$.  The secular perturbations scale linearly with eccentricity and the higher-order mean motion perturbations also increase with eccentricity.  Thus secular perturbations from the giant planets are significantly reduced in the Nice model.  \ref{sec:colres} shows the amplified importance of these interactions for those simulations with $a_S=1.165-1.170$.

With Jupiter more distant and the giant planet AMD reduced by an order of magnitude, we also expect a lower inner planet AMD after the collision due to the ineffectiveness of the giant planets to pump the inner planet eccentricities throughout the simulation.  \cite{Brasser2009} showed that a set of circular, coplanar inner Solar System planets would experience significant pumping of their eccentricities as a result of the giant planet resonant crossings (650 Myr after CAIs) in the Nice model, so the Nice model would require lower $\left\langle{\rm AMD_{tp}}\right\rangle$ at the end of the era of early Solar System evolution that we are probing in this study. 

\begin{figure}[!ht]
\centering
\includegraphics[width = \linewidth]{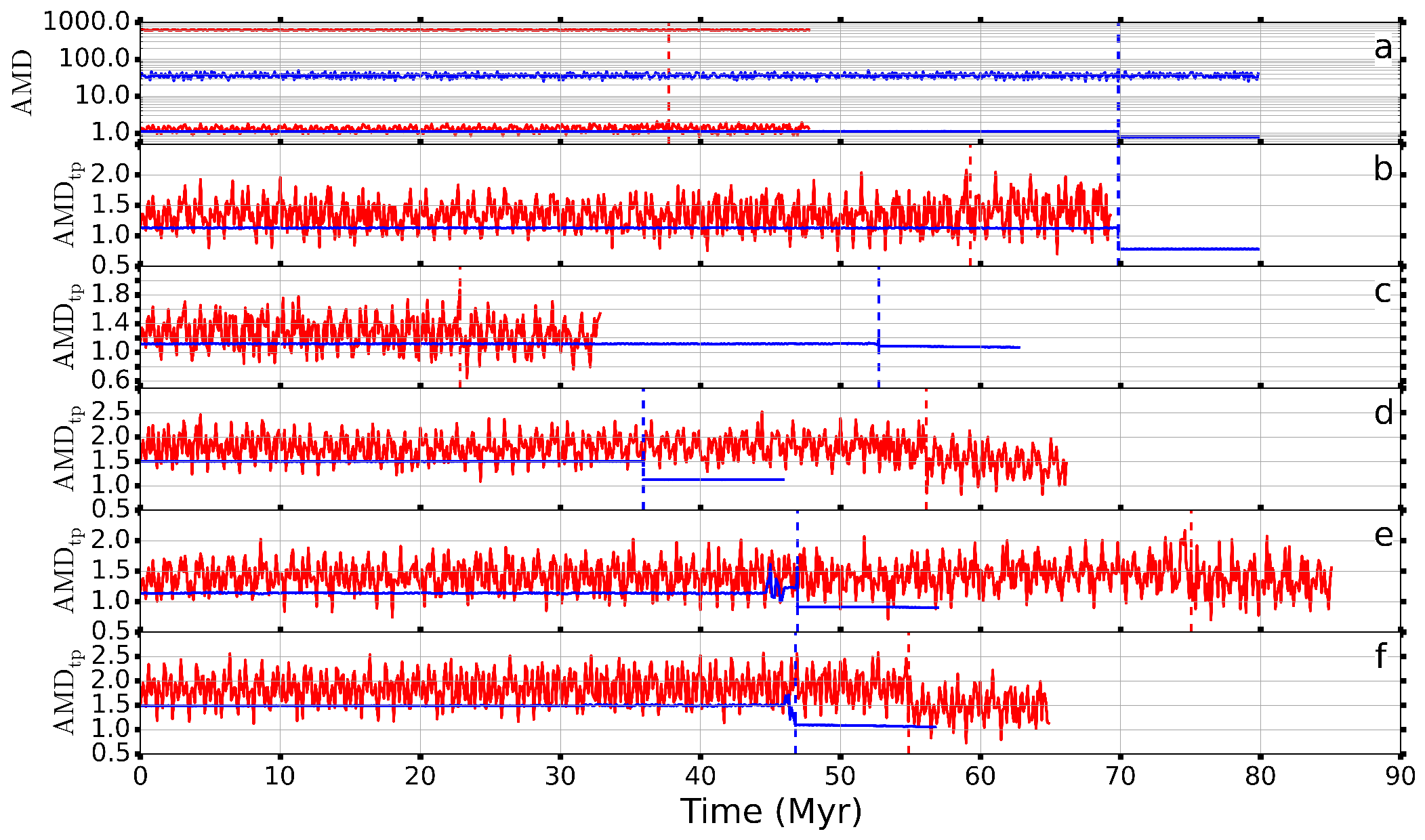}
\caption{The evolution of the summed AMD of the terrestrial and giant planets have been indicated in (a) for both the SS4 (red) and Nice4 (blue) runs with $a_S=0.920$ and $e_S=0.020$; note the order-of-magnitude difference in AMD.  The evolution has been evaluated up to the collision time (vertical dashed lines), which have been color coded to match the respective parent set, and continued for an additional 10 Myr.  The summed AMD evolution of the giants in the other runs look the same as in (a) and thus have been omitted.  The evolution of ${\rm AMD_{tp}}$ for the cases indicated in (b) $a_S=0.940$; $e_S=0.020$, (c) $a_S=0.945$; $e_S=0.014$, (d) $a_S=1.060$; $e_S=0.076$, (e) $a_S=1.085$; $e_S=0.022$, and (f) $a_S=1.125$; $e_S=0.074$ illustrate that variations with ${\rm AMD_{tp}}$ depend on the giant planet configuration assumed.  These cases were chosen because the respective SS4 and Nice4 run each result in a ``success'' within the restriction of a 20 -- 80 Myr time window and $\left\langle{\rm AMD_{tp}}\right\rangle < 1.5$, and they are depicted in Figure \ref{fig:EManalogs} by overlapping symbols.}
\label{fig:AMD}
\end{figure}

In Figure \ref{fig:AMD}, the evolution of the ${\rm AMD_{tp}}$ is given for six cases comparing the SS4 run (red) with the corresponding Nice4 (blue) run.  These cases were chosen as they were characterized as a ``success'' in the 20 -- 80 Myr collision time window and $\left\langle{\rm AMD_{tp}}\right\rangle < 1.5$ for both giant planet configurations.  Fig. \ref{fig:AMD}a also illustrates the corresponding sum of the AMD for the giant planets.  The evolution of the summed AMD of the giants looks the same in the other cases considered and thus have been omitted.  This comparison explains some of the trends apparent in Fig. \ref{fig:AMDgm}, specifically the mechanisms behind the changes between the instantaneous ${\rm AMD_{tp}}$ and $\left\langle{\rm AMD_{tp}}\right\rangle$, but the details of the individual cases presented are diverse as the result of chaotic variations.  

Overall in Fig. \ref{fig:AMD}, the ${\rm AMD_{tp}}$ the Nice4 runs are calmer with less variation as compared to the SS4 runs, where this difference is attributed to the much lower scaled AMD being transferred between the giant and terrestrial planets.  Another distinguishing feature is the evolution of the systems after the collision.  There is an equilibration (flatness) in the ${\rm AMD_{tp}}$ after the collision in the Nice4 cases, whereas the SS4 cases continue to display variations.  This demonstrates the perturbative effects of the giants in these two scenarios.  This is manifest in the more general picture of Fig. \ref{fig:AMDgm} with the Nice4 runs having ${\rm AMD_{tp}}$ relatively unchanged over an additional 10 Myr of simulation whereas the ${\rm AMD_{tp}}$ of SS4 cases vary more substantially.  Table \ref{tab:gmcomp} demonstrates that a correlation exists between the collision time and the giant planet architecture.  This likely results from a greater excitation of the terrestrial planets by the giant planets in the SS4 runs than in the Nice4 runs.

\subsection{Comparison of the Four Scenarios}
\label{sec:stat_gm}

By performing performing a statistical analysis of our results, we can estimate how the number of events per category scales with the mass ratio.  Table \ref{tab:gmcomp} shows the number of events per category and when normalized by the total number of runs per mass ratio.  We have included a second set of results using only those SS1 runs that have $a_L>0.76$ and also excluded the extended range simulations (SS8I and SS8E) to allow for a fairer comparison.  Upon inspection of the full sets (SS8M, SS4, and SS1), we can see that the percentage of runs that survive 200 Myr and eject due to instability both remain relatively flat with respect to the choice of mass ratio.  When we consider the range of SS1 in the final column, the 1/1 runs are more stable and have fewer ejections than the other SS runs.  There are also differences between the 1/1 and higher mass ratios for the categories of collisions due to the modified classification scheme for the 1/1 because of the reduced number of ways to make an Earth-like mass from bodies other than the progenitors. Thus we expect to see a depletion of ``pseudo-success'' counts, which is evident in the SS1 runs compared to an equal percentage in the higher mass ratio runs (see $\S$\ref{sec:Sys_char}).

\begin{table*}[ht]
\scriptsize
\centering
\begin{tabular}{|l|cr|cr|cr|cr|cr|}
\hline 
\hline
Category  & \multicolumn{2}{c|}{SS8M}	& \multicolumn{2}{c|}{SS4}  &  \multicolumn{2}{c|}{Nice4} &  \multicolumn{2}{c|}{SS1} &  \multicolumn{2}{c|}{SS1 ($a_L>0.76$)}	   \\
\cline{2-11}
					&  Counts & Percent &  Counts & Percent &  Counts & Percent &  Counts & Percent  &  Counts & Percent\\
\hline
Survived 200 Myr  				&  195 	& 22.2  &	1921  & 23.7 & 3001	& 37.0	& 	211 & 24.0  &		184 & 34.2	\\
Ejection 									&   59	&  6.7  &	 459  &  5.7 &   43	&  0.5  & 	 45	&  5.1  &		7 & 	1.3\\
Any Collision             &  626  & 71.1  & 5729  & 70.6 & 5065 & 62.5  &   624 & 70.9  &		347 & 64.5	\\
\hline
``Early''						      &  232  & 26.4  & 2108  & 26.1 & 1980 & 24.4  &   397 &	45.1  &		204 & 37.9	\\
``Success'' 							&   90 	& 10.2  &  612  &  7.5 &  516 &  6.4  &    33	&  3.8  &		32 & 	5.9\\
\hspace{0.5cm}``SS-like''     &   (3)  &  (0.34)  &  (49)  &  (0.60) &  (111) &  (1.4)  &     (1) &  (0.11) &		1 & 	(0.19)  \\
``Pseudo-success'' 				&  138 	& 15.7  & 1238  & 15.3 & 1005 & 12.4  &     8	&  0.9  &		4 & 	0.74\\
``Non SS-mass''						&  166	& 18.9  & 1769  &	21.7 & 1564 & 19.3  &   186 & 21.1  &		107 & 	19.9\\
\hline
Mean survival time & \multicolumn{2}{c|}{73.82} & \multicolumn{2}{c|}{76.21} & \multicolumn{2}{c|}{97.06} & \multicolumn{2}{c|}{62.32} & \multicolumn{2}{c|}{83.39}\\
Median survival time & \multicolumn{2}{c|}{35.83} & \multicolumn{2}{c|}{40.60} & \multicolumn{2}{c|}{63.86} & \multicolumn{2}{c|}{11.00} & \multicolumn{2}{c|}{21.82}\\
Standard deviation &  \multicolumn{2}{c|}{78.59} & \multicolumn{2}{c|}{79.03} & \multicolumn{2}{c|}{87.41} &  \multicolumn{2}{c|}{83.26} & \multicolumn{2}{c|}{90.15}\\

\hline
\end{tabular}
\caption[Comparison of the results of this work for the 1/1 mass ratio to \cite{Rivera2002}]	% The class file doesn't do 
										% anything with the square 
										% bracketed short caption, 
										% but I always put one in.
	{Counts of the results in the intermediate semimajor axis range, 0.76 $\leq$ $a_S$ $\leq$ 1.55 AU.  The SS8M and SS1 cases consider 880 total runs, and the SS4 and Nice4 cases each consider 8109 total runs.  The subcharacterization of ``SS-like'' requires that the time of collision occur between 20-80 Myr from the start of the simulation and the $\left\langle{\rm AMD_{tp}}\right\rangle$ be less than 1.5 times the mean value of the current Solar System terrestrial planets.  The mean and median survival times (in Myr) are given considering all outcomes from each mass ratio.  An additional column has been provided for the SS1 runs considering only the cases which $a_L>0.76$ ($a_S < 1.25$), and there are 538 total runs after making this selection.}
	%\label{lasttable}		% notice the second label for counting
	
	\label{tab:gmcomp}
	
\end{table*} 

We provide the mean, median, and standard deviation of the survival times for the various groupings of runs in Table \ref{tab:gmcomp}.  The associated event time corresponds to each ejection or collision; a value of 200 Myr is used for the NC case.  The SS4 and SS8M cases show similar values.  The SS1 full sample has shorter times, but the reduced SS1 set has longer characteristic survival times. The Nice4 set has longer times than SS4, which motivates a dynamical study that we present in  \ref{sec:colres}.  Table \ref{tab:gm8comp} demonstrates an increase in the median survival time with increasing heliocentric distance, where dynamical timescales are longer and the proto-Moon is farther from the most massive terrestrial planets.

\begin{table}[!ht]
\centering

\begin{tabular}{|r|l|ccc|}
\hline 
\hline
&          & \multicolumn{3}{|c|}{\bf Nice4}\\
\hline
&  & NC & Late &  Early   \\
\hline
 \parbox[t]{2mm}{\multirow{3}{*}{\rotatebox[origin=c]{90}{\textbf{SS4}}}}&
NC  				  &  1669 & 	 214  &   38   \\
&Late 				&  1234	& 	2137  &  694	  \\
&Early				&  98  	& 	 777  & 1248	  \\
\hline

\hline
\end{tabular}
\caption[Comparison of the results of this work to \cite{Rivera2002}]	% The class file doesn't do 
										% anything with the square 
										% bracketed short caption, 
										% but I always put one in.
	{
	Comparison of the results of the SS4 runs against the Nice4 runs.  Outcomes for each combination of $a_S,e_S$ of individual runs are placed into one of the broad categories: all of the planets survive for the full 200 Myr simulation (NC), a collision/ejection occurs before 8 Myr (Early), or a collision/ejection occurs after 8 Myr (Late).  The values shown here demonstrate how many simulations fell within a given category in the SS4 runs and the resulting category in the Nice4 runs (e.g., 98 Early outcomes in SS4 changed to NC in Nice4).
	%\label{lasttable}		% notice the second label for counting
	}
	\label{tab:SSvNice}
  
\end{table}

Table \ref{tab:SSvNice} shows a comparison of the SS4 and Nice4 runs with respect to the general outcomes: NC, early, and late.  The early category here corresponds to the grouping of the ``early'' collisions with the ejections within the same time regime ($<$ 8 Myr).  The late category describes a grouping of any collision or ejection that occurs after 8 Myr but prior to 200 Myr.  From this comparison we can see how the change in the giant planet configuration affects the outcomes.  For instance, there were 98 early outcomes in the SS4 set that became NC in the Nice4 set.  A clear majority of the runs ($62.5\%$) lie along the upper left to lower right diagonal that indicates similar lifetimes, which implies that the terrestrial planets are primarily responsible for their own dynamics.  Substantially more runs lie within the lower left triangle (2109) than in the upper right triangle (946), indicating the importance of giant planet perturbations.  But the fact that the upper right region still has a significant fraction of the outcomes ($\sim$$11.7\%$) suggests that chaotic variations are of comparable importance to the difference in giant planet perturbations between the two models.

\begin{table*}[ht]
\centering
\begin{tabular}{|l|crr|crr|cr|cr|cr|}
\hline 
\hline
Collision scenario   & \multicolumn{3}{c|}{SS8M} & \multicolumn{3}{c|}{SS8I/SS8E} & \multicolumn{2}{c|}{SS4} & \multicolumn{2}{c|}{Nice4} & \multicolumn{2}{c|}{SS1}\\
& All & S & P$_{\rm ML}$ & All & S & P$_{\rm ML}$ & All & S & All & S & All & S \\
\hline
small impactor &   2&0 &  0 &   2&0  &  0 &    9&1 &    7&0  &   0&0\\
large impactor &  74&20& 5 &  63&3  & 3 &  762&122 &  720&133  &  37&7\\
hit-and-run    &  11&1 &  1 &   7&0  &  0 &   78&8 &   47&6  &   3&0\\
canonical      &  21&8 &  1 &  16&0  &  0 &  224&53 &  241&54  &  29&6\\
other					 & 241&57&37 & 159&23& 15 & 2263&397 & 1981&311  & 146&18\\
\hline
\end{tabular}
\caption{Counts for the late combined collision categories (``success'', ``pseudo-success'', and ``non-SS mass'') that exhibit impacts consistent with the ($b_{col}/r_{col}$, $v_{col}/v_{esc}$) results of SPH simulations for each of the collision scenario domains.  These counts are derived from the same data that produced Figs. \ref{fig:coljd4}c and \ref{fig:col}a-d, where the S and P$_{\rm ML}$ columns denote the counts for the ``success'' and ``pseudo-success'' wherein Mars collides with the proto-Earth, respectively.  We have presented an additional row (``other'') to show the number of runs with a relatively low collision velocity ($v_{col}/v_{esc}<2$) that did not fall within each of the prescribed domains.}  
%Note that the area for the small impactor, large impactor, hit-and-run, and canonical impact scenarios should be used for fair comparisons that are 0.0675, 0.12, 0.015, and 0.01, respectively.}
\label{tab:colsc}
\end{table*}

{Table \ref{tab:colsc} provides the number of late collisions that are consistent with the domains of each impact scenario (Figs. \ref{fig:coljd4}c and \ref{fig:col}a-d) for low collision velocity ($v_{col}/v_{esc}<2$).  Each impact scenario has a different mass ratio assumption, so we limit the comparisons to reflect those assumptions.  The SS8 cases can correspond to the small impactor or canonical impact scenario and represent either the ``success'' or ``psuedo-success'' outcomes.  The ``pseudo-success'' outcomes that involve a collision with Mars and the proto-Earth are given because of the interesting possibility of a Mars-Theia swap.  In both the SS8M and extended regions (SS8I/SS8E), canonical impacts occur more often than the small impactor (extremely rare for all mass ratios).  The hit-and-run impact scenario roughly corresponds to our SS4 and Nice4 runs, which indicate that it is a possible impact scenario but not overly preferred occurring only for a few percent of all ``success'' cases.  The large impactor is consistent with our SS1 runs which account for $\sim$22\% of the ``success'' cases, which is by far the largest fraction by mass ratio obtained.  The SS1 cases have far fewer total counts of late collisions and small number statistics may affect this result.  However the ``other'' row shows that a collision can occur outside of the domain for any of the impact scenarios considered a large fraction of the time.}

\begin{figure}[!ht]
\centering
\includegraphics[width=\linewidth]{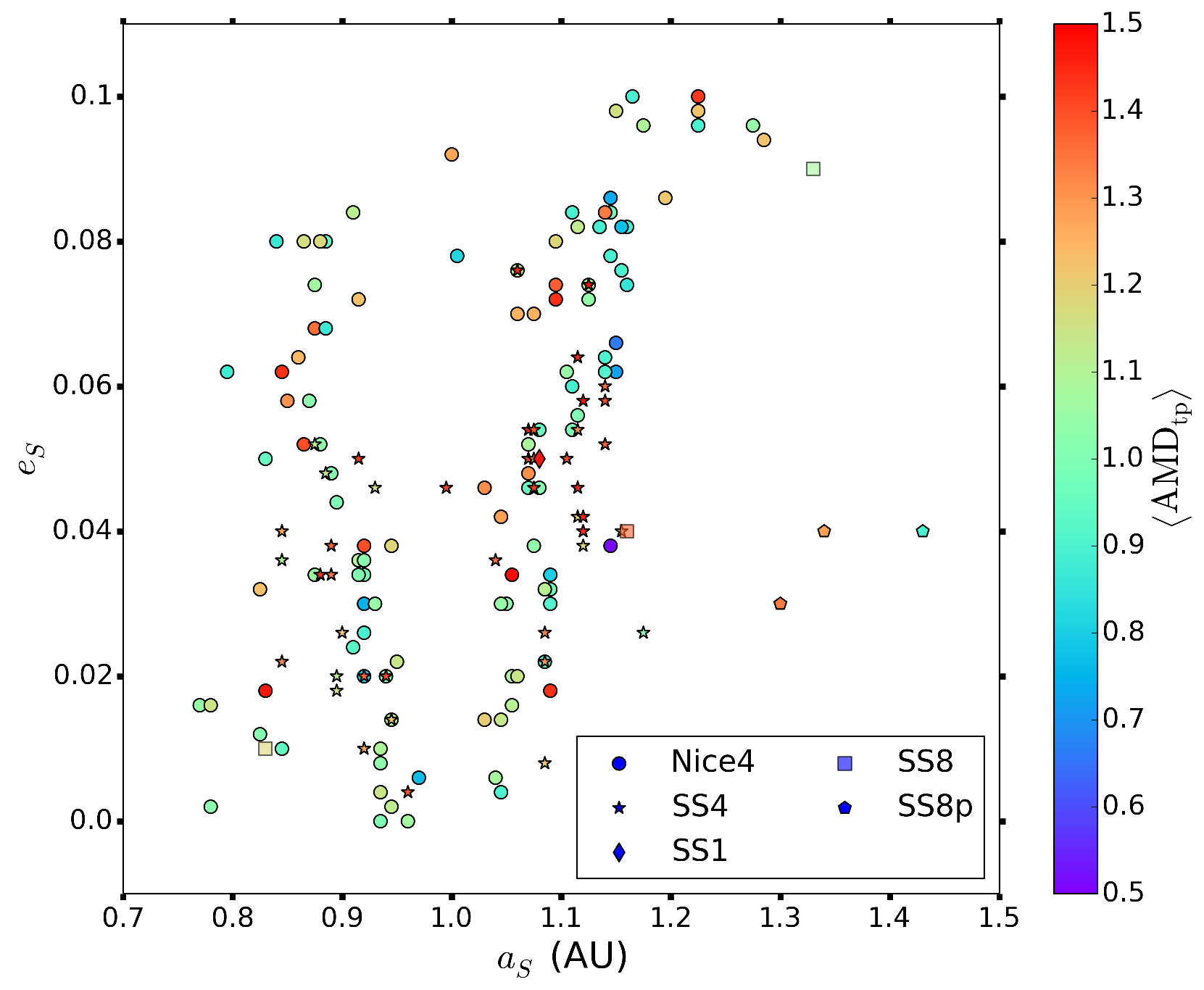}
\caption{Resulting sample of runs that had collisions thought to resemble the evolution of the early Solar System ($20-80$ Myr and $\left\langle{\rm AMD_{tp}}\right\rangle < 1.5$), as denoted by the shaded region in Figure \ref{fig:AMDgm}b, plotted with respect to the starting ($a_S,e_S$) parameter space.  The points are color coded with respect to the scaled $\left\langle{\rm AMD_{tp}}\right\rangle$.  The symbol shapes denote the set of runs from which the point was taken.}
\label{fig:EManalogs}
\label{lastfig}
\end{figure}

We present a distribution of candidates likely to resemble the Solar System in Figure \ref{fig:EManalogs} using the resultant merged mass, collision time, and $\left\langle{\rm AMD_{tp}}\right\rangle$ within the ($a_S,e_S$) parameter space.  These candidates were chosen with the requirements that the time of collision was $20-80$ Myr after formation of the bulk proto-Earth and the mean post-collision AMD was $<1.5$ times the reference scaled AMD.  Additionally we require that the resulting system match the heliocentric ordering and mass distribution of the terrestrial planets, as do all ``success'' outcomes and a particular subset of ``pseudo-success'' results within the 8/1 mass ratio.  Note that in most cases $a_S$ is close to 1 AU, with larger departures from this location typically found for higher $e_S$.  The high density of MMRs within the region may provide enough chaos due to their overlap to induce instabilities on a long ($20-80$ Myr) timescale.

Tables \ref{tab:ColSS_super} and \ref{tab:ColNice_super1} give the collision parameters for select simulations for each mass ratio with Solar System-like outcomes.  These parameters can serve as initial conditions for future SPH simulations.  For the SS cases we provide the parameters for the three lowest $\left\langle{\rm AMD_{tp}}\right\rangle$ simulations in each of the low resolution (8/1 and 1/1) mass ratios and for the 20 lowest $\left\langle{\rm AMD_{tp}}\right\rangle$ simulations in the high resolution (4/1) SS case.  As there are substantially more low $\left\langle{\rm AMD_{tp}}\right\rangle$ cases in the Nice4 runs, we provide parameters for only the 41 cases with $\left\langle{\rm AMD_{tp}}\right\rangle$ less than 1.0.  These simulations were selected in this manner because it has been shown that the $\left\langle{\rm AMD_{tp}}\right\rangle$ will increase as a result of a giant planet rearrangement that occurs at an epoch subsequent to our simulations according to the Nice model.  

Previous studies \citep{Agnor2012,Brasser2013} have demonstrated that it is most likely to obtain the current ${\rm AMD_{tp}}$ if the inner planets were at least 30\% calmer prior to the giant planet rearrangement, implying  $\left\langle{\rm AMD_{tp}}\right\rangle$ of 0.7.  Furthermore, the results of \cite{Laskar1997} showed up to 20$\%$ variation in ${\rm AMD_{tp}}$ relative to the current Solar System for $\sim$4 Gyr into the past and future. Thus, we believe these candidates are likely to best represent the post-collision state of the early Solar System for which we are probing.

{
\renewcommand{\baselinestretch}{1.0}
\small\normalsize
\begin{table*}[ht]
\small
\centering

\begin{tabular}{|r|ccccccccr|}
\hline 
\hline
&$a_S$ &$e_S$ 	& $b_{col}/r_{col}$ & $v_{col}/v_{esc}$	& $b_{\infty}/r_{col}$ & $v_{\infty}/v_{esc}$ & $L_{col}$ & $\left\langle{\rm AMD_{tp}}\right\rangle$ & $t_{col}$ \\
&(AU) &  & & & & & & &  (Myr)\\
\hline \parbox[t]{2mm}{\multirow{20}{*}{\rotatebox[origin=c]{90}{\textbf{4/1}}}}
&	0.845	&	0.022	&	0.975	&	1.069	&	2.755	&	0.378	&	3.188	&	1.318	&	39.97296	\\
&	0.845	&	0.036	&	0.761	&	1.044	&	2.638	&	0.301	&	2.433	&	1.101	&	60.77479	\\
&	0.845	&	0.040	&	0.901	&	1.049	&	2.973	&	0.318	&	2.893	&	1.275	&	37.27665	\\
&	0.875	&	0.052	&	0.990	&	1.020	&	5.004	&	0.202	&	3.088	&	1.099	&	26.77981	\\
&	0.885	&	0.048	&	0.722	&	1.016	&	4.116	&	0.178	&	2.245	&	1.137	&	31.69784	\\
&	0.895	&	0.018	&	0.689	&	1.024	&	3.171	&	0.223	&	2.159	&	1.149	&	30.02089	\\
&	0.895	&	0.020	&	0.534	&	1.024	&	2.493	&	0.219	&	1.671	&	1.118	&	20.49392	\\
&	0.900	&	0.026	&	0.988	&	1.011	&	6.784	&	0.147	&	3.054	&	1.221	&	57.67595	\\
&	0.920	&	0.010	&	0.574	&	1.012	&	3.709	&	0.157	&	1.778	&	1.267	&	56.49750	\\
&	0.930	&	0.046	&	0.789	&	1.027	&	3.452	&	0.235	&	2.478	&	1.165	&	54.77748	\\
&	0.945	&	0.014	&	0.886	&	1.025	&	4.058	&	0.224	&	2.779	&	1.231	&	22.83292	\\
&	1.075	&	0.050	&	0.703	&	1.047	&	2.365	&	0.311	&	2.252	&	1.337	&	75.47788	\\
&	1.085	&	0.008	&	0.581	&	0.999	&	--	&	--	&	1.775	&	1.213	&	70.62102	\\
&	1.085	&	0.022	&	0.822	&	1.000	&	--	&	--	&	2.513	&	1.300	&	75.05928	\\
&	1.085	&	0.026	&	0.811	&	1.009	&	6.118	&	0.134	&	2.501	&	1.332	&	79.18826	\\
&	1.115	&	0.042	&	0.418	&	1.002	&	7.174	&	0.058	&	1.280	&	1.203	&	49.73579	\\
&	1.115	&	0.054	&	0.897	&	1.011	&	6.078	&	0.149	&	2.773	&	1.307	&	69.64702	\\
&	1.120	&	0.038	&	0.281	&	0.999	&	--	&	--	&	0.857	&	1.178	&	34.66217	\\
&	1.155	&	0.040	&	0.814	&	1.007	&	6.829	&	0.120	&	2.508	&	1.200	&	55.95764	\\
&	1.175	&	0.026	&	0.626	&	1.012	&	4.035	&	0.157	&	1.937	&	1.009	&	74.22902	\\
\hline \parbox[t]{2mm}{\multirow{3}{*}{\rotatebox[origin=c]{90}{\textbf{1/1}}}}
&	1.04	&	0.01	&	0.096	&	1.009	&	0.740	&	0.131	&	0.476	&	1.748	&	28.16045	\\
&	1.08	&	0.05	&	0.855	&	1.008	&	6.889	&	0.125	&	4.232	&	1.471	&	64.33009	\\
&	1.10	&	0.07	&	0.579	&	1.054	&	1.826	&	0.334	&	2.998	&	1.811	&	24.07112	\\
\hline \parbox[t]{2mm}{\multirow{3}{*}{\rotatebox[origin=c]{90}{\textbf{8/1} s}}}
&	0.83	&	0.01	&	0.846	&	1.088	&	2.143	&	0.429	&	1.692	&	1.178	&	20.92872	\\
&	1.16	&	0.04	&	0.907	&	1.254	&	1.504	&	0.756	&	2.092	&	1.373	&	29.92395	\\
&	1.33	&	0.09	&	0.999	&	1.102	&	2.375	&	0.464	&	2.025	&	1.076	&	28.53637	\\
\thickhline \parbox[t]{2mm}{\multirow{3}{*}{\rotatebox[origin=c]{90}{\textbf{8/1} p}}}
& 1.30	&	0.03	& 0.953	& 1.016	& 5.458	& 0.177	& 1.703 & 1.336 &	63.12962 \\
& 1.34	& 0.04	& 0.358	& 1.259	& 0.590	& 0.764	& 0.794	& 1.283	& 39.33927 \\
& 1.43	& 0.04	& 0.907	& 1.206	& 1.621	& 0.675	& 1.925 & 0.893 &	20.36825 \\

\hline
\end{tabular}
\caption[Collision properties for simulations with $M_L:M_S = 8/1$, $e_L=0.05$, $e_S \approx 0.14$, $i_L={2 \over 3}^\circ$.]	% The class file doesn't do 
										% anything with the square 
										% bracketed short caption, 
										% but I always put one in.
	{
	Collision properties are given for simulations with the current Solar System gas giant planet architecture that have a collision within the time interval of $20-80$ Myr and result in the lowest $\left\langle{\rm AMD_{tp}}\right\rangle$.  We show the results for 3 simulations for low resolution 1/1 ``success'' set, 6 simulations for the low resolution 8/1 set (3 ``success'' and 3 ``pseudo-success'' in which Mars and L collided producing the third planet from the Sun), and 20 simulations for the high resolution 4/1 ``success'' set.  The first two columns give the starting semimajor axis ($a_S$) and eccentricity ($e_S$) of the proto-Moon.   The following four columns show the impact parameter ($b_{col}/r_{col}$) and velocity ($v_{col}/v_{esc}$) at the time of the collision ($t_{col}$) and these variables ``prior'' to the terminal approach, ($b_{\infty}/r_{col}$) and ($v_{\infty}/v_{esc}$).  The final three columns give the spin angular momentum ($L_{col}$), average summed terrestrial planet AMD $\left(\left\langle{\rm AMD_{tp}}\right\rangle\right)$, and collision epoch ($t_{col}$).  The 8/1, 1/1, and 4/1 mass ratios have a different collision radius ($r_{col}$) for each set of runs that are 9577, 10666, and 10101 km, respectively.  The associated escape velocity ($v_{esc}$) values are 9.16, 8.7, and 8.9 km s$^{-1}$.  The escape velocity for the ``pseudo-success'' (Mars + L) cases given is slightly reduced to 9.14 km s$^{-1}$  and the collision radius is 9559 km.
	%\label{lasttable}		% notice the second label for counting
	}
	\label{tab:ColSS_super}

\end{table*}
}

{
\renewcommand{\baselinestretch}{1.0}

\begin{table*}[ht]
\small
\centering

\begin{tabular}{|r|ccccccccr|}
\hline 
\hline
&$a_S$ &$e_S$ 	& $b_{col}/r_{col}$ & $v_{col}/v_{esc}$	&    $b_{\infty}/r_{col}$ & $v_{\infty}/v_{esc}$ & $L_{col}$ & $\left\langle{\rm AMD_{tp}}\right\rangle$ & $t_{col}$\\
&(AU) & & & & & & & &  (Myr)\\
\hline \parbox[t]{2mm}{\multirow{41}{*}{\rotatebox[origin=c]{90}{\textbf{4/1}}}}
&	0.795	&	0.062	&	0.586	&	1.139	&	1.223	&	0.546	&	2.042	&	0.874	&	63.95583	\\
&	0.830	&	0.050	&	0.984	&	1.045	&	3.387	&	0.304	&	3.146	&	0.953	&	71.41782	\\
&	0.840	&	0.080	&	0.458	&	1.031	&	1.882	&	0.251	&	1.443	&	0.867	&	64.70716	\\
&	0.845	&	0.010	&	0.471	&	1.013	&	2.908	&	0.164	&	1.460	&	0.948	&	41.69330	\\
&	0.885	&	0.068	&	0.678	&	1.008	&	5.249	&	0.130	&	2.090	&	0.868	&	21.07309	\\
&	0.885	&	0.080	&	0.565	&	1.047	&	1.899	&	0.312	&	1.811	&	0.954	&	23.59008	\\
&	0.895	&	0.044	&	0.883	&	1.042	&	3.151	&	0.292	&	2.814	&	0.995	&	68.09084	\\
&	0.910	&	0.024	&	0.430	&	1.049	&	1.428	&	0.316	&	1.379	&	0.922	&	26.14729	\\
&	0.920	&	0.020	&	0.623	&	1.021	&	3.097	&	0.205	&	1.946	&	0.783	&	69.83626	\\
&	0.920	&	0.026	&	0.551	&	1.003	&	7.247	&	0.076	&	1.690	&	0.891	&	26.02022	\\
&	0.920	&	0.030	&	0.984	&	1.022	&	4.760	&	0.211	&	3.076	&	0.748	&	53.08803	\\
&	0.920	&	0.034	&	0.485	&	1.002	&	7.337	&	0.066	&	1.488	&	0.989	&	22.58135	\\
&	0.935	&	0.000	&	0.587	&	1.026	&	2.632	&	0.229	&	1.841	&	0.997	&	74.14782	\\
&	0.940	&	0.020	&	0.772	&	1.000	&	42.819	&	0.018	&	2.361	&	0.990	&	29.78183	\\
&	0.970	&	0.006	&	0.687	&	1.020	&	3.514	&	0.199	&	2.142	&	0.766	&	39.14276	\\
&	1.005	&	0.078	&	0.568	&	1.159	&	1.123	&	0.586	&	2.013	&	0.816	&	27.20748	\\
&	1.045	&	0.004	&	0.376	&	0.999	&	--	&	--	&	1.148	&	0.900	&	61.30051	\\
&	1.070	&	0.046	&	0.615	&	1.021	&	3.021	&	0.208	&	1.923	&	0.958	&	21.50781	\\
&	1.080	&	0.054	&	0.506	&	1.000	&	17.932	&	0.028	&	1.548	&	0.956	&	61.90243	\\
&	1.085	&	0.022	&	0.628	&	1.000	&	--	&	--	&	1.920	&	0.909	&	46.93830	\\
&	1.090	&	0.030	&	0.130	&	1.016	&	0.732	&	0.180	&	0.404	&	0.916	&	22.38280	\\
&	1.090	&	0.032	&	0.980	&	0.997	&	--	&	--	&	2.990	&	0.964	&	34.70711	\\
&	1.090	&	0.034	&	0.549	&	1.019	&	2.836	&	0.197	&	1.711	&	0.802	&	45.49049	\\
&	1.110	&	0.054	&	0.934	&	1.005	&	9.464	&	0.099	&	2.872	&	0.992	&	32.52596	\\
&	1.110	&	0.060	&	0.570	&	1.037	&	2.162	&	0.273	&	1.807	&	0.887	&	28.36091	\\
&	1.110	&	0.084	&	0.387	&	1.211	&	0.685	&	0.684	&	1.433	&	0.894	&	21.62221	\\
&	1.135	&	0.082	&	0.580	&	1.077	&	1.565	&	0.399	&	1.908	&	0.890	&	77.08674	\\
&	1.140	&	0.062	&	0.815	&	1.007	&	7.054	&	0.116	&	2.509	&	0.907	&	40.66331	\\
&	1.140	&	0.064	&	0.046	&	1.016	&	0.258	&	0.180	&	0.142	&	0.898	&	45.77411	\\
&	1.145	&	0.038	&	0.683	&	1.016	&	3.823	&	0.181	&	2.122	&	0.515	&	78.32024	\\
&	1.145	&	0.078	&	0.238	&	1.055	&	0.748	&	0.335	&	0.766	&	0.892	&	77.72215	\\
&	1.145	&	0.084	&	0.679	&	1.039	&	2.515	&	0.280	&	2.156	&	0.965	&	28.67391	\\
&	1.145	&	0.086	&	0.570	&	1.049	&	1.894	&	0.315	&	1.827	&	0.732	&	58.82919	\\
&	1.150	&	0.062	&	0.682	&	1.009	&	5.178	&	0.133	&	2.106	&	0.721	&	68.68296	\\
&	1.150	&	0.066	&	0.470	&	1.010	&	3.418	&	0.139	&	1.451	&	0.663	&	42.46626	\\
&	1.155	&	0.076	&	0.392	&	0.997	&	--	&	--	&	1.197	&	0.899	&	56.59923	\\
&	1.155	&	0.082	&	0.787	&	1.013	&	4.978	&	0.160	&	2.438	&	0.770	&	40.85851	\\
&	1.160	&	0.074	&	0.900	&	1.005	&	9.420	&	0.096	&	2.766	&	0.855	&	30.35833	\\
&	1.160	&	0.082	&	0.829	&	0.997	&	--	&	--	&	2.526	&	0.884	&	56.37437	\\
&	1.165	&	0.100	&	0.954	&	1.009	&	7.272	&	0.132	&	2.943	&	0.891	&	72.27950	\\
&	1.225	&	0.096	&	0.960	&	1.023	&	4.513	&	0.218	&	3.005	&	0.896	&	38.51745	\\

\hline
\end{tabular}
\caption[Collision properties for simulations with $M_L:M_S = 8/1$, $e_L=0.05$, $e_S \approx 0.14$, $i_L={2 \over 3}^\circ$.]	% The class file doesn't do 
										% anything with the square 
										% bracketed short caption, 
										% but I always put one in.
	{
	Collision properties for 41 select simulations in the Nice4 set with the lowest $\left\langle{\rm AMD_{tp}}\right\rangle$ values and within the collision time window of $20-80$ Myr.  See Table \ref{tab:ColSS_super} for symbol definitions.  The collision radius ($r_{col}$) and escape velocity ($v_{esc}$) values for the Nice4 runs are 10101 km and 8.9 km s$^{-1}$, respectively.
	%\label{lasttable}		% notice the second label for counting
	}
	\label{tab:ColNice_super1}
	\label{lasttable}		% notice the second label for counting
\end{table*}
}

\section{Conclusions}
\label{sec:conc}
{This study has investigated the possible orbital parameters of the Earth-Moon progenitors at an era starting $\sim$$30-50$ Myr after the formation of CAIs.}  In doing so, we have probed a parameter space that depends on the assumed mass ratio of the progenitors ($m_L/m_S$), starting semimajor axis of the proto-Moon ($a_S$), its eccentricity ($e_S$), and the orbits of the giant planets within the system.  Our primary constraint {for solutions to be considered realistic} comes from the time of collision between the progenitors, as it is our best known observable from radioactive dating of the Apollo lunar samples.  The dating measurements have limited the formation of the Moon to occur 70 -- 110 Myr after the formation of the CAIs, which implies a time interval of $20-80$ Myr after the bulk formation of the (proto-)Earth {(which is estimated to have occurred $30-50$ Myr after the formation of the CAIs).  Assuming that the vast majority of planetesimals have been accreted}, we have shown in our model that a significant fraction of orbital configurations allow a fifth terrestrial planet to collide with the proto-Earth on a $20-80$ Myr timescale.  Within this subset of configurations, we consider the cases in which the terrestrial planets end up with low AMD (i.e., small orbital eccentricities) to be Solar System-like and show that they largely correspond to a closely spaced pairing of the Earth-Moon progenitors at the epoch when our integrations began.  

Some parameters allow for switching of the planet ordering in semimajor axis to occur among the two initially outermost terrestrial planets and lead to a collision between a Mars-sized body with the proto-Earth.  But in that case, the outermost terrestrial planet after the Giant Impact is farther from the Sun than was the outer most terrestrial planet at the beginning epoch.  Thus, all five of the terrestrial planets would have been interior to the present orbit of Mars at the beginning of our simulations in order to account for the current state of the Solar System.  Note that this scenario applies \emph{only} if the the mass of the proto-Moon is approximately equal to the mass of Mars.

Many large-scale trends in the evolution of our simulated system as a function of the starting position of the impactor are independent of the mass of the impactor and of the configuration of the giant planets (contemporary Solar System versus the Nice model).  However the extent of the stability region in our models is dependent on the choice of mass ratio, where the Solar System 1/1 (SS1) and Nice model 4/1 (Nice4) cases both demonstrate larger stable zones than those of the Solar System 4/1 (SS4) runs and the Solar System 8/1 (SS8M) runs have a smaller stable zone.

Following \cite{Rivera2002}, we have described simulations where the Earth-Moon progenitors collide within 8 -- 200 Myr of evolution as our ``success'' outcomes.  The location of the proto-Moon at the beginning of the epoch that we are simulating in many of these ``success'' cases lies on the border of an unstable region approximately symmetric about $a_S = 1.0$ AU {(spanning roughly from 0.8 to 1.2 AU)}.  The possibility that the proto-Moon arose from this region is further supported by the post-collision angular momentum deficit averaged for 10 Myr, $\left\langle{\rm AMD_{tp}}\right\rangle$, which describes the degree of dynamical excitation of the resulting systems.  {These dynamical results are consistent with recent empirical evidence that indicates the putative giant impactor was not carbonaceous chondritic and contained only a slightly higher $\Delta^{17}$O value than that of Earth \citep{Herwartz2014,Hartmann2014}, implying that the material from which it was comprised condensed in a similar region as the proto-Earth.}

Our results show far more systems with low $\left\langle{\rm AMD_{tp}}\right\rangle$ after a collision between the Earth-Moon progenitors for the Nice configuration of giant planets than for the SS configuration.  However, the Nice model requires calmer, more ordered systems after the last Giant Impact since these systems typically have their $\left\langle{\rm AMD_{tp}}\right\rangle$ increased by eccentricity pumping during the giant planet rearrangement \citep{Brasser2009,Brasser2013,Agnor2012}.  Therefore, our results cannot be used to determine a preference for either the SS or Nice configurations without simulating the effect of the giant planet rearrangement in the Nice model on the ${\rm AMD_{tp}}$ of the systems formed.  Note also that a higher percentage of ``success'' outcomes in the Nice model extend to a broader range in starting $a_S$ as compared to the 4/1 mass ratio with the current giant planet architecture.

{We have computed the scaled impact parameter ($b_{col}/r_{col}$) and collision velocity ($v_{col}/v_{esc}$) for the collisions in our simulations.  The combination of a slow and grazing impact (with $b_{col}/r_{col} \gtrsim 0.7$ and $v_{col}/v_{esc} \lesssim 1.2$) appears most often.  Different impact scenarios determined by smooth particle hydrodynamic (SPH) models assume different mass ratios and require different combinations of impact parameter and collision velocity.  Through this characterization, we find the collision parameters appropriate for the hit-and-run  scenario \citep{Reufer2012} to occur less frequently that those appropriate to the canonical impact scenario \citep{Canup2001}.  However, our results do not discriminate strongly between the canonical and large impactor scenarios.   Our results show weak gravitational focusing manifested in the impact parameter, which implies a preference against head-on collisions, especially those that are fast ($v_{col}/v_{esc} > 1.2$).  As a result, we find the small impactor scenario \citep{Cuk2012} to be less likely in our current parameter space as fast collisions are also expected to produce higher values of $\left\langle{\rm AMD_{tp}}\right\rangle$.}  

{Future SPH investigations are required to reconcile the mass and compositional constraints, especially noting that a large fraction result of our simulations result in collision parameters outside the given impact scenarios.  These SPH studies could use collision parameters given in Tables \ref{tab:ColSS_super} and \ref{tab:ColNice_super1}, which list results for runs having an impact with the prescribed time interval that yields a dynamically calm system analogous to the actual Solar System.  Other recent studies \citep{Elser2011} have investigated in detail which collisions are likely to form satellites, and \cite{Nakajima2014} showed that the canonical and 7:3 scenarios are more likely to produce results commensurate with the expected thermodynamics of the circumplanetary disk to form the Moon.  Furthermore, \citep{Meier2014} have recently indicated that the hit-and-run and canonical scenarios to be the most likely based on the published isotopic predictions of each model.  Much of the uncertainty in these models lies in only having samples of the mantle from the Earth-Moon system and Mars, where measurements from the mantles of Venus and Mercury would help to fully address the assumptions made for terrestrial embryo compositions.}

\section*{Acknowledgements}
B. Q. gratefully acknowledges a Fellowship from the NASA Postdoctoral Program.  The authors thank N. Haghighipour for stimulating conversations over the course of this work.  B. Q. acknowledges S. Satyal for his assistance with computational resources.  We thank T. Dobrovolskis, D. Jontof-Hutter, and A. Morbidelli for helpful comments on the manuscript.


\label{lastpage}
{
\appendix

\section{Resonant Case Study}
\label{sec:colres}
From Figure \ref{fig:cm}, we can see features indicative of resonant dynamics, especially in the SS4 and Nice4 simulations, for which we have high-resolution plots.  The 1:1 resonances have been noted in previous sections, but additional resonances are also evident, including the neighboring first-order mean motion resonances that are roughly symmetric about 1.0 AU. 

We next consider the narrow unstable zone in the range of $a_S=1.165-1.170$ AU in the SS4 runs (Fig. \ref{fig:cm}b) because this feature is missing in the Nice4 runs (Fig. \ref{fig:cm}d).  This unstable zone stands out in the SS4 case as it persists for all eccentricities considered.  In order to investigate this region, we display the variation in the specific elements related to the mean motion and secular components $\left\{e,\varpi_{\rm Jup}-\varpi_S\right\}$ in Figure \ref{fig:secres} for two beginning values of $a_S$ and for both the SS4 and Nice4 cases.  This sample of the parameter space is used to explore the possible chaos induced by the 4L:3 MMR that is located at $a_S\approx1.1672$ AU.  Through this comparison, a nearby secular resonance with Jupiter \citep{Froeschle1989,Batygin2011,Batygin2013a} in the Solar System case is found to enhance eccentricity pumping and destabilize the region.  No comparable secular resonance exists in our results for the Nice Model (Fig. \ref{fig:cm}d).

Figure \ref{fig:secres} illustrates the variation for the first 1 Myr of evolution in the orbital elements $\left\{e,\varpi_{\rm Jup}-\varpi_S\right\}$ for the SS4 (Figs. \ref{fig:secres}a, \ref{fig:secres}b, \ref{fig:secres}e, and \ref{fig:secres}f) and Nice4 (Figs. \ref{fig:secres}c, \ref{fig:secres}d, \ref{fig:secres}g, and \ref{fig:secres}h) cases considering two nearby starting semimajor axes, $a_S = 1.160$ AU (top four panels) and $a_S = 1.165$ AU (bottom four panels).  These results are categorized by the scale of variation in $\varpi_{\rm Jup}-\varpi_S$ and its corresponding effect on the evolution of eccentricity.  Figure \ref{fig:secres}a demonstrates a slow variation and near resonant behavior that leads to pumping of eccentricity and a larger value of maximum eccentricity, $e_{\rm max}$, in Figure \ref{fig:secres}b.  Figs. \ref{fig:secres}c and \ref{fig:secres}g show rapid variation in $\varpi_{\rm Jup}-\varpi_S$, within the Nice model, leading to a lower value of $e_{\rm max}$ in the corresponding Figs. \ref{fig:secres}e and \ref{fig:secres}h.  

Figure \ref{fig:secres}e begins with circulation in $\varpi_{\rm Jup}-\varpi_S$, but achieves a  temporary secular resonance with Jupiter after 0.5 Myr.  The switching between circulation and libration indicates a source of chaos \citep{Wisdom1980,Tsiganis2010,Batygin2013b,Deck2013}, which becomes manifest in a rapidly varying eccentricity and the largest value of $e_{\rm max}$.  This leads this particular simulation to eventually go unstable.

From the comparison of nearby initial conditions, the evidence suggests that a weak MMR resides in the region $a_S = 1.165-1.170$ and induces some excitation of eccentricity independent of the chosen architecture of the giant planets in our 5 terrestrial planet model.  However, the addition of a secular resonance from Jupiter in the SS4 configuration changes the landscape from stable (for 200 Myr) to unstable due to the additional chaos.

\begin{figure}[!ht]
\includegraphics[width=\linewidth]{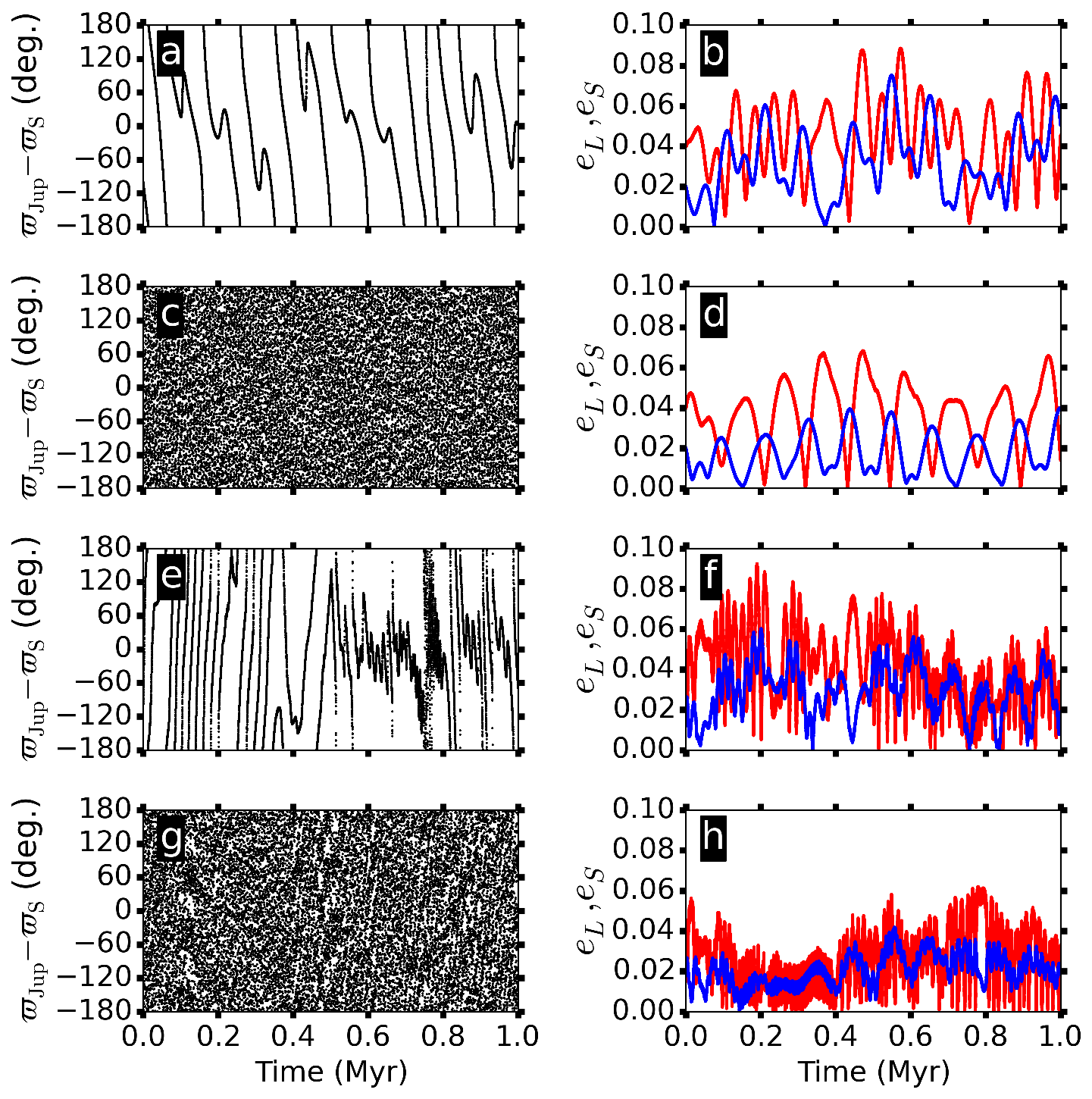}
\caption{Comparison of elements indicative of a 4L:3 mean motion resonance for the 4/1 mass ratio in the Solar System (a,b,e,f) and Nice (c,d,g,h) architectures with $a_S=1.160$ (a-d) and $a_S=1.165$ (e-h).  All four cases begin with the same eccentricity $e_S=0.04$ and $e_L=0.02$ for the proto-Moon and proto-Earth, respectively. The panels on the right (b,d,f,h) illustrate the eccentricity variations of both the proto-Earth (blue) and the proto-Moon (red) for the first 1 Myr of evolution.  The panels on the left (a,c,e,g) demonstrate a test of the possible commensurability of precession with respect to Jupiter.  We note the variations present in the SS4 case with $a_S=1.165$ (e,f) lead to a collision within 200 Myr while the 3 other cases do not.}
\label{fig:secres}

\end{figure}
}

% Bibliographic references with the natbib package:
% Parenthetical: \citep{Bai92} produces (Bailyn 1992).
% Textual: \citet{Bai95} produces Bailyn et al. (1995).
% An affix and part of a reference:
%   \citep[e.g.][Ch. 2]{Bar76}
%   produces (e.g. Barnes et al. 1976, Ch. 2).-

\bibliography{bibliography}

%% Use the plainnat style for ``Icarus'' mode to display DOI numbers
%% among other things.  However, revert to the Elsevier elsart-harv
%% mode for ``Elsevier'' mode.
\bibliographystyle{plainnat}

\beginsupplement

\clearpage	% Make sure things don't run together.
\begin{landscape}
{
\renewcommand{\baselinestretch}{1}
\small\normalsize
\begin{table}
\small
\centering

\begin{tabular}{|r|cccccccccccr|}
\hline 
\hline
&$a_S$  & $r_{col}$	& $v_{esc}$	& $b_{col}/r_{col}$ & $v_{col}/v_{esc}$	&    $b_{\infty}/r_{col}$ & $v_{\infty}/v_{esc}$ & $L_{col}$ & $m_{merged}$ & $P_{rot}$ & $\phi$ & $t_{col}$\\
&(AU) & (km) & (km s$^{-1}$) & & & & & & ($M_\oplus$) & (hr) & ($^\circ$) & (Myr)\\
\hline \parbox[t]{2mm}{\multirow{8}{*}{\rotatebox[origin=c]{90}{\textbf{Circular}}}}
&0.76	&	11386	&	9.6	&	0.985	&	1.034	&	3.879	&	0.262	&	7.282	&	1.321	&	1.374	&	88.35	&	0.56972	\\
&0.77	&	11386	&	9.6	&	0.53	&	1.007	&	4.551	&	0.117	&	3.816	&	1.321	&	2.622	&	87.80	&	0.09300	\\
&0.78	&	11386	&	9.6	&	0.941	&	1.011	&	6.370	&	0.149	&	6.806	&	1.321	&	1.470	&	151.52	&	10.85945	\\
&0.87	&	11386	&	9.6	&	0.658	&	1.004	&	7.511	&	0.088	&	4.727	&	1.321	&	2.117	&	114.92	&	37.70558	\\
&0.98	&	10666	&	8.7	&	0.976	&	1.000	&	91.061	&	0.011	&	4.796	&	1.012	&	1.392	&	101.93	&	0.00453	\\
&0.99	&	10666	&	8.7	&	0.653	&	0.998	&	--	&	--	&	3.203	&	1.012	&	2.084	&	54.37	&	0.00087	\\
&1.01	&	10666	&	8.7	&	0.121	&	0.998	&	--	&	--	&	0.592	&	1.012	&	11.271	&	1.44	&	0.00002	\\
&1.02	&	10666	&	8.7	&	0.734	&	0.996	&	--	&	--	&	3.591	&	1.012	&	1.859	&	59.17	&	0.07565	\\

%\hline
%\end{tabular}
%\caption[Collision properties for simulations with $M_L:M_S = 1:1$, $e_L=i_L=0^\circ$.]	% The class file doesn't do 
										%% anything with the square 
										%% bracketed short caption, 
										%% but I always put one in.
	%{
	%Collision properties for simulations with $M_L:M_S = 1:1$, $e_L=i_L=0^\circ$.	
	%%\label{lasttable}		% notice the second label for counting
	%}
	%\label{tab:Col11All0}
%
%\end{table}
%}
%\end{landscape}
%
%
%\begin{landscape}
%{
%\renewcommand{\baselinestretch}{1}
%\small\normalsize
%\begin{table}
%\small
%\centering
%
%\begin{tabular}{|r|cccccccccccr}
%\hline 
%\hline
%&$a_S$  & $r_{col}$	& $v_{esc}$	& $b_{col}/r_{col}$ & $v_{col}/v_{esc}$	&    $b_{\infty}/r_{col}$ & $v_{\infty}/v_{esc}$ & $L_{col}$ & $m_{merged}$ & $P_{rot}$ & $\phi$ & $t_{col}$\\
%&(AU) & (km) & (km s$^{-1}$) & & & & & & ($M_\oplus$) & (hr) & ($^\circ$) & (Myr)\\
\hline \parbox[t]{2mm}{\multirow{21}{*}{\rotatebox[origin=c]{90}{\textbf{Eccentric}}}}
&0.76	&	11386	&	9.6	&	0.203	&	1.016	&	1.154	&	0.178	&	1.472	&	1.321	&	6.795	&	27.80	&	0.52157	\\
&0.77	&	8724	&	7.5	&	0.706	&	1.057	&	2.177	&	0.343	&	0.904	&	0.613	&	3.285	&	164.94	&	1.06842	\\
&0.78	&	11386	&	9.6	&	0.762	&	1.033	&	3.021	&	0.261	&	5.631	&	1.321	&	1.777	&	139.25	&	0.71870	\\
&0.79	&	8724	&	7.5	&	0.154	&	1.119	&	0.342	&	0.503	&	0.208	&	0.613	&	14.266	&	154.10	&	4.99831	\\
&0.80	&	11386	&	9.6	&	0.855	&	1.030	&	3.589	&	0.245	&	6.298	&	1.321	&	1.589	&	120.87	&	8.64452	\\
&0.81	&	11386	&	9.6	&	0.294	&	1.027	&	1.285	&	0.235	&	2.161	&	1.321	&	4.630	&	120.83	&	4.99254	\\
&0.84	&	11386	&	9.6	&	0.204	&	1.017	&	1.110	&	0.187	&	1.482	&	1.321	&	6.751	&	26.26	&	116.58322	\\
&0.92	&	10666	&	8.7	&	0.638	&	1.179	&	1.203	&	0.625	&	3.692	&	1.012	&	1.808	&	11.46	&	89.76271	\\
&0.93	&	11386	&	9.6	&	0.978	&	1.130	&	2.102	&	0.526	&	7.906	&	1.321	&	1.266	&	105.76	&	7.34083	\\
&0.94	&	11386	&	9.6	&	0.847	&	1.087	&	2.159	&	0.427	&	6.588	&	1.321	&	1.519	&	58.43	&	13.81439	\\
&0.95	&	11386	&	9.6	&	0.474	&	1.042	&	1.689	&	0.292	&	3.530	&	1.321	&	2.835	&	100.75	&	7.06466	\\
&0.96	&	10666	&	8.7	&	0.680	&	0.997	&	--	&	--	&	3.328	&	1.012	&	2.006	&	15.95	&	0.01935	\\
&0.97	&	10666	&	8.7	&	0.649	&	1.000	&	35.721	&	0.018	&	3.187	&	1.012	&	2.094	&	105.17	&	0.21892	\\
&0.98	&	10666	&	8.7	&	0.761	&	1.015	&	4.398	&	0.176	&	3.795	&	1.012	&	1.758	&	54.11	&	6.08506	\\
&0.99	&	8724	&	7.5	&	0.904	&	1.413	&	1.280	&	0.998	&	1.547	&	0.613	&	1.919	&	112.49	&	5.38715	\\
&1.01	&	10666	&	8.7	&	0.379	&	1.017	&	2.112	&	0.183	&	1.894	&	1.012	&	3.524	&	108.00	&	0.01463	\\
&1.02	&	10666	&	8.7	&	0.550	&	1.006	&	4.915	&	0.113	&	2.717	&	1.012	&	2.456	&	128.27	&	0.04161	\\
&1.03	&	11386	&	9.6	&	0.508	&	0.998	&	--	&	--	&	3.628	&	1.321	&	2.758	&	34.68	&	0.28982	\\
&1.04	&	10666	&	8.7	&	0.883	&	1.010	&	6.173	&	0.145	&	4.383	&	1.012	&	1.523	&	80.15	&	1.33044	\\
&1.05	&	11386	&	9.6	&	0.716	&	1.021	&	3.560	&	0.205	&	5.230	&	1.321	&	1.913	&	40.20	&	2.65015	\\
&1.06	&	11386	&	9.6	&	0.798	&	1.082	&	2.089	&	0.413	&	6.176	&	1.321	&	1.620	&	125.35	&	2.64230	\\
&1.07	&	10666	&	8.7	&	0.746	&	1.009	&	5.468	&	0.138	&	3.697	&	1.012	&	1.805	&	95.83	&	7.31536	\\

\hline
\end{tabular}
\caption[Collision properties for simulations with $M_L/M_S = 1/1$, $e_L=e_S = 0.05$, $i_L={2 \over 3}^\circ$.]	% The class file doesn't do 
										% anything with the square 
										% bracketed short caption, 
										% but I always put one in.
	{
	%Collision properties for simulations with $M_L:M_S = 1:1$, $e_L=e_S = 0.05$, $i_L={2 \over 3}^\circ$.
	Collision properties for simulations with $m_L/m_S = 1/1$ for the circular ($e_S = 0.0$, $i_S=0.0$) and eccentric ($e_S=0.05$, $i_S=2/3^\circ$) cases whose ranges of inquiry have been provided in Table \ref{tab:IC2} (QL1).  The starting semimajor axis ($a_S$) of the proto-Moon is provided to connect with the collisional outcomes as defined by the two-body collision radius ($r_{col}$), escape velocity ($v_{esc}$), scaled collision parameters ($b_{col}/r_{col}$ and $b_{\infty}/r_{col}$), scaled collision velocities ($v_{col}/r_{col}$ and $v_{\infty}/r_{col}$), spin angular momentum ($L_{col}$), merged mass ($m_{merged}$), rotational period ($P_{rot}$), obliquity ($\phi$), and collision epoch ($t_{col}$).
	%\label{lasttable}		% notice the second label for counting
	}
	%\label{tab:Col11Ecc05Inc23}
	\label{tab:Col11All0}

\end{table}
}
\end{landscape}

\clearpage	% Make sure things don't run together.
\begin{landscape}
{
\renewcommand{\baselinestretch}{0.75}
\small\normalsize
\begin{table}
\small
\centering

\begin{tabular}{|r|cccccccccccr|}
\hline 
\hline
&$a_S$  & $r_{col}$	& $v_{esc}$	& $b_{col}/r_{col}$ & $v_{col}/v_{esc}$	&    $b_{\infty}/r_{col}$ & $v_{\infty}/v_{esc}$ & $L_{col}$ & $m_{merged}$ & $P_{rot}$ & $\phi$ & $t_{col}$\\
&(AU) & (km) & (km s$^{-1}$) & & & & & & ($M_\oplus$) & (hr) & ($^\circ$) & (Myr)\\
\hline \parbox[t]{2mm}{\multirow{14}{*}{\rotatebox[origin=c]{90}{\textbf{Circular}}}}
&0.76	&	10200	&	8.9	&	0.887	&	1.015	&	5.145	&	0.175	&	2.779	&	1.017	&	2.359	&	122.72	&	0.05826	\\
&0.78	&	10200	&	8.9	&	0.904	&	1.183	&	1.692	&	0.633	&	3.302	&	1.017	&	1.986	&	37.20	&	49.50380	\\
&0.96	&	10101	&	8.9	&	0.720	&	0.996	&	--	&	--	&	2.192	&	1.012	&	2.903	&	148.35	&	1.69996	\\
&0.98	&	10101	&	8.9	&	0.781	&	0.997	&	--	&	--	&	2.382	&	1.012	&	2.671	&	24.75	&	0.01810	\\
&1.02	&	10101	&	8.9	&	0.852	&	0.998	&	--	&	--	&	2.599	&	1.012	&	2.448	&	37.25	&	0.00137	\\
&1.04	&	10101	&	8.9	&	0.898	&	1.132	&	1.918	&	0.530	&	3.107	&	1.012	&	2.048	&	117.31	&	80.89603	\\
&1.22	&	7538	&	5.7	&	0.979	&	1.252	&	1.628	&	0.753	&	0.777	&	0.310	&	1.340	&	87.31	&	113.01316	\\
&1.26	&	7538	&	5.7	&	0.621	&	1.035	&	2.421	&	0.265	&	0.407	&	0.310	&	2.556	&	87.94	&	70.51328	\\
&1.28	&	7538	&	5.7	&	0.745	&	1.019	&	3.847	&	0.197	&	0.481	&	0.310	&	2.164	&	79.83	&	133.32056	\\
&1.30	&	7538	&	5.7	&	0.472	&	1.099	&	1.138	&	0.456	&	0.329	&	0.310	&	3.165	&	86.38	&	196.87659	\\
&1.34	&	7538	&	5.7	&	0.862	&	1.080	&	2.287	&	0.407	&	0.589	&	0.310	&	1.766	&	75.40	&	14.09427	\\
&1.36	&	10101	&	8.9	&	0.917	&	1.347	&	1.369	&	0.902	&	3.779	&	1.012	&	1.684	&	71.35	&	51.08611	\\
&1.38	&	7538	&	5.7	&	0.703	&	1.646	&	0.885	&	1.308	&	0.734	&	0.310	&	1.419	&	126.76	&	46.28822	\\
&1.48	&	7538	&	5.7	&	0.540	&	1.051	&	1.751	&	0.324	&	0.360	&	0.310	&	2.893	&	106.14	&	1.69104	\\

%\hline
%\end{tabular}
%\caption[Collision properties for simulations with $M_L:M_S = 4:1$, $e_L=i_L=0^\circ$.]	% The class file doesn't do 
										%% anything with the square 
										%% bracketed short caption, 
										%% but I always put one in.
	%{
	%Collision properties for simulations with $M_L:M_S = 4:1$, $e_L=i_L=0^\circ$.	
	%%\label{lasttable}		% notice the second label for counting
	%}
	%\label{tab:Col41All0}
%
%\end{table}
%}
%\end{landscape}
%
%\clearpage	% Make sure things don't run together.
%\begin{landscape}
%{
%\renewcommand{\baselinestretch}{1}
%\small\normalsize
%\begin{table}
%\small
%\centering
%
%\begin{tabular}{cccccccccccr}
%\hline 
%\hline
%$a_S$  & $r_{col}$	& $v_{esc}$	& $b_{col}/r_{col}$ & $v_{col}/v_{esc}$	&    $b_{\infty}/r_{col}$ & $v_{\infty}/v_{esc}$ & $L_{col}$ & $m_{merged}$ & $P_{rot}$ & $\phi$ & $t_{col}$\\
%(AU) & (km) & (km s$^{-1}$) & & & & & & ($M_\oplus$) & (hr) & ($^\circ$) & (Myr)\\
\hline \parbox[t]{2mm}{\multirow{35}{*}{\rotatebox[origin=c]{90}{\textbf{Eccentric}}}}
&0.76	&	9445	&	8.8	&	0.975	&	1.100	&	2.339	&	0.459	&	1.776	&	0.922	&	3.096	&	56.93	&	4.34370	\\
&0.78	&	10200	&	8.9	&	0.952	&	1.098	&	2.309	&	0.453	&	3.225	&	1.017	&	2.033	&	102.58	&	3.56843	\\
&0.80	&	9346	&	8.8	&	0.785	&	1.133	&	1.672	&	0.532	&	1.459	&	0.917	&	3.646	&	28.79	&	5.10044	\\
&0.82	&	10200	&	8.9	&	0.465	&	1.434	&	0.649	&	1.028	&	2.058	&	1.017	&	3.186	&	101.25	&	26.91425	\\
&0.88	&	10101	&	8.9	&	0.456	&	1.128	&	0.985	&	0.522	&	1.574	&	1.012	&	4.044	&	45.93	&	1.26131	\\
&0.90	&	10200	&	8.9	&	0.958	&	1.237	&	1.627	&	0.728	&	3.656	&	1.017	&	1.793	&	62.66	&	7.60210	\\
&0.92	&	10200	&	8.9	&	0.071	&	1.006	&	0.653	&	0.110	&	0.221	&	1.017	&	29.716	&	113.26	&	15.04478	\\
&0.94	&	10101	&	8.9	&	0.452	&	1.053	&	1.441	&	0.330	&	1.457	&	1.012	&	4.368	&	87.53	&	0.06394	\\
&0.96	&	10200	&	8.9	&	0.955	&	1.047	&	3.223	&	0.310	&	3.087	&	1.017	&	2.124	&	92.94	&	0.57139	\\
&0.98	&	10200	&	8.9	&	0.954	&	1.096	&	2.334	&	0.448	&	3.224	&	1.017	&	2.033	&	83.50	&	6.28991	\\
&1.00	&	10101	&	8.9	&	0.143	&	1.099	&	0.344	&	0.456	&	0.479	&	1.012	&	13.274	&	77.61	&	43.11912	\\
&1.02	&	9445	&	8.8	&	0.378	&	1.953	&	0.440	&	1.678	&	1.222	&	0.922	&	4.498	&	163.18	&	9.23646	\\
&1.04	&	10200	&	8.9	&	0.496	&	1.120	&	1.101	&	0.504	&	1.713	&	1.017	&	3.828	&	24.55	&	2.18081	\\
&1.06	&	10101	&	8.9	&	0.353	&	1.063	&	1.039	&	0.361	&	1.148	&	1.012	&	5.543	&	163.74	&	1.76777	\\
&1.08	&	12008	&	10.4	&	0.840	&	1.195	&	1.534	&	0.655	&	10.642	&	1.624	&	1.276	&	80.01	&	37.42025	\\
&1.10	&	10200	&	8.9	&	0.119	&	1.048	&	0.396	&	0.314	&	0.384	&	1.017	&	17.081	&	86.76	&	3.38930	\\
&1.12	&	10101	&	8.9	&	0.961	&	0.997	&	--	&	--	&	2.933	&	1.012	&	2.170	&	91.57	&	0.07625	\\
&1.14	&	9346	&	8.8	&	0.627	&	1.196	&	1.142	&	0.657	&	1.230	&	0.917	&	4.322	&	47.44	&	36.87121	\\
&1.16	&	6586	&	5.6	&	0.524	&	2.509	&	0.571	&	2.301	&	0.439	&	0.258	&	1.667	&	117.16	&	54.57086	\\
&1.18	&	10101	&	8.9	&	0.385	&	1.271	&	0.623	&	0.785	&	1.496	&	1.012	&	4.254	&	41.64	&	26.76659	\\
&1.20	&	10200	&	8.9	&	0.175	&	1.094	&	0.430	&	0.444	&	0.589	&	1.017	&	11.128	&	113.61	&	45.97811	\\
&1.22	&	9346	&	8.8	&	0.932	&	1.100	&	2.241	&	0.458	&	1.682	&	0.917	&	3.162	&	112.93	&	67.77095	\\
&1.24	&	12008	&	10.4	&	0.275	&	1.254	&	0.457	&	0.756	&	3.660	&	1.624	&	3.710	&	16.99	&	67.13215	\\
&1.26	&	10101	&	8.9	&	0.421	&	1.124	&	0.924	&	0.512	&	1.448	&	1.012	&	4.394	&	144.40	&	101.71220	\\
&1.28	&	7538	&	5.7	&	0.409	&	1.150	&	0.829	&	0.567	&	0.298	&	0.310	&	3.493	&	123.47	&	25.20295	\\
&1.30	&	7538	&	5.7	&	0.733	&	1.213	&	1.295	&	0.686	&	0.563	&	0.310	&	1.849	&	64.79	&	1.51370	\\
&1.32	&	12008	&	10.4	&	0.609	&	1.081	&	1.607	&	0.410	&	6.977	&	1.624	&	1.946	&	148.51	&	85.48865	\\
&1.36	&	7538	&	5.7	&	0.371	&	1.261	&	0.610	&	0.768	&	0.297	&	0.310	&	3.509	&	91.87	&	2.72535	\\
&1.38	&	9445	&	8.8	&	0.634	&	1.087	&	1.617	&	0.426	&	1.141	&	0.922	&	4.820	&	91.53	&	15.60265	\\
&1.40	&	12008	&	10.4	&	0.990	&	1.021	&	4.919	&	0.205	&	10.706	&	1.624	&	1.268	&	31.40	&	40.90143	\\
&1.44	&	12008	&	10.4	&	0.485	&	1.046	&	1.662	&	0.305	&	5.377	&	1.624	&	2.526	&	56.37	&	121.01339	\\
&1.46	&	7538	&	5.7	&	0.300	&	1.000	&	9.942	&	0.030	&	0.190	&	0.310	&	5.478	&	46.64	&	180.02197	\\
&1.48	&	9346	&	8.8	&	0.803	&	1.210	&	1.427	&	0.680	&	1.593	&	0.917	&	3.339	&	98.78	&	7.35633	\\
&1.50	&	7538	&	5.7	&	0.935	&	1.003	&	11.840	&	0.079	&	0.594	&	0.310	&	1.751	&	47.50	&	43.54981	\\
&1.54	&	12008	&	10.4	&	0.255	&	1.045	&	0.882	&	0.302	&	2.819	&	1.624	&	4.817	&	58.08	&	73.64233	\\

\hline
\end{tabular}
\caption[Collision properties for simulations with $M_L:M_S = 4/1$, $e_L=0.05$, $i_L={2 \over 3}^\circ$.]	% The class file doesn't do 
										% anything with the square 
										% bracketed short caption, 
										% but I always put one in.
	{
	Collision properties for simulations with $m_L/m_S = 4/1$ for the circular ($e_S = 0.0$, $i_S=0.0$) and eccentric ($e_S=0.10$, $i_S=2/3^\circ$) cases whose ranges of inquiry have been provided in Table \ref{tab:IC2} (QL4).  See Table \ref{tab:Col11All0} for symbol definitions.
	%\label{lasttable}		% notice the second label for counting
	}
	\label{tab:Col41Ecc05}

\end{table}
}
\end{landscape}

\clearpage	% Make sure things don't run together.
\begin{landscape}

{
\renewcommand{\baselinestretch}{1}
\small\normalsize
\begin{table}
\small
\centering

\begin{tabular}{|r|cccccccccccr|}
\hline 
\hline
&$a_S$  & $r_{col}$	& $v_{esc}$	& $b_{col}/r_{col}$ & $v_{col}/v_{esc}$	&    $b_{\infty}/r_{col}$ & $v_{\infty}/v_{esc}$ & $L_{col}$ & $m_{merged}$ & $P_{rot}$ & $\phi$ & $t_{col}$\\
&(AU) & (km) & (km s$^{-1}$) & & & & & & ($M_\oplus$) & (hr) & ($^\circ$) & (Myr)\\
\hline \parbox[t]{2mm}{\multirow{21}{*}{\rotatebox[origin=c]{90}{\textbf{Circular}}}}
&0.76	&	9462	&	8.8	&	0.173	&	1.006	&	1.539	&	0.113	&	0.301	&	0.927	&	18.419	&	104.17	&	0.03712	\\
&0.78	&	9462	&	8.8	&	0.828	&	1.204	&	1.486	&	0.671	&	1.724	&	0.927	&	3.212	&	144.44	&	26.96370	\\
&0.86	&	9577	&	9.2	&	0.452	&	1.093	&	1.118	&	0.442	&	0.909	&	1.012	&	6.864	&	113.62	&	4.83383	\\
&0.92	&	9577	&	9.2	&	0.581	&	1.001	&	15.393	&	0.038	&	1.069	&	1.012	&	5.835	&	120.47	&	3.99954	\\
&0.96	&	9577	&	9.2	&	0.978	&	0.998	&	--	&	--	&	1.794	&	1.012	&	3.479	&	76.71	&	0.28256	\\
&0.98	&	9577	&	9.2	&	0.730	&	0.998	&	--	&	--	&	1.340	&	1.012	&	4.658	&	138.95	&	0.01285	\\
&1.02	&	9577	&	9.2	&	0.359	&	0.997	&	--	&	--	&	0.659	&	1.012	&	9.468	&	165.43	&	0.00421	\\
&1.04	&	9577	&	9.2	&	0.233	&	0.998	&	--	&	--	&	0.427	&	1.012	&	14.613	&	79.07	&	2.21106	\\
&1.20	&	9577	&	9.2	&	0.450	&	1.088	&	1.145	&	0.428	&	0.901	&	1.012	&	6.928	&	128.88	&	40.50028	\\
&1.30	&	9577	&	9.2	&	0.511	&	1.098	&	1.239	&	0.452	&	1.031	&	1.012	&	6.055	&	99.00	&	138.95936	\\
&1.32	&	6800	&	5.1	&	0.724	&	1.090	&	1.821	&	0.434	&	0.313	&	0.220	&	1.880	&	86.78	&	85.53283	\\
&1.34	&	6800	&	5.1	&	0.842	&	1.747	&	1.027	&	1.432	&	0.584	&	0.220	&	1.009	&	173.46	&	77.67254	\\
&1.36	&	9462	&	8.8	&	0.834	&	1.341	&	1.252	&	0.893	&	1.934	&	0.927	&	2.864	&	78.32	&	63.20767	\\
&1.38	&	6800	&	5.1	&	0.752	&	1.076	&	2.038	&	0.397	&	0.321	&	0.220	&	1.836	&	64.53	&	3.62907	\\
&1.40	&	6800	&	5.1	&	0.882	&	2.488	&	0.964	&	2.278	&	0.871	&	0.220	&	0.676	&	34.28	&	55.21185	\\
&1.44	&	6800	&	5.1	&	0.645	&	1.151	&	1.301	&	0.570	&	0.294	&	0.220	&	2.001	&	52.16	&	30.93605	\\
&1.46	&	9559	&	9.2	&	0.943	&	1.183	&	1.765	&	0.632	&	1.962	&	1.007	&	3.159	&	43.87	&	36.21679	\\
&1.48	&	9559	&	9.2	&	0.080	&	1.218	&	0.140	&	0.695	&	0.171	&	1.007	&	36.322	&	85.75	&	164.48464	\\
&1.50	&	9577	&	9.2	&	0.734	&	1.165	&	1.431	&	0.598	&	1.572	&	1.012	&	3.970	&	99.86	&	29.25898	\\
&1.52	&	12221	&	10.6	&	0.476	&	1.028	&	2.066	&	0.237	&	5.659	&	1.714	&	2.623	&	69.18	&	145.52244	\\
&1.54	&	9577	&	9.2	&	0.909	&	1.145	&	1.868	&	0.557	&	1.912	&	1.012	&	3.263	&	105.17	&	84.74566	\\

%\hline
%\end{tabular}
%\caption[Collision properties for simulations with $M_L:M_S = 8:1$, $e_L=i_L=0^\circ$.]	% The class file doesn't do 
										%% anything with the square 
										%% bracketed short caption, 
										%% but I always put one in.
	%{
	%Collision properties for simulations with $M_L:M_S = 8:1$, $e_L=i_L=0^\circ$.	
	%%\label{lasttable}		% notice the second label for counting
	%}
	%\label{tab:Col81All0}
%
%\end{table}
%}
%\end{landscape}
%
%\clearpage	% Make sure things don't run together.
%\begin{landscape}
%
%{
%\renewcommand{\baselinestretch}{1}
%\small\normalsize
%\begin{table}
%\small
%\centering
%
%\begin{tabular}{cccccccccccr}
%\hline 
%\hline
%$a_S$  & $r_{col}$	& $v_{esc}$	& $b_{col}/r_{col}$ & $v_{col}/v_{esc}$	&    $b_{\infty}/r_{col}$ & $v_{\infty}/v_{esc}$ & $L_{col}$ & $m_{merged}$ & $P_{rot}$ & $\phi$ & $t_{col}$\\
%(AU) & (km) & (km s$^{-1}$) & & & & & & ($M_\oplus$) & (hr) & ($^\circ$) & (Myr)\\
\hline \parbox[t]{2mm}{\multirow{15}{*}{\rotatebox[origin=c]{90}{\textbf{Inclined}}}}
&0.78	&	9577	&	9.2	&	0.781	&	1.011	&	5.315	&	0.149	&	1.452	&	1.012	&	4.298	&	44.63	&	34.54802	\\
&0.86	&	9577	&	9.2	&	0.239	&	1.104	&	0.563	&	0.468	&	0.485	&	1.012	&	12.872	&	78.45	&	8.76137	\\
&0.96	&	9462	&	8.8	&	0.988	&	1.284	&	1.576	&	0.805	&	2.194	&	0.927	&	2.524	&	66.54	&	2.73556	\\
&0.98	&	9577	&	9.2	&	0.359	&	0.998	&	--	&	--	&	0.659	&	1.012	&	9.472	&	41.97	&	0.03136	\\
&1.02	&	6800	&	5.1	&	0.790	&	1.514	&	1.052	&	1.137	&	0.475	&	0.220	&	1.241	&	33.42	&	13.74298	\\
&1.04	&	9577	&	9.2	&	0.994	&	1.334	&	1.501	&	0.884	&	2.439	&	1.012	&	2.558	&	120.03	&	18.48778	\\
&1.20	&	9577	&	9.2	&	0.464	&	1.046	&	1.574	&	0.308	&	0.892	&	1.012	&	6.993	&	159.78	&	149.72290	\\
&1.30	&	9462	&	8.8	&	0.614	&	1.017	&	3.346	&	0.187	&	1.080	&	0.927	&	5.126	&	85.77	&	101.67901	\\
&1.32	&	6800	&	5.1	&	0.695	&	1.208	&	1.238	&	0.678	&	0.333	&	0.220	&	1.769	&	47.36	&	30.25171	\\
&1.34	&	6800	&	5.1	&	0.886	&	1.210	&	1.574	&	0.681	&	0.426	&	0.220	&	1.385	&	132.49	&	8.75704	\\
&1.36	&	6800	&	5.1	&	0.894	&	1.026	&	3.967	&	0.231	&	0.364	&	0.220	&	1.619	&	92.58	&	6.79075	\\
&1.40	&	6800	&	5.1	&	0.905	&	1.152	&	1.822	&	0.572	&	0.414	&	0.220	&	1.423	&	111.90	&	23.51694	\\
&1.44	&	9577	&	9.2	&	0.896	&	1.284	&	1.428	&	0.805	&	2.114	&	1.012	&	2.951	&	67.55	&	72.08816	\\
&1.46	&	6800	&	5.1	&	0.100	&	1.150	&	0.203	&	0.567	&	0.046	&	0.220	&	12.911	&	89.03	&	34.43338	\\
&1.48	&	9577	&	9.2	&	0.168	&	1.080	&	0.443	&	0.408	&	0.333	&	1.012	&	18.754	&	64.76	&	143.02573	\\

\hline
\end{tabular}
\caption[Collision properties for simulations with $M_L:M_S = 8/1$, $e_L=0.0$, $i_L={2 \over 3}^\circ$.]	% The class file doesn't do 
										% anything with the square 
										% bracketed short caption, 
										% but I always put one in.
	{
	Collision properties for simulations with $m_L/m_S = 8/1$ for the circular ($e_S = 0.0$, $i_S=0.0$) and inclined ($e_S=0.00$, $i_S=2/3^\circ$) cases whose ranges of inquiry have been provided in Table \ref{tab:IC2} (QL8).  See Table \ref{tab:Col11All0} for symbol definitions.	
	%\label{lasttable}		% notice the second label for counting
	}
	\label{tab:Col81Inc23}

\end{table}
}
\end{landscape}

\clearpage	% Make sure things don't run together.
\begin{landscape}

{
\renewcommand{\baselinestretch}{1}
\small\normalsize
\begin{table}
\small
\centering

\begin{tabular}{|r|cccccccccccr|}
\hline 
\hline
&$a_S$  & $r_{col}$	& $v_{esc}$	& $b_{col}/r_{col}$ & $v_{col}/v_{esc}$	&    $b_{\infty}/r_{col}$ & $v_{\infty}/v_{esc}$ & $L_{col}$ & $m_{merged}$ & $P_{rot}$ & $\phi$ & $t_{col}$\\
&(AU) & (km) & (km s$^{-1}$) & & & & & & ($M_\oplus$) & (hr) & ($^\circ$) & (Myr)\\
\hline \parbox[t]{2mm}{\multirow{33}{*}{\rotatebox[origin=c]{90}{\textbf{Inclined}}}}
&0.76	&	9462	&	8.8	&	0.601	&	1.319	&	0.921	&	0.861	&	1.371	&	0.927	&	4.039	&	28.39	&	38.70518	\\
&0.78	&	9462	&	8.8	&	0.703	&	1.112	&	1.611	&	0.485	&	1.352	&	0.927	&	4.097	&	108.18	&	1.13417	\\
&0.80	&	9577	&	9.2	&	0.955	&	1.029	&	4.064	&	0.242	&	1.805	&	1.012	&	3.456	&	73.93	&	6.90832	\\
&0.82	&	9462	&	8.8	&	0.621	&	1.512	&	0.827	&	1.134	&	1.623	&	0.927	&	3.413	&	86.34	&	20.93428	\\
&0.84	&	12221	&	10.6	&	0.774	&	1.127	&	1.676	&	0.521	&	10.090	&	1.714	&	1.471	&	92.26	&	59.53256	\\
&0.86	&	9559	&	9.2	&	0.420	&	1.450	&	0.580	&	1.050	&	1.071	&	1.007	&	5.790	&	132.25	&	14.30560	\\
&0.88	&	9462	&	8.8	&	0.603	&	1.028	&	2.586	&	0.240	&	1.073	&	0.927	&	5.161	&	53.38	&	0.71785	\\
&0.90	&	9577	&	9.2	&	0.503	&	1.043	&	1.780	&	0.295	&	0.965	&	1.012	&	6.466	&	8.63	&	0.02739	\\
&0.92	&	9577	&	9.2	&	0.725	&	1.207	&	1.295	&	0.676	&	1.609	&	1.012	&	3.878	&	97.82	&	0.80805	\\
&0.94	&	9462	&	8.8	&	0.775	&	1.629	&	0.981	&	1.286	&	2.183	&	0.927	&	2.537	&	60.94	&	5.39288	\\
&0.96	&	9577	&	9.2	&	0.734	&	1.255	&	1.215	&	0.759	&	1.695	&	1.012	&	3.682	&	158.44	&	6.02780	\\
&0.98	&	9577	&	9.2	&	0.887	&	1.096	&	2.163	&	0.449	&	1.787	&	1.012	&	3.491	&	124.26	&	0.02423	\\
&1.00	&	9559	&	9.2	&	0.804	&	1.560	&	1.047	&	1.197	&	2.205	&	1.007	&	2.811	&	44.94	&	8.69315	\\
&1.04	&	9462	&	8.8	&	0.994	&	1.755	&	1.209	&	1.443	&	3.017	&	0.927	&	1.836	&	103.67	&	4.43843	\\
&1.06	&	9577	&	9.2	&	0.945	&	1.082	&	2.481	&	0.412	&	1.879	&	1.012	&	3.320	&	161.87	&	25.50972	\\
&1.08	&	9462	&	8.8	&	0.408	&	1.384	&	0.590	&	0.957	&	0.975	&	0.927	&	5.678	&	92.13	&	29.74847	\\
&1.10	&	9462	&	8.8	&	0.790	&	1.272	&	1.279	&	0.786	&	1.738	&	0.927	&	3.186	&	73.81	&	1.25386	\\
&1.12	&	9462	&	8.8	&	0.424	&	1.884	&	0.500	&	1.597	&	1.382	&	0.927	&	4.008	&	95.27	&	7.56787	\\
&1.14	&	9577	&	9.2	&	0.279	&	1.044	&	0.967	&	0.301	&	0.535	&	1.012	&	11.655	&	116.46	&	0.11474	\\
&1.16	&	9445	&	8.8	&	0.763	&	1.071	&	2.133	&	0.383	&	1.352	&	0.922	&	4.068	&	97.54	&	17.02054	\\
&1.18	&	9445	&	8.8	&	0.575	&	1.634	&	0.727	&	1.293	&	1.555	&	0.922	&	3.537	&	49.00	&	15.81986	\\
&1.20	&	9462	&	8.8	&	0.860	&	1.011	&	5.951	&	0.146	&	1.502	&	0.927	&	3.687	&	50.28	&	2.71507	\\
&1.22	&	9577	&	9.2	&	0.915	&	1.052	&	2.935	&	0.328	&	1.770	&	1.012	&	3.525	&	44.27	&	20.94605	\\
&1.28	&	9462	&	8.8	&	0.978	&	1.742	&	1.194	&	1.427	&	2.946	&	0.927	&	1.880	&	114.53	&	47.74907	\\
&1.32	&	9577	&	9.2	&	0.267	&	1.139	&	0.557	&	0.546	&	0.559	&	1.012	&	11.162	&	68.75	&	7.49073	\\
&1.34	&	6800	&	5.1	&	0.161	&	1.278	&	0.259	&	0.797	&	0.082	&	0.220	&	7.201	&	99.65	&	15.33821	\\
&1.36	&	6800	&	5.1	&	0.798	&	1.183	&	1.492	&	0.633	&	0.375	&	0.220	&	1.573	&	71.71	&	196.57797	\\
&1.38	&	9462	&	8.8	&	0.900	&	1.348	&	1.342	&	0.904	&	2.097	&	0.927	&	2.641	&	83.79	&	92.97132	\\
&1.40	&	9559	&	9.2	&	0.589	&	1.058	&	1.808	&	0.345	&	1.096	&	1.007	&	5.659	&	71.04	&	26.38767	\\
&1.44	&	12221	&	10.6	&	0.655	&	1.347	&	0.977	&	0.903	&	10.207	&	1.714	&	1.454	&	119.30	&	123.48015	\\
&1.46	&	6800	&	5.1	&	0.206	&	1.182	&	0.387	&	0.629	&	0.097	&	0.220	&	6.097	&	84.29	&	14.39429	\\
&1.48	&	6800	&	5.1	&	0.787	&	1.348	&	1.174	&	0.904	&	0.421	&	0.220	&	1.399	&	91.21	&	14.25047	\\
&1.50	&	12221	&	10.6	&	0.877	&	0.997	&	--	&	--	&	10.111	&	1.714	&	1.468	&	94.92	&	139.60955	\\
&1.54	&	9462	&	8.8	&	0.272	&	1.069	&	0.771	&	0.377	&	0.502	&	0.927	&	11.021	&	62.65	&	156.18477	\\

\hline
\end{tabular}
\caption[Collision properties for simulations with $m_L:m_S = 8/1$, $e_L=0.05$, $e_S \approx 0.14$, $i_L={2 \over 3}^\circ$.]	% The class file doesn't do 
										% anything with the square 
										% bracketed short caption, 
										% but I always put one in.
	{
	Collision properties for simulations with $m_L/m_S = 8/1$ for the eccentric, inclined ($e_L=0.05$, $e_S\approx0.14$, $i_S=2/3^\circ$) case whose starting conditions where provided in \cite{Rivera2002} but results where not shown.  See Table \ref{tab:Col11All0} for symbol definitions.
	%\label{lasttable}		% notice the second label for counting
	}
	\label{tab:Col81Ecc05Inc23}

\end{table}
}
\end{landscape}

\end{document}